%% file: main.tex
\RequirePackage{fix-cm}

\documentclass[twocolumn]{svjour3}

\pdfoutput=1

\usepackage{hyperref}
\hypersetup{breaklinks}

\usepackage{placeins}
\usepackage{listings}
\usepackage{xcolor}
\usepackage{xspace}
\usepackage{subcaption}

\definecolor{code-gray}{gray}{0.95}
\usepackage[frozencache,cachedir=.,
  newfloat
]{minted} 
\setminted{
  bgcolor=code-gray,
  samepage=true,
  autogobble,
}

\AtBeginDocument{%

}

\usepackage{tikz}
\usetikzlibrary{calc,positioning,decorations.text,backgrounds,shapes.geometric,positioning,fit}
\usepackage{etoolbox}
\usepackage{standalone}

\usepackage[
    backend=biber,
    doi=true,
    sorting=nyt,
    giveninits=true,
]{biblatex}
\DeclareSourcemap{
 \maps[datatype=bibtex,overwrite=true]{
  \map{
    \step[fieldsource=Collaboration, final=true]
    \step[fieldset=usera, origfieldval, final=true]
  }
 }
}
\renewbibmacro*{author}{%
  \iffieldundef{usera}{%
    \printnames{author}%
  }{%
    \printfield{usera} Collaboration
  }%
}%

\usepackage{url}
\newcommand{\acts}{ACTS\xspace}

\newcommand{\pu}{$\big<\mu\big>$\xspace}
\newcommand*{\pT}{\ensuremath{p_{ \mathrm{T}}}\xspace}
\newcommand*{\ttbar}{\ensuremath{\mathrm{t}\bar{\mathrm{t}}}\xspace}
\newcommand*{\TeV}{\ensuremath{ \mathrm{Te\kern -0.1em V}}}
\newcommand*{\GeV}{\ensuremath{ \mathrm{Ge\kern -0.1em V}}}
\newcommand*{\MeV}{\ensuremath{ \mathrm{Me\kern -0.1em V}}}

\begin{document}

\title{A Common Tracking Software Project}

\input{authors.tex}

\journalname{Preprint submitted to \textit{Computing and Software for Big Science}}
\date{\today}


\maketitle{}

\begin{abstract}
The reconstruction of the trajectories of charged particles, or track reconstruction, is a key computational challenge for particle and nuclear physics experiments. While the tuning of track reconstruction algorithms can depend strongly on details of the detector geometry, the algorithms currently in use by experiments share many common features. At the same time, the intense environment of the High-Luminosity LHC accelerator and other future experiments is expected to put even greater computational stress on track reconstruction software, motivating the development of more performant algorithms. We present here A Common Tracking Software (\acts) toolkit, which draws on the experience with track reconstruction algorithms in the ATLAS experiment and presents them in an experiment-independent and framework-independent toolkit. It provides a set of high-level track reconstruction tools which are agnostic to the details of the detection technologies and magnetic field configuration and tested for strict thread-safety to support multi-threaded event processing. We discuss the conceptual design and technical implementation of \acts, selected applications and performance of \acts, and the lessons learned.

\end{abstract}


\clearpage

\clearpage
\input{introduction}
\FloatBarrier
\input{conceptual_design}
\FloatBarrier
\input{implementation}
\FloatBarrier
\input{applications}

\FloatBarrier
\input{experience}
\FloatBarrier
\input{conclusion}
\FloatBarrier

\section*{Acknowledgments}
This work was supported by the CERN Strategic R\&D Program on Technologies for Future Experiments (CERN-OPEN-2018-006), the National Science Foundation under Cooperative Agreement OAC-1836650,
the Office of Nuclear Physics within the U.S. DOE Office of Science under Contract No. DE-SC0012704
and the German Federal Ministry of Education and Research (BMBF).

\section*{Declarations}
\paragraph{Funding}
This work was supported by the CERN Strategic R\&D Program on Technologies for Future Experiments (CERN-OPEN-2018-006), the National Science Foundation under Cooperative Agreement OAC-1836650,
the Office of Nuclear Physics within the U.S. DOE Office of Science under Contract No. DE-SC0012704
and the German Federal Ministry of Education and Research (BMBF).

\paragraph{Conflict of interest}
The authors declare that they have no conflict of interest.

\paragraph{Availability of data and material} 
Not applicable. No associated data except for code.

\paragraph{Code availability}
The code used for this research is available open source~\cite{acts_on_github}.

\begingroup
\setlength{\emergencystretch}{4em}
\printbibliography
\endgroup

\end{document}

%% file: authors.tex

\author{
Xiaocong Ai
  \and 
Corentin Allaire
  \and 
Noemi Calace
  \and 
Angéla Czirkos
  \and 
Markus Elsing
  \and 
Irina Ene
  \and 
Ralf Farkas
  \and 
Louis-Guillaume Gagnon
  \and 
Rocky Garg
  \and 
Paul Gessinger
  \and 
Hadrien Grasland
  \and 
Heather M. Gray
  \and 
Christian Gumpert
  \and 
Julia Hrdinka
  \and 
Benjamin Huth
  \and 
Moritz Kiehn
  \and 
Fabian Klimpel
  \and 
Bernadette Kolbinger
  \and 
Attila Krasznahorkay
  \and 
Robert Langenberg
  \and 
Charles Leggett
  \and 
Georgiana Mania
  \and 
Edward Moyse
  \and 
Joana Niermann
  \and 
Joseph D. Osborn
  \and 
David Rousseau
  \and 
Andreas Salzburger \thanks{Corresponding author. \email{andreas.salzburger@cern.ch}}
  \and 
Bastian Schlag
  \and 
Lauren Tompkins
  \and 
Tomohiro Yamazaki
  \and 
Beomki Yeo
  \and 
Jin Zhang
}

\institute{
  Angéla Czirkos
  \at
  \textbf{Eötvös Loránd University, 1053 Budapest, Hungary}
  \and
  Bastian Schlag
  \at
  \textbf{Institut für Physik, Johannes Gutenberg-Universität Mainz, 55128 Mainz, Germany}
  \and
  Xiaocong Ai \and Georgiana Mania
  \at
  \textbf{Deutsches Elektronen Synchrotron, 22607 Hamburg, Germany}
  \and
  Rocky Garg \and Lauren Tompkins
  \at
  \textbf{Stanford University, CA 94305, Stanford, USA}
  \and
  Corentin Allaire \and Noemi Calace \and Markus Elsing \and Paul Gessinger \and Moritz Kiehn \and Fabian Klimpel \and Bernadette Kolbinger \and Attila Krasznahorkay \and Joana Niermann \and Andreas Salzburger \and Bastian Schlag
  \at
  \textbf{CERN, 1211, Geneva, Switzerland}
  \and
  Hadrien Grasland \and David Rousseau
  \at
  \textbf{Université Paris-Saclay, CNRS/IN2P3, IJCLab, 91405 Orsay, France}
  \and
  Jin Zhang
  \at
  \textbf{Institute of High Energy Physics, Chinese Academy of Sciences 100 039, Beijing, China}
  \and
  Irina Ene \and Louis-Guillaume Gagnon \and Heather M. Gray \and Tomohiro Yamazaki \and Beomki Yeo
  \at
  \textbf{Department of Physics, University of California, CA 94720, Berkeley, USA}
  \and
  Joseph D. Osborn
  \at
  \textbf{Oak Ridge National Laboratory, TN 37831, Oak Ridge, USA}
  \and
  Robert Langenberg \and Edward Moyse
  \at
  \textbf{University of Massachussets, MA 01003, Amherst, USA}
  \and
  Ralf Farkas
  \at
  \textbf{Universität Bonn, 53012 Bonn, Germany}
  \and
  Georgiana Mania
  \at
  \textbf{Universität Hamburg, FB Informatik, 20148 Hamburg, Germany}
  \and
  Heather M. Gray \and Charles Leggett
  \at
  \textbf{Lawrence Berkeley National Laboratory, CA 94720, Berkeley, USA}
  \and
  Benjamin Huth
  \at
  \textbf{Universität Regensburg, 93053 Regensburg, Germany}
  \and
  Fabian Klimpel
  \at
  \textbf{Technische Universität München, 80333 München, Germany}
}

%% file: introduction.tex
\section{Introduction}
\label{sec:introduction}


Track reconstruction will become the most computationally intensive component of event reconstruction because it scales combinatorially with increasing number of charged particles. At proton-proton ($pp$) colliders such as the Large Hadron Collider (LHC), the increasing multiplicity is usually due to an increase in the simultaneous $pp$ interactions per event, or pile-up ($\mu$). For heavy-ion collisions, on the other hand, the particle multiplicity is primarily determined by the centrality of the event, which depends on the number of nucleon participants in each collision. For most tracking algorithms, the execution time scales approximately quadratically with the charged particle multiplicity. 

\begin{figure}[ht]
\centering
\includegraphics[width=1.0\linewidth]{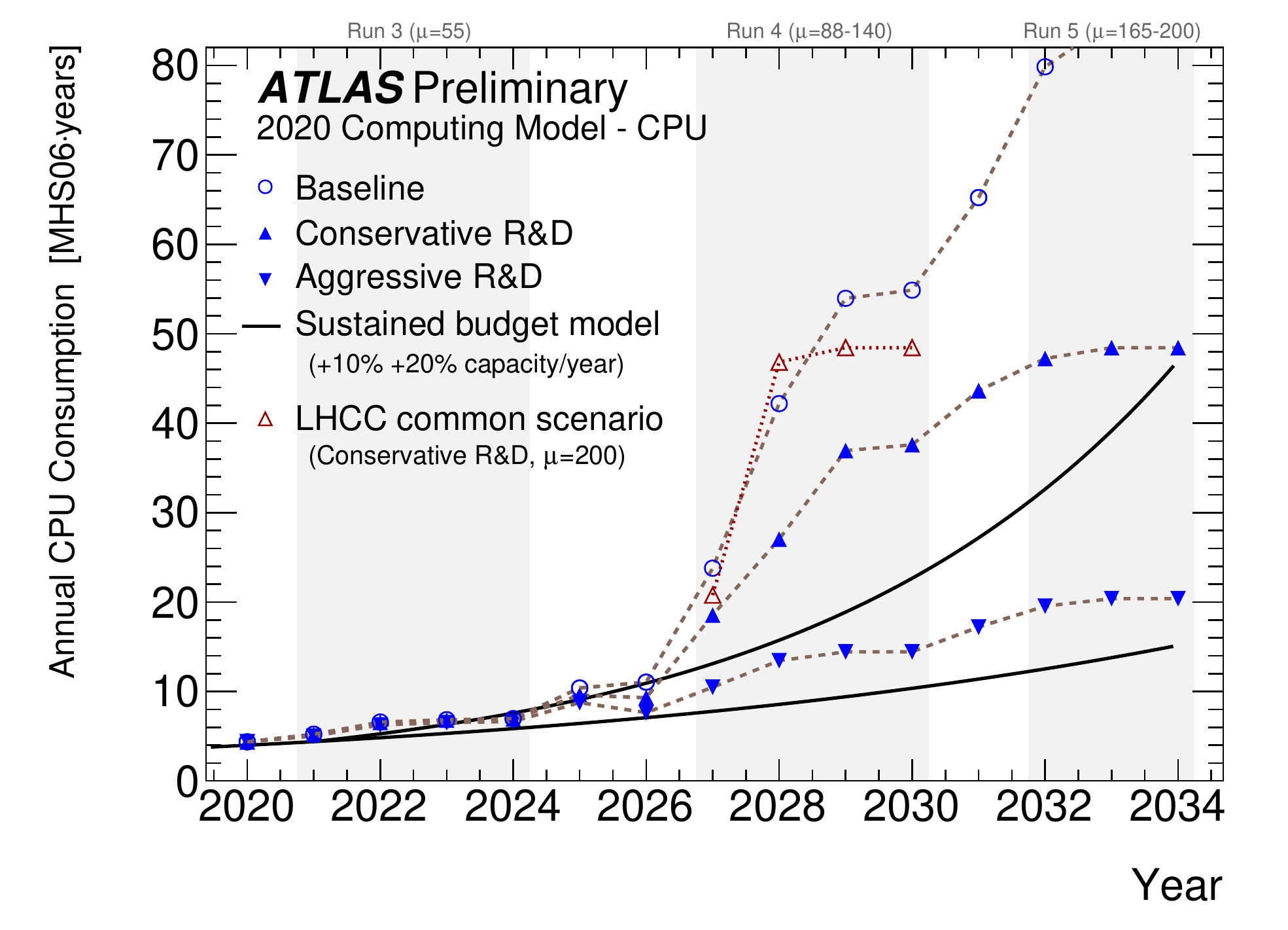}
\caption{
Estimated CPU resources (in MHS06~\cite{hepspec}) needed for the 2020 to 2032 time frame for both data and simulation processing for the ATLAS experiment. Three different scenarios considered by ATLAS are shown ranging from the baseline to that in which the aggressive R\&D program is successful (blue points).  The common scenario agreed between the different experiments as a reference is shown with red triangles. The black lines indicate the amount of CPU that can be expected based on current budget models.  From Ref.~\cite{Calafiura:2729668}\label{fig:cpuHLLHC}}
\end{figure}

In the general-purpose detector at the LHC, ATLAS~\cite{Aad:2008zzm}, for example, there are currently an average of approximately 500 charged particles with sufficient momentum to be reconstructed within the detector acceptance. 
However, the upgrade of the LHC, the High-Luminosity LHC (HL-LHC)~\cite{ApollinariG.:2017ojx}, which is expected to begin data-taking in 2027 will increase the instantaneous luminosity by a factor of five. The higher luminosity will result in an increase of the pile-up from the current average of 34 to 140-200 in ATLAS and the second general-purpose detector at the LHC, CMS~\cite{cms_2008}. The acceptance of the upgraded detectors will approximately double and additional detector layers will be added increasing the number of read-out channels. This means that there will be an average of 4000 charged particles within the upgraded detector acceptance and current minimum momentum requirements~\footnote{As the number of charged particles decreases rapidly with transverse momentum, the transverse momentum requirement can be raised to decrease the CPU time of track reconstruction algorithms. However, this must be balanced by its impact on the physics program of the experiment.}. The rates at which the detectors are read-out will increase by an order of magnitude.  In total, there are expected to be approximately 300,000 individual detector measurements in each event. Furthermore, additional funding for computing resources is expected to be limited in the HL-LHC era~\cite{Software:2751565,Calafiura:2729668}. \autoref{fig:cpuHLLHC} shows that the CPU resources needed for event reconstruction are expected to exceed the available computing budget by at least a factor of two. Future $pp$ colliders, such as the hadron-hadron option for the Future Circular Collider (FCC-hh), are anticipated to have an even larger number of up to 1000 simultaneous $pp$ collisions~\cite{Benedikt:2018csr}.

Future collider-based nuclear physics experiments will accumulate several thousands of charged particles from heavy-ion collisions that occur both in the nominal interaction region and farther down the beam pipe. This leads to high occupancy and also out-of-time pile-up that creates a challenging track reconstruction environment, similar to expectations for the HL-LHC.

Historically, particle and nuclear physics have relied on Moore's Law~\cite{moore}, which is the observation that the number of transistors on an integrated circuit approximately doubles every two years. However, in the last decade, the current processor technologies have become limited in terms of the clock speeds that can be obtained due to the power density. Therefore, recent increases in speed have been achieved by adding processing cores instead of increasing the speed of individual cores. Further throughput increases are expected to be achieved through the use of different computing architectures such as Graphics Processing Units (GPUs), Field Programmable Gate Arrays (FPGAs) or integrated System on a Chip (SoC) circuits. See Ref.~\cite{10.1145/3282307} for a recent discussion about the evolution of these technologies. Exploiting these architectures demands increasingly parallelized code and changes to programming paradigms. In addition, the rapid advances in the fields of artificial intelligence and machine learning, have resulted in a wide range of new ideas for tracking algorithms. These include cellular automata~\cite{Pantaleo:2293435}, graph neural networks~\cite{Ju:2020xty}, similarity hashing~\cite{Amrouche:2019yxv} amongst many others. While no algorithm has yet emerged to displace existing track reconstruction methods, it is still early in the development cycle for such algorithms and the field of machine learning is undergoing rapid evolution.

During event processing, the raw signals from the detectors are processed to obtain the reconstructed objects used for physics analysis. Using information from dedicated tracking detectors, sophisticated algorithms are used to reconstruct the trajectories of charged particles from the energy they deposit in the detector elements, including solid state detectors with segmented readout, gas tubes or other tracking devices. Such track reconstruction algorithms can be considered to be part of a more general class of pattern recognition algorithms. Track reconstruction algorithms have been used in particle and nuclear physics experiments for more than half a century. 

Track reconstruction methods~\cite{RevModPhys.82.1419} can be categorized as global and local methods, although the two categories cannot always be strictly separated. Global methods find trajectories using the entire detector’s measurement ensemble, often through conformal mapping or transform methods, such as the Hough transform~\cite{Hough:1959qva,Duda1972UseOT}. Other global approaches use neural networks~\cite{Hopfield:1982pe} to find connected sets of measurements. Local methods generate track seeds and search for additional hits to complete them. Local methods include the track road and track following methods such as the Kalman filter (KF)~\cite{kalman1960,Billoir:1983mz,Fruhwirth:1987fm}. 

A Common Tracking Software (\acts{}) entered this rapidly evolving ecosystem in 2016. It began with a small team at CERN and has since grown into an international collaboration with approximately 15 regular contributors. \acts has its origins in the track reconstruction algorithms developed for and extensively used by the ATLAS experiment~\cite{Gumpert:2243297}. \acts is an attempt to develop community driven track reconstruction software, where community contributions and extensions are explicitly encouraged. \acts provides algorithms for track reconstruction within a generic, framework- and experiment-independent open-source software toolkit~\cite{Ai:2019kze,Ai:2020jbw,Gessinger:2020nne}. \acts includes data structures and algorithms for performing track reconstruction in addition to a tool for fast track simulation. The \acts code is designed to be inherently thread-safe to support parallel code execution, and data structures are optimized for vectorization, which will speed up linear algebra operations. The implementation is designed to be fully agnostic to detection technologies, detector design, and the event processing framework to allow it to be used by a range of experiments. However, tuning of the algorithms for specific detectors is required to achieve the ultimate physics performance. Experiment-specific adaptions and tuning of the toolkit, including contextual data such as detector conditions and alignment, are made possible in \acts through C++ compile-time specializations. In addition, \acts is designed to be highly customizable and extendable to provide an R\&D platform for the development and study of novel algorithms and techniques.

An early version of \acts has been used to simulate the dataset for the Tracking Machine Learning (TrackML) challenge~\cite{Amrouche:2019wmx,Kiehn:2019tbl,Amrouche:2021tio}, which was performed in two stages to invite collaborators from within and external to particle physics to stimulate the development of new ideas for track reconstruction. The dataset produced for this challenge has subsequently been used to explore a range of novel track reconstruction algorithms~\cite{Tuysuz:2020eaa,Fox:2020hfm,Heintz:2020soy,Amrouche:2019yxv,Bapst:2019llh}. We use this dataset to demonstrate the current performance of the \acts{} algorithms, although no rigorous performance tuning has been done. This document describes the concepts, design and implementation of the \acts toolkit and does not attempt to quantify its ultimate performance on any specific detector setup. \acts has been explored for a range of different detectors including Belle~II~\cite{abe2010belle}, CEPC~\cite{thecepcstudygroup2018cepc_v1,thecepcstudygroup2018cepc_v2},  sPHENIX~\cite{Adare:2015kwa,Osborn:2020soo,Osborn:2021zlr}, PANDA~\cite{panda_vtx_tdr:2012,Bettoni:2007ms}, FASER~\cite{Ariga:2019ufm}, and the future ATLAS Inner Tracking system (ITk)~\cite{ATLAS_ITPixel_Phase2_TDR,ATLAS_ITStrip_Phase2TDR,ATL-PHYS-PUB-2019-014} for the HL-LHC data taking era.

The concepts, design and implementation of the \acts project are presented here. For further details of the implementation, see the current release, Ref.~\cite{xiaocong_ai_2020_3937454}. \autoref{sec:concdes} discusses the concepts and design of the \acts software. The technical implementation is discussed in \autoref{sec:techimp}.  Selected applications and early performance studies of tracking and vertexing are discussed in \autoref{sec:appperf}. \autoref{sec:experience} highlights some lessons learned from the experience. The conclusion and a brief outlook are covered in \autoref{sec:conc}.

%% file: conceptual_design.tex
\section{Conceptual Design}
\label{sec:concdes}

The \acts project was initiated to serve three primary goals. Firstly, to preserve and advance the well-tested code bases from the LHC experiments, while enabling preparation for the HL-LHC era and other future particle and nuclear physics experiments. This requires a state-of-the-art software development environment that allows the contributors to work with modern programming language standards and development workflows. Secondly, to provide an R\&D test bed for algorithmic research (including machine learning techniques) and portability to accelerated hardware. Thirdly, to ultimately provide a mature track reconstruction toolkit, that can be used as a platform for rapid development of tracking applications for future tracking detectors.

Software development for particle and nuclear physics experiments is subject to a number of constraints: an event processing framework steers the execution of algorithmic blocks, and a well-defined event data model (EDM) holds the event information and defines the communication between different components. Examples of event processing frameworks include Gaudi~\cite{Barrand:2001ny} used by the LHCb experiment~\cite{lhcb_2008}, the Athena~\cite{athena} extension of Gaudi for the ATLAS experiment, CMSSW~\cite{CMSSW} for the CMS experiment, and the ROOT~\cite{Brun:1997pa} event processing loop. In recent years, many of these processing frameworks have been adapted and extended to enable multi-process (MP) or multi-threaded (MT) workflows to accommodate different types of hardware and to optimize the usage of memory and computing cores. The details of the implementation of MP or MT workflows differ between the various frameworks and experiments, but the overall concepts are the same. In a typical MP workflow, the concept of a single event is typically preserved, i.e. a single process acts on one encapsulated event. In collider physics, an event is a single beam-beam collision or a single simulated event. In MT-capable applications, multiple events can be processed simultaneously, which requires the function calls to be independent of the current event or be provided with the appropriate event context, as discussed in \autoref{subsec:concurrency}. In this case, the method call is fully controlled and defined by the input and output data, and the algorithmic module is a stateless engine that has no memory of previous calls, configurations and operations.
Despite the complex steering and brokering of the event processing, the actual work load is performed by smaller modules or tools, which are not necessarily controlled by the framework's public interface. \acts aims to provide such a toolkit for track and vertex reconstruction, together with a high-level EDM definition that can be directly included in experiment-specific applications, extended by adding additional functionality, and rearranged and adapted to the specific needs of an experiment.

To prepare the \acts toolkit for such general use, its design has the following central concepts:
\begin{itemize}
    \item minimal dependency of the core components on external software packages
    \item abstraction of the EDM and geometry description from the specific details of any experiment
    \item general mathematical formulations of algorithms independent of specific detector geometry, magnetic field, or detector technology
    \item a customizable connection to the algorithm configuration
    \item transparent import and handling of experiment-specific contextual conditional data, such as detector calibration and detector alignment
    \item facilitation of the integration of core functionality typically governed by the event processing framework, e.g.~message logging
    \item a plugin mechanism for extending the toolkit with external software packages
\end{itemize}

Several of the key concepts of the design of \acts are described in further detail in the following. The implementation is discussed in \autoref{sec:techimp}.

\subsection{Concurrent code execution}
\label{subsec:concurrency}

\begin{figure*}[!ht]
\centering
\input{acts_fig_design_parallel_execution}
\caption{Illustration of multi-threaded event processing with the sequence proceeding from left to right, in the context of an experimental software framework. Two threads execute different experiment-specific algorithms, which are illustrated by different shapes. The algorithms are distributed across threads by a scheduler. Execution occurs out of order for the three events indicated by different colors. Data flow integrity, drawn as arrows connecting algorithms,
is respected. \acts{} components can be used inside the algorithms, shown 
as loops attached to individual algorithms instances. They can optionally
increase concurrency by running on parts of the event simultaneously, as shown
for algorithm 4.}
\label{fig:parallel_execution}
\end{figure*}
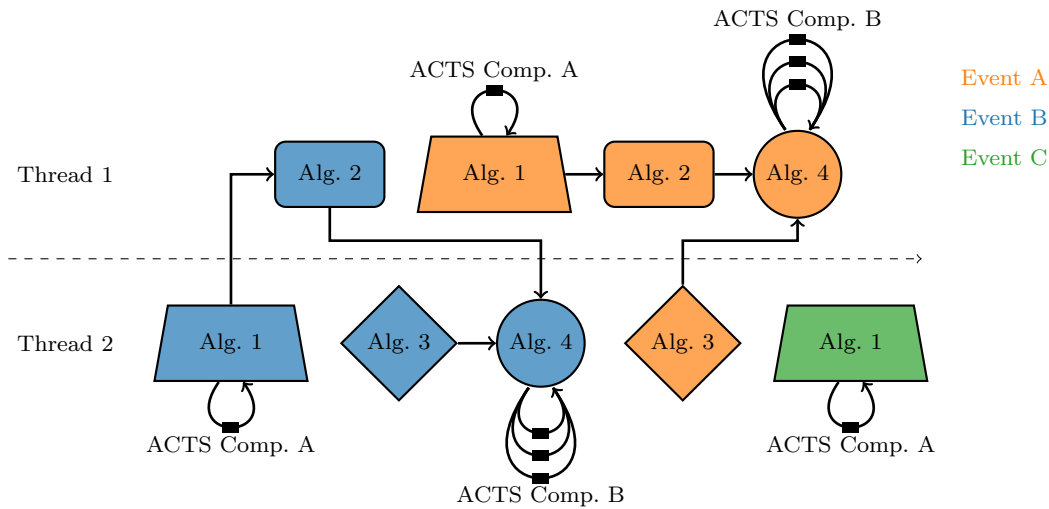

\acts is designed to accommodate the heterogeneous computing landscape with parallel code execution paths. Therefore all algorithmic modules can be called in parallel while processing an event and between the processing of multiple events without interference as illustrated in \autoref{fig:parallel_execution}. The contextual and conditional data is handled transparently as described in \autoref{subsec:contextuality}. To avoid restricting the caller code to any predefined pattern, all \acts modules are designed such that each function call has to be fully controlled by the data input and output flow, and back channel communication to caller functions is forbidden\footnote{In C++, this is enforced by restricting methods to follow a const-correct signature and by forbidding mutable data members.}. If caching is required, e.g.~for performance reasons, the cache must be provided as part of the input data, as discussed in \autoref{sec:techimp}. The correct and reproducible behavior of the code in sequential and concurrent code execution paths is tested within unit and integration test suites. These tests include checks for identical results when running in single threaded and multi threaded mode. More advanced examples test the correct behavior with multiple alignment or magnetic field conditions during a single execution run.

The actual code execution pattern, e.g.~event parallelism in an MP or MT framework, or intra-event parallelism at different stages of track reconstruction, is the responsibility of the caller application and thus no technology, language, nor dedicated library for parallel code execution is provided in the \acts core modules. 
However, example applications in the repository rely on the Intel Threading Building Blocks (TBB)~\cite{tbb} multi-threading library to demonstrate how concurrent execution can be implemented. \acts allows core modules to be wrapped in callable kernel structures that can be used on accelerators with dedicated technology back ends. First demonstrators of such an approach have been successfully deployed~\cite{Ai:2021kzk}, however further development and simplification of the code base is needed for \acts to run efficiently on different types of computing hardware. This is one of the dedicated R\&D lines of the project as discussed in \autoref{subsec:rnd}.

\subsection{Contextual data handling} 
\label{subsec:contextuality}

\begin{figure*}[!ht]
\centering
\input{acts_fig_design_contextual_geometry}
\caption{Illustration of contextual geometry handling. At job initialization time, only a nominal (or initially aligned) version of the \acts geometry is built.
Three threads execute on events in parallel. All threads request details of the \acts{} geometry by providing their event context, which fully defines the alignment of the detector in the current call context. The method to perform the alignment can be experiment-specific. \label{fig:contextual_geometry}}
\end{figure*}
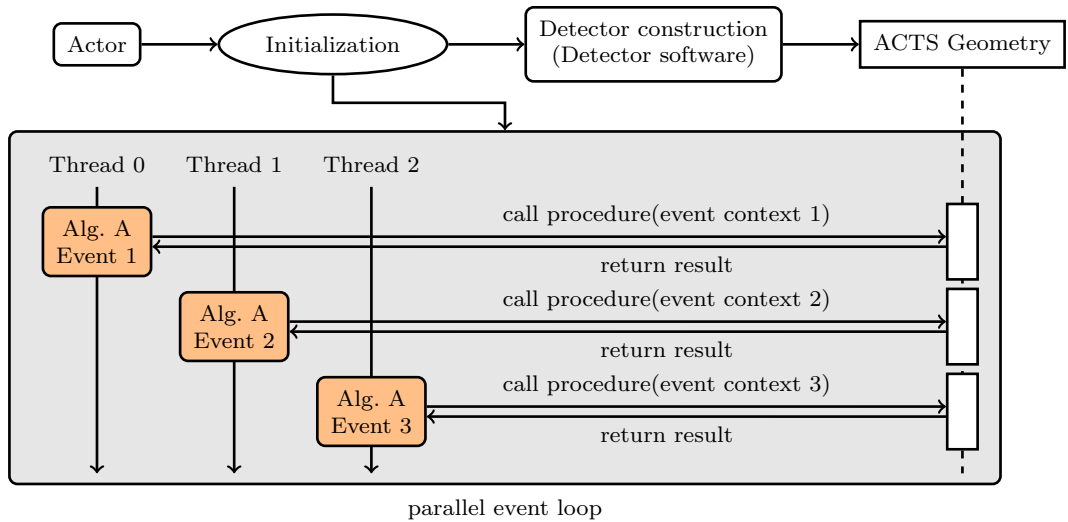

A general track reconstruction toolkit that serves different experiments must be able to handle a contextual experimental environment. Detectors may have temporary or permanent imperfections, suffer from changing alignment and data-taking conditions, and, in general, operate in a time-dependent manner. Track reconstruction uses high-precision measurements and every effect must be accounted for in order to achieve optimal results. Detector conditions, on the other hand, are one of the most specific aspects of any experimental setup, and a general solution or implementation for such a diverse problem would be very challenging. Therefore, a transparent handling schema for all contextual data has been applied throughout the \acts code base: a set of contextual objects, defined and implemented in the experiment's software stack, are handed through the entire call structure of \acts (see \autoref{fig:contextual_geometry}). This ensures that each geometry call that relies on detector information is aware of the geometry context of that particular call and allows the correct detector alignment to be applied within that specific call context. Other conditional data, such as the magnetic field status or detector calibration data are implemented in the same way. In all cases, the caller code can be assured that contextuality will be respected with minimal computational overhead, because the context information is unpacked and correctly interpreted. The choice about whether the contextual object carries either a parameter to identify the context to be applied or the full contextual data is left to the implementation within a particular experiment.

See \autoref{subsec:application} for details of a concrete implementation of a contextual environment.

\subsection{Research and Development projects}
\label{subsec:rnd}

Recent technology advances in both hardware and software have transformed the computing landscape in the scientific and private sector. Machine learning is a rapidly growing field, and hardware-based acceleration becomes increasingly prominent due to the growing use of high performance computing centers and limitations in increases in processor speed. While both areas have already been explored in particle and nuclear physics, additional R\&D is needed to fully exploit these advances in future data processing, particularly in the domain of track and vertex reconstruction. The tracking machine learning challenge has demonstrated that machine learning algorithms can reach the same order of magnitude in both physics performance and execution speed compared to the current track reconstruction algorithms. End-to-end solutions based on machine learning are expected to require significant development time. However, certain aspects of track reconstruction such as track classification or data segmentation~\cite{Amrouche:2021tlm} have already shown promising results. Such smaller components, however, need to be tested in a realistic data flow. A key element in the design of \acts is to provide a playground to facilitate prototyping, development, and testing of such new ideas.
The plugin mechanism of \acts allows the core track reconstruction code to be coupled with external libraries from the machine learning and data science sectors, or with code with different language backends, which is needed for code execution on accelerators. The \texttt{ONNX}~\cite{onnx} library for the deployment of machine learning based tracking solutions and the \texttt{autodiff} ~\cite{autodiff} library for automated compiler based differentiation have both been demonstrated within \acts. Furthermore, the \texttt{CUDA}~\cite{cuda} and \texttt{SYCL}~\cite{sycl} have been integrated for GPU-based seed finding algorithms. 

%% file: acts_fig_design_parallel_execution.tex
  \begin{tikzpicture}[scale=0.8]

    \definecolor{tab1}{RGB}{31,119,180}
    \definecolor{tab2}{RGB}{255,127,14}
    \definecolor{tab3}{RGB}{44,160,44}
    \definecolor{tab4}{RGB}{214,39,40}
    \definecolor{tab5}{RGB}{148,103,189}
    \definecolor{tab6}{RGB}{140,86,75}
    \definecolor{tab7}{RGB}{227,119,194}
    \definecolor{tab8}{RGB}{127,127,127}
    \definecolor{tab9}{RGB}{188,189,34}
    \definecolor{tab10}{RGB}{23,190,207}

    \tikzset{trap/.append style={
        trapezium, trapezium angle=80, 
        minimum width=0cm, minimum height=1cm, draw, thick,
        inner sep=3mm,
      }
    }

    \tikzset{circ/.append style={circle,draw,thick}}
    \tikzset{diam/.append style={diamond,draw,thick}}
    \tikzset{rect/.append style={rectangle,rounded corners,draw,thick,inner sep=3mm}}

    \tikzset{ea/.append style={fill=tab1!70!white}}
    \tikzset{eb/.append style={fill=tab2!70!white}}
    \tikzset{ec/.append style={fill=tab3!70!white}}

    \begin{scope}[node distance=5mm]
      \begin{scope}[shift={(0,1.4cm)}]
        \node (t1l) at (-2cm, 0) {Thread 1};
        \node [rect,ea,right=2cm of t1l] (t1a1ea) {Alg. 2};
        \node [trap,right=of t1a1ea,eb] (t1a2eb) {Alg. 1};
        \node [rect,right=of t1a2eb,eb] (t1a1eb) {Alg. 2};
        \node [circ,right=of t1a1eb,eb] (t1a3eb) {Alg. 4};
      \end{scope}

      \begin{scope}[shift={(0,-1.4cm)}]
        \node (t2l) at (-2cm, 0) {Thread 2};
        \node [trap,ea,right=5mm of t2l] (t2a2ea) {Alg. 1};
        \node [diam,right=of t2a2ea,ea] (t2a4ea) {Alg. 3};
        \node [circ,right=of t2a4ea,ea] (t2a3ea) {Alg. 4};
        \node [diam,right=of t2a3ea,eb] (t2a4eb) {Alg. 3};
        \node [trap,right=of t2a4eb,ec] (t2a2ec) {Alg. 1};
      \end{scope}
    \end{scope}

    \coordinate (leftend) at ($(t1l.west)!0.5!(t2l.west)$);
    \coordinate (rightend) at (leftend-|t2a2ec.east);
    \draw [dashed,->] (leftend) -- (rightend);

    \begin{scope}[->,line width=1pt]
      \draw (t2a2ea) |- (t1a1ea);
      \draw (t2a4ea) -- (t2a3ea);
      \draw (t1a1ea) --(0,3mm-|t1a1ea) -| (t2a3ea);
      
      \draw (t1a2eb) -- (t1a1eb);
      \draw (t1a1eb) -- (t1a3eb);
      \draw (t2a4eb) --(0,3mm-|t2a4eb) -| (t1a3eb);
    \end{scope}

    \foreach \n/\s in {t2a2ea/1, t2a2ec/1, t1a2eb/-1} {
        \ifdim\s pt < 0.0pt
            \def\l{above}
            \def\a{north}
        \else
            \def\l{below}
            \def\a{south}
        \fi
        
        \draw[->,line width=1pt] 
        (\n.\a)++(-0.2,0) 
            ..controls ++(-0.7,{-1*\s}) and ++(0.7,{-1*\s}).. 
            node[\l] {ACTS Comp. A} 
            node[pos=0.5,fill,rectangle,inner sep=0,minimum height=4pt,minimum width=6pt] {}
        ++(0.4,0) ;
        
    }
    
    \foreach \n/\s in {t2a3ea/1, t1a3eb/-1} {
        \ifdim\s pt < 0.0pt
            \def\l{above}
            \def\a{north}
        \else
            \def\l{below}
            \def\a{south}
        \fi
        
        \foreach \i [count=\ii] in {1,1.5,2} {
            \ifdim\i pt = 1.0pt
                \def\m{->}
            \else
                \def\m{-}
            \fi
            
            \draw[\m,line width=1pt] 
                (\n.\a)++(-0.2,0) 
                    ..controls ++({-0.7*\i*1.0},{-1*\s*\i}) and ++({0.7*\i*1.0},{-1*\s*\i}).. 
                    node[pos=0.5,fill,rectangle,inner sep=0,minimum height=4pt,minimum width=6pt] (n\ii) {}
                ++(0.4,0) ;
        }
        \node[\l] at (n3) {ACTS Comp. B};
    }

    \begin{scope}[shift={($(rightend) + (5mm,3cm)$)}]
      \node [anchor=west,tab2] (ea) at (0,0) {Event A};
      \node [anchor=west,tab1,below=1mm of ea] (eb) {Event B};
      \node [anchor=west,tab3,below=1mm of eb] (ec) {Event C};
    \end{scope}

  \end{tikzpicture}

%% file: acts_fig_design_contextual_geometry.tex
  \begin{tikzpicture}[scale=0.7,inner sep=5pt,line width=1pt]

    \definecolor{tab1}{RGB}{31,119,180}
    \definecolor{tab2}{RGB}{255,127,14}
    \definecolor{tab3}{RGB}{44,160,44}
    \definecolor{tab4}{RGB}{214,39,40}
    \definecolor{tab5}{RGB}{148,103,189}
    \definecolor{tab6}{RGB}{140,86,75}
    \definecolor{tab7}{RGB}{227,119,194}
    \definecolor{tab8}{RGB}{127,127,127}
    \definecolor{tab9}{RGB}{188,189,34}
    \definecolor{tab10}{RGB}{23,190,207}
  
    \tikzset{roundrect/.append style={rectangle,rounded corners=3pt}}
  
    \node[roundrect,draw] (actor) at (0,0) {Actor};
    
    \node[ellipse,draw,right=of actor] (init) {Initialization};
    
    \node[roundrect,draw,right= of init,align=center] (detconst) {Detector construction \\ (Detector software)};

    \node[rectangle,draw,right=of detconst] (actsg) {ACTS Geometry};

    \draw[->] (actor) -- (init);
    \draw[->] (init) -- (detconst);
    \draw[->] (detconst) -- (actsg);

    \node[inner ysep=5pt,inner xsep=0,below=1.0cm of actor] (t0) {Thread 0};
    \node[inner ysep=5pt,inner xsep=0,right=5mm of t0] (t1) {Thread 1};
    \node[inner ysep=5pt,inner xsep=0,right=5mm of t1] (t2) {Thread 2};

    \begin{scope}[scale=0.9]
      
      \def\dist{6.5cm}
      \foreach \i in {0,1,2} {
          \draw[->] (t\i) --++(0,{-1*\dist}) coordinate (low);
      }
      
      \draw[dashed] (actsg) -- (low-|actsg) coordinate (br);
      
      \coordinate (clab) at ($(t2)!0.5!(t2-|actsg)$);
      
      \foreach \f/\n [count=\i] in {0.25/t0,0.525/t1,0.8/t2} {
          \node[roundrect,draw,fill=tab2!50!white,align=center] (t\i a) at ($(\n)-(0,{\f*\dist})$) {Alg. A \\Event \i};
          
          \node[rectangle,draw,minimum height=1cm,minimum width=4mm,fill=white] 
          (g\i) at (t\i a-|actsg) {};
          
          \draw[->] (t\i a.east)++(0,3pt) -- ($(g\i.west) + (0,3pt)$);
          \draw[<-] (t\i a.east)++(0,-3pt) -- ($(g\i.west) + (0,-3pt)$);
          
          \node[anchor=south] at (t\i a-|clab) {call procedure(event context \i)};
          \node[anchor=north] at (t\i a-|clab) {return result};
          
      }
      
      \begin{scope}[on background layer]
          \node[fit=(t0)(br),roundrect,draw,inner sep=5mm,inner ysep=4pt,fill=black!10!white,line width=1pt]
          (exec) {};
          \node[below] at (exec.south) {parallel event loop};
      \end{scope}
      
      \draw[->] (init.south) -- ($(init.south)!0.5!(init.south|-exec.north)$) -| (exec.north);

    \end{scope}

  \end{tikzpicture}

%% file: implementation.tex
\section{Technical Implementation}
\label{sec:techimp}

\subsection{Basic Technology Choices}
\label{sec:tech}
\acts targets modern many-core, general purpose CPUs, which are widely available and the default computing architecture currently used by the LHC experiments~\cite{Duckeck:2005rb,CMS:2005aa} and other experiments in particle and nuclear physics. Both x86 and ARM architectures have been demonstrated to work with \acts. All recent CPUs have vector units and significant performance improvements can be obtained from vectorizable code. While hardware accelerators such as GPUs and FPGAs are not necessarily part of most baseline architectures, they are actively explored by the \acts developers and the larger particle and nuclear physics community, particularly for online software and when looking ahead towards the HL-LHC. 

\acts is implemented in C++\,17~\cite{ISO14882}, which is widely used in the particle and nuclear physics community, and thus can be easily integrated with existing software. As a compiled programming language with minimal implicit runtime facilities and a high degree of low-level hardware control, C++ enables achieving excellent execution performance. However, it is difficult to learn and use correctly, particularly regards to memory management. This is mitigated through guidelines and implementation choices in \acts, which include strict ownership handling via movable types and value-like semantics as well as the adoption of best practices such as unit tests and continuous integration, which ensure code quality.

\begin{figure}
    \centering
    \input{acts_fig_impl_integration}
    \caption{Example of the integration of \acts into an experiment's software framework. The experiment- and detector-specific code (green) is expected to handle low-level data preparation and provide \texttt{SourceLink}s and \texttt{Measurement}s as input to \acts algorithms. \acts provides tracks and vertices as output for further experiment-specific reconstruction and analysis.}
    \label{fig:integration-into-experiment software}
\end{figure}
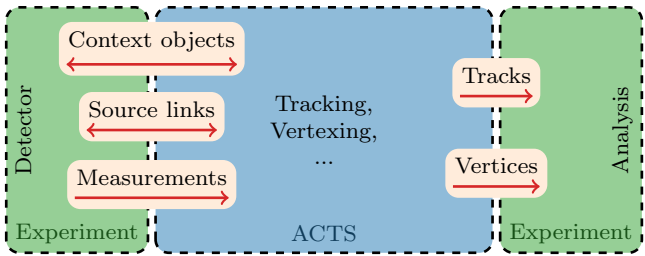

The \acts code is designed to have minimal dependencies on external packages. Only two third-party libraries are required: \texttt{Eigen}~\cite{Guennebaud:2010aa} for linear algebra and \texttt{Boost}~\cite{boost} for unit testing, file system handling, and a few key containers. In addition, \texttt{CMake}~\cite{cmake} is used both as a dependency management tool and as the build system.

The general strategy for algorithmic development in \acts draws on the experience of previous particle and nuclear physics software efforts and, in particular, the existing ATLAS offline tracking software~\cite{Cornelissen:2008zzc}. A key choice is to favor small compile-time interfaces and templates over virtual interfaces for better performance and greater implementation flexibility. \acts favors data-oriented programming over object-oriented programming, which means that the communication between different parts of the code occurs through sharing common data structures rather than predefined interfaces. 

\subsection{Code Organization}
\label{sec:code}
The \acts code~\cite{xiaocong_ai_2020_3937454} is a single open-source repository hosted on GitHub~\cite{acts_on_github}. A single repository allows for the easy development and integration of components and avoids version mismatches and accidental incompatibilities. The test and validation code provides examples of end-to-end tests and allows development with realistic reconstruction chains. No additional client application is necessary. The use of GitHub facilitates collaboration independent of affiliation. Key components within the repository are the core library, the plugins, the fast tracking detector simulation, Fatras, based on the original ATLAS fast track simulation~\cite{Edmonds:1091969} and the test and validation code. \autoref{fig:integration-into-experiment software} illustrates how \acts can be integrated into an experimental framework.

The core library implements the basic tools and key algorithms with minimal external dependencies. The plugins directory contains core-like functionality that requires additional external dependencies. The source code of the core library and the plugins extensions are located in the \texttt{Core} and \texttt{Plugins} directories. Examples available in the plugins directory include geometry tools based on the external \texttt{TGeo}~\cite{BRUN2003676} package from the \texttt{ROOT} toolkit, which is currently used in particle and nuclear physics experiments to describe detector geometries, and \texttt{CUDA} or \texttt{SYCL} which enables code to run on GPUs.

The source code of the Fatras simulation is not located within the \texttt{Core} folder because it is not required for reconstruction. It can be built on demand. Fatras makes heavy use of the core functionality and therefore can be maintained more easily as part of the same repository, e.g.~to adapt to core interface changes.

Releases of \acts{} follow semantic versioning~\cite{sv}, where a subset of the interface is considered when determining the major version. The software is provided under the Mozilla Public License, v. 2.0 (\texttt{MPLv2})~\cite{mplv2}. Common code formatting is ensured by requiring submitted code to the repository to pass a formatting check using the \texttt{clang-format}~\cite{clang-format} LLVM~\cite{llvm} extension.

\subsection{Core Components}
\label{sec:core}

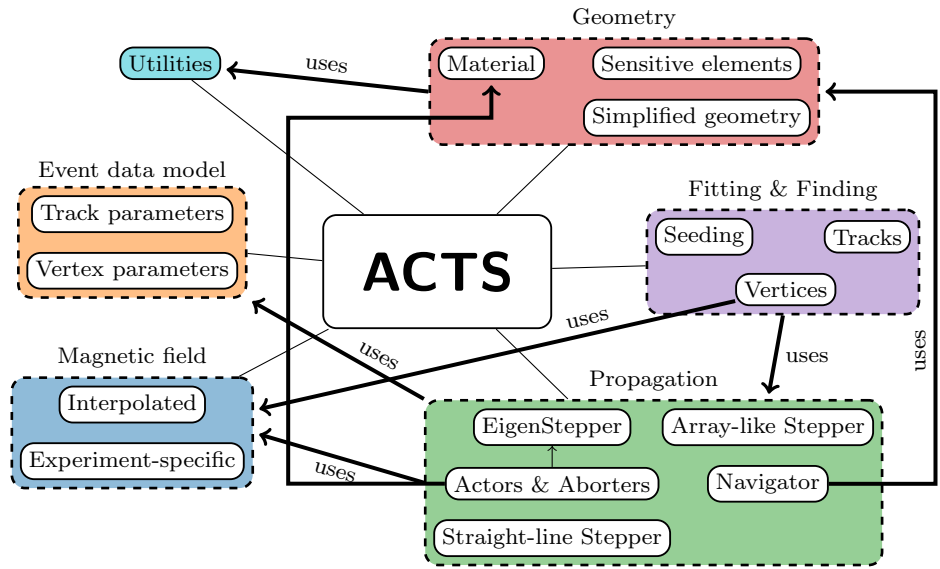
\begin{figure*}[!ht]
    \centering
    \input{acts_fig_impl_components}
    \caption{Overview of selected components in the \acts repository and their interactions. The components are categorized into modules, such as geometry, propagation or event data. A non-exhaustive number of relationships where one component "uses" other components in different modules are indicated by arrows. The stepper components are connected to the magnetic field module, because they are used to retrieve the magnetic field information.}
    \label{fig:components}
\end{figure*}

The core library of \acts is organized into modules and each module groups tools and algorithms with similar functionality. An overview of key modules is shown in \autoref{fig:components}.
The communication between algorithms occurs via common event data structures defined in the \texttt{EventData} module as described in \autoref{sec:edm}.
The \texttt{Geometry} module handles the tracking geometry, which is the logical and geometric grouping of detector surfaces into layers and volumes. The tracking geometry uses the \texttt{Surfaces} component, which implements different surfaces for detectors and boundaries. The related \texttt{Material} component contains tools to describe surface and volume based material and the algorithms to create such a geometrical mapping. See \autoref{sec:geometry} for further details about both modules.

The \texttt{Propagator} module provides tools to propagate particle states along their trajectories in different magnetic fields (see \autoref{sec:propagator}). The \texttt{TrackFinding} and \texttt{TrackFitting} modules use both the \texttt{Geometry} and the \texttt{Propagator} modules. The \texttt{Vertexing} module is largely standalone but relies on output from other modules as input and the propagation infrastructure. The \texttt{Seeding} module contains a geometry independent seeding algorithm that acts purely on global three-dimensional points.

\subsection{Configuration, State, and Context}
\label{sec:config}
\begin{listing}
\begin{minted}{cpp}
class Algorithm {
public:
    // Necessary and sufficient 
    // configuration variables.
    struct Config {
        double aValue = 0.25;
        std::string name = "something";
    };
    
    // Construct algorithm from its configuration.
    Algorithm(const Config& cfg) : m_cfg(cfg) {}

private:
    Config m_cfg;
    // ... e.g. values derived from 
    // the configuration
};
\end{minted}
\caption{Example of an algorithm implementation with a nested configuration type. The configuration type allows settings with default values to be addressed by name. Plain types mean that no explicit \texttt{Config} constructor is required. This pattern is used throughout the code base to support configuration and construction. This listing also demonstrates the naming convention for private member variables and the \texttt{CamelCase} coding style. \label{listing:config}}
\end{listing}

\begin{listing}
\begin{minted}{cpp}
class StatefulAlgorithm {
public:
    // Cached or intermediate values that
    // persist between calls.
    struct State {
        double lastValue = 0.;
    };
    
    // `const` implementation since all mutable 
    // states are contained within the state object.
    void doWork(State& state) const;
};
\end{minted}
\caption{Example of a stateful algorithm implementation with a nested state type. All state information, i.e. either cached or intermediate values, are contained in the nested state type. The algorithm type itself remains stateless and const-correct. \label{listing:state}}
\end{listing}

Components in \acts{} are typically highly configurable. To enable this flexibility, without being bound to any specific configuration environment, patterns using a nested C++ structure are used. \autoref{listing:config} provides an example of such a pattern, where a \texttt{Config} structure contains all configuration parameters as members. The constructor of the outer type takes an instance of the configuration structure as an argument, and runs its setup accordingly.

\acts{} supports both inter- and intra-event parallelization without expliciting implementing either. Instead, the explicit state objects for potentially stateful algorithms must be provided by the user as demonstrated in listing~\ref{listing:state}. An example of a stateful algorithm would be e.g.~a track finding algorithm that uses information about previously found tracks in the event (provided by the state) to prevent unnecessary or duplicated track search. By creating these state objects within their own framework, experiments must explicitly decide how and at which levels execution is parallelized and where synchronization might need to occur.

A similar problem exists for detector-related structures including the geometry, magnetic field, or calibrations that vary between events. During parallel execution, these structures cannot be handled as global states. Similar to the handling of the algorithm state, all algorithms that might require varying context data take an explicit context object. These objects are then passed through the full execution chain and handled by the experiment-specific code where necessary. An example of an application with contextual data, changing the detector alignment, is demonstrated in \autoref{subsubsec:cpuutil}.

\subsection{Event Data Model}
\label{sec:edm}

The event data model binds all modules together by providing shared data structures. The EDM is used to communicate between different steps of the reconstruction chain. Thus, it needs to be both generic enough to hold all possible event data types, but also minimal enough to avoid overheads, as it will be used extensively throughout the code base. Event data consists of measurements, track parameters, and vertex parameters, which can be represented as vectors.

The two different track parameter spaces in \acts are
bound and free track parameters. Bound track parameters describe a track bound to a surface. The surface can be a real detector surface such as the planar surface of a silicon detector or a virtual surface, such as the straw surface and the perigee~\footnote{The perigee refers to the point or surface of closest approach.} surface used to describe a track near an anode wire in a gaseous tracking detector and a vertex, respectively. The bound track parameters have six dimensions and comprise of a two-dimensional position on the local surface, two momentum direction angles (or angle-like parameters), a curvature parameter and time. The bound parameters can only be defined with reference to a surface and the interpretation of the two local position components are surface-dependent. At the perigee surface, the bound track parameters are 
\begin{equation}
 \mathbf{b} = (d_0, z_0, \phi, \theta, \frac{q}{p}, t), 
\end{equation}
where the $d_0$ and $z_0$ represent the transverse and longitudinal impact parameters, respectively. The remaining parameters are the azimuthal angle $\phi$, the polar angle $\theta$, the charged signed inverse momentum and the time $t$. This parameterization exists for charged and neutral particles. In the latter case the inverse momentum representation is changed to $1/p$.  The time $t$ is transparently respected in track propagation and potential measurement inclusion.

In contrast, free parameters require no reference surface and use the same definition everywhere. Within \acts they are described by the 3D position and direction vectors, time, and a curvature parameter. Therefore, they are 8-dimensional and are used throughout track propagation
\begin{equation}
  \mathbf{f} = ( x, y, z, d_x, d_y, d_z, \frac q p, t),
\end{equation}
and also referred to as free track parameters.

Measurements are treated as vectors in a sub-space of a (bound) track parameter vector space. Measurements are typically associated to a surface and only measure a subset of the available track parameters; most often at least one local position. Many track reconstruction methods, such as the Kalman filter (see \autoref{subsec:kalmanfilter}), include a projection from a subset of the track parameters to the measurement space. For \acts, the measurement space is assumed to always be consistent with the bound track parameter space defined by the surface. The inclusion of time information directly in the track fit is a novel feature of the \acts algorithms. For example, a pixel detector measurement with time information $\mathbf{m}_i = (m_x, m_y, m_t)$ can be compared to the estimated track parameterization $\mathbf{b}_i = (l_x, l_y, \phi, \theta, \frac{q}{p}, t)$ on the same surface $i$ using a projection matrix $\mathrm{\mathbf{H}}_i$ to form a three dimensional residual vector: $\mathbf{r}_i = \mathbf{m}_i - \mathrm{\mathbf{H}}_i \mathbf{b}_i$. Time is treated in the same way as the other track parameters.

Compile-time programming via template substitutions is used to dispatch execution into highly-optimized code paths for each dimensional measurement type. A separate data-structure provides an optimized collection of measurements. This data-structure can also store a tree of \emph{track states}, each potentially containing a measurement and/or the estimated track parameters.

The dedicated event data model used by the vertexing components is designed
to be as flexible as possible and the input tracks can be of any user-defined type. This approach facilitates experiment-specific integration while
keeping overhead minimal at the same time.

\subsection{Geometry}
\label{sec:geometry}

The geometry description used for reconstruction is a simplified version of the detailed detector description used in detailed Monte Carlo simulation programs such as \texttt{Geant4}~\cite{Agostinelli:2002hh}. The description of the sensitive detectors (including misalignment and other contextual information) needs to be as precise as possible. However several approximations to the detector description for the non-sensitive detector elements are made. During reconstruction, the noise from the detector material is accounted for either deterministically or stochastically.

In \acts, the reconstruction geometry is entirely built from surface objects. Compound layer objects and volume objects are based on the surface class. A volume shape is built from the boundary surfaces. The boundary surfaces are also referred to as portal surfaces as they connect the volumes.  Layers are defined by their bounding and contained surfaces. The contained surfaces can either be declared \textit{sensitive} when they represent detection elements or be passive material surfaces.

Navigation through the detector proceeds either by using portal surfaces that connect volumes with other volumes or by performing a local search of layer surfaces after entering a layer object through its bounding surfaces. All surfaces can be propagated to, carry material, or refer to sensitive detector elements, and are thus suitable for both reconstruction and fast simulation.

\subsubsection{Layer geometry and plugin mechanism for detector elements}

Tracking detectors are frequently built from physical layer structures that support the modules, the on-detector electronics, power cabling and cooling units, and often feature stave structures. The logical division into layer structures is used in \acts to restrict the local navigation to an area of interest instead of attempting to navigate the full detector. 

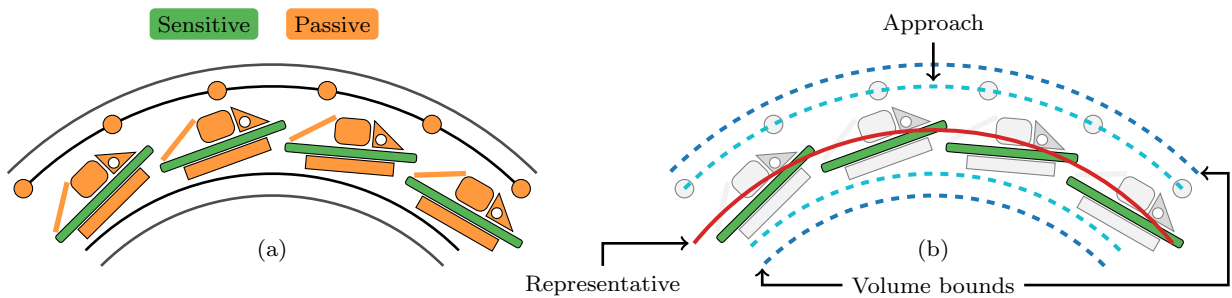
\begin{figure*}[!ht]
\centering
\input{acts_fig_impl_layer_geometry}
\caption{Illustration of the layer geometry for planar detection modules.
(a) shows the highly detailed geometry, in which both sensitive and passive elements are present. (b) shows the simplified version, where all passive elements are discarded (grayed out). Instead, various virtual surface approximations of the detailed structure are shown and used in the modeling.
The representative surface is closest to the sensor locations, while the approach surfaces form an envelope around them. A volume surrounds the layer, which also features boundary surfaces.}
\label{fig:layer_navigation}
\end{figure*}

Each layer has a set of approach surfaces, as well as a representative surface, which is a single surface representing the layer in a fast navigation search. The approach surfaces describe the boundary of the layer and are the entry point into the local layer navigation. In track propagation the intersection of the approach surface is used for finding possible surface candidates within the layer object that are then tested for intersection with the trajectory. The different types of surfaces are illustrated in \autoref{fig:layer_navigation}.

\acts allows this generic geometry description to be supplemented with experiment-specific information. Each sensitive surface can have an associated object containing specific information of the particular experiment. For example, this can be used to interface with an experiment's geometry library. In addition, \acts ships with plugins which can be used to translate a geometry from an existing representation, such as \texttt{DD4Hep} ~\cite{Petri__2017}, \texttt{TGeo} or \texttt{GeoModel} ~\cite{geomodel}.

\subsubsection{Surface and Volume-based Material}

In addition to defining the exact positions and shapes of the measurement devices, the detector geometry description must provide an adequate description of the detector material. Because passive and active material is the main source of uncertainty in track reconstruction, a precise description of the amount, type and location of the material in the detector volume is required. The passive material can be handled as either deterministic changes to the trajectory estimate or stochastic addition to the covariance matrix. 

While a precise description is required for the simulation of individual interactions of the particles with the detector material, it can be simplified for track reconstruction. The material can usually be approximated as average material mixtures, described by an effective amount of traversed radiation length for evaluating the multiple scattering and bremsstrahlung contributions, and an effective ionization loss can be applied. Furthermore, small structures in the full simulation geometry can be merged into close-by approximate material slabs. This simplification speeds up the track reconstruction algorithms, because navigating and propagating through a simplified geometry requires fewer CPU cycles, predominantly due to the reduction of surface candidate intersections and fewer calls to the material integration calculations. The relative importance of accuracy and speed must be optimized for each experimental setup. \acts deploys a highly configurable approach to this problem: every surface and every volume can carry an attached material description, including the auxiliary layer surfaces and volume boundary surfaces.
Depending on the environment, corrections need to be applied during track propagation as described in \autoref{sec:propagator}, which require a precise description of the material. Therefore, the dedicated mapping algorithm in \acts projects the detailed material description onto a selected set of surfaces or into a selected set of volumes. An example of a material mapping application is discussed in \autoref{sec:appperf}.
The material description on surfaces or within volumes can be either homogeneous or binned, using the \acts grid infrastructure. When the propagation reaches a surface that carries material, the appropriate material integration methods will be called. Similarly, if the propagation proceeds within a volume that carries a material description, the corresponding extension for the transport equations become active and query the volume material.

\subsection{Propagator}
\label{sec:propagator}

A core module of \acts is the propagation engine, which carries out the task of transporting track parameters through the detector. Minimum requirements for the propagator include reliable navigation through all the detector components and the mathematical transportation of the track parameters and their associated covariance matrices. Additional actions can be performed during track parameter transport in both track reconstruction and fast simulation. An example of such an action is the intersection with additional sensitive modules. This can be used to count the number of missed sensitive detector element on a track, log several parameters or execute any particular action that can be performed along a particle's trajectory. The propagation engine therefore consists of two components:
\begin{enumerate}
    \item a \emph{Stepper} module which performs the mathematical transport through the magnetic field
    \item a \emph{Navigator} module which predicts the potential candidate surfaces in the detector geometry and regulates the associated step size for the stepper
\end{enumerate}

The propagator is steered by a dedicated options object, which is provided for each propagation call. It contains two lists of structures: a list of \emph{actors} and a list of \emph{aborters}. Both lists can be extended by the client code at compile time and are called after each propagation step and can contain surface material interaction logic (as part of the \emph{actors}), target conditions, or restrictions on the maximum allowed path length.

\begin{listing}
\begin{minted}{cpp}
// Pre-Stepping: target setting
navigator.target(state);

// Propagation loop: stepping
while (/* step */){
    // Perform a step & check the result
    stepper.step(state);
    navigator.status(state);
    // Apply the actors
    actionList(state, result);
    // Check for abort condition
    if (abortList(result, state)) break;
    // Target after stepping
    navigator.target(state);
}
\end{minted}
\caption{Simplified listing of the propagation loop showing the interplay of the \emph{Navigator}, the \emph{Stepper}, the \emph{actors} and the \emph{aborters}. The \texttt{state} object holds the cached track parameterization, while the \emph{actors} collect information in the \texttt{result} object, which in turn can be interpreted together with the \texttt{state} object for eventual abort conditions. \label{listing:propagator}}
\end{listing}

\acts includes two steppers based on a fourth-order Runge-Kutta-Nyström algorithm~\cite{Myrheim:1979ng}. One has an array-like math implementation and the other is based on the \texttt{Eigen} math library.
These steppers receive the magnetic field as an input.
For the \texttt{Eigen}-based stepper, an extension for propagation through non-vacuum material based on the simultaneous track and error propagation (STEP) algorithm~\cite{Lund:2009zzc} exists and is invoked in presence of a volume material description.
A straight-line stepper also exists, which can be used in the absence of a magnetic field.
A purely helical stepper is not implemented, but both
Runge-Kutta-Nyström based steppers can provide helical stepping behavior for a 
constant magnetic field. \autoref{listing:propagator} provides a simplified listing of the propagation loop showing the interplay between the different components.

\subsubsection{Magnetic Field Access}

\begin{figure}[ht]
\centering
\input{acts_fig_impl_magfield}
\caption{Illustration of the magnetic field cell implementation. A two-dimensional field map in the $xy$-plane is shown. The colored
circles represent propagation steps where magnetic field lookup is
performed. Step locations that fall inside each lookup field cell are indicated with the same color. Before crossing the boundary into the next cell, each step
reuses the previously retrieved field cell.
\label{fig:magfield}}
\end{figure}
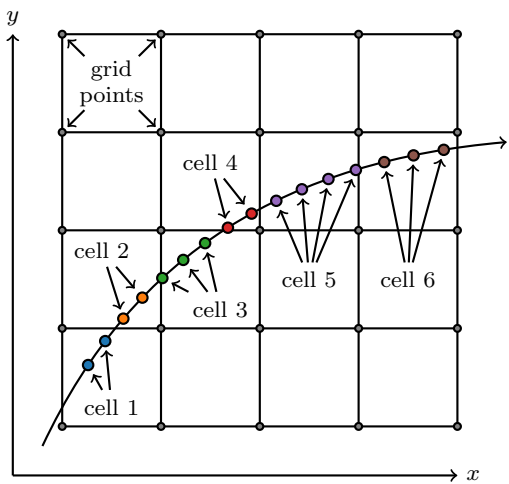

The magnetic field is accessed via a dedicated provider that is passed to the stepper modules. The implementation of
the magnetic field (both in memory and in conceptual design) can be changed and a few standard implementations are provided. An
interpolated magnetic field map, which implements an internal caching
mechanism, is also provided and can be used to approximate any inhomogeneous
magnetic field by supplying suitable input. When following a particle through
the detector, calls to the magnetic field are often made in short succession. Therefore, to optimize the
lookup or potential re-use of the magnetic field information, the steppers
access the field via a thread-local cache type. In
the implementation of the interpolated magnetic field map, this cache
contains the current field interpolation cell. A successive call to the field
interface either results in a renewed interpolation if the call remains
within the same field cell, or the retrieval of a neighboring field cell.
The field cell concept is visualized in \autoref{fig:magfield}, where a
particle trajectory is shown in the $xy$-plane with the propagation step locations color-coded according to their respective field cell.

\begin{figure}[ht]
  \centering
  \includegraphics[width=\linewidth]{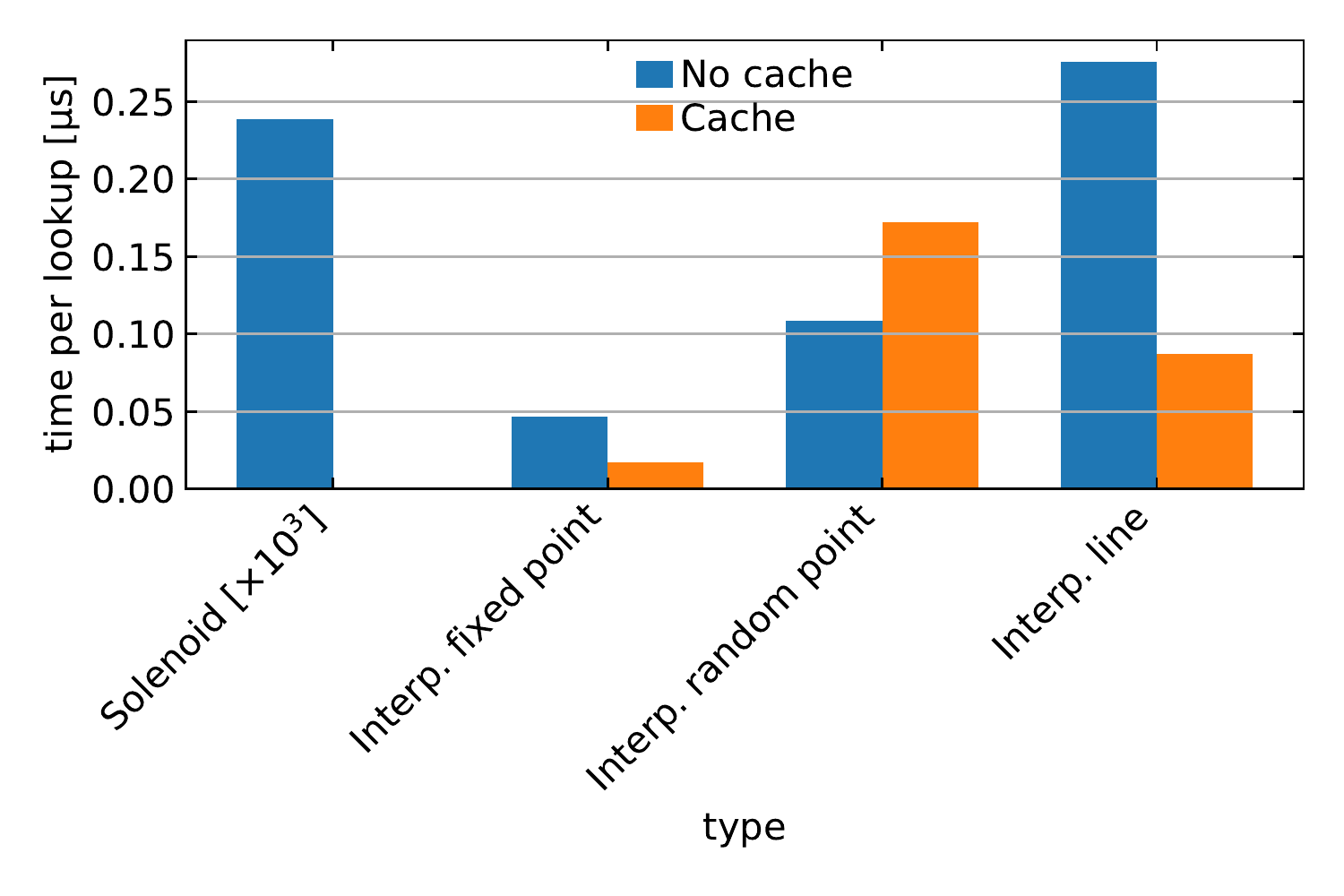}
  \caption{Performance of the magnetic field lookup for a number of different scenarios. Results for the analytical solenoid field, and the interpolated magnetic field map are shown. Field queries at a fixed point, at a sequence of random points, and a sequence along a straight line are measured. Performance with and without field interpolation cell caching are shown.}
  \label{fig:bfieldbench}
\end{figure}

\autoref{fig:bfieldbench} shows the performance of the \acts magnetic field
interpolation for different scenarios. A dynamically calculated solenoidal field is shown as a baseline. From that field, an interpolation grid is derived during initialization, and its field lookup performance is measured for a number of access patterns: a fixed point, random points, and a sequence of points along a straight line. The last emulates the typical access pattern of particle propagation. All interpolated field query strategies are approximately three orders of magnitudes faster than the solenoidal field calculation. The impact of the field cell cache is also shown. For the fixed point, the caching results in significant performance improvements, while for fully random points, it degrades the performance. This is expected because random point access will almost always result in a cache miss, while for fixed point, a cache hit is guaranteed. For the straight line access pattern, the cache again improves performance.

\subsection{Track Seed Finder}
\label{subsec:seedfinder}

Track seed finding algorithms are used as the first step in track reconstruction to obtain a coarse estimation of the possible track candidates and their properties, which are then used by track following algorithms.
The current implementation of the track seed finder in \acts creates triplets of measurements, with the goal of identifying the triplet of measurements corresponding to a single particle. In track seed finding, the goal is to maximize the efficiency while minimizing the number of seeds that do not correspond to a particle, or \textit{fakes}, and duplicates. Maximizing the efficiency is the highest priority, because particles without a seed will never be reconstructed as tracks, while fakes and duplicates can be eliminated in subsequent steps of the track reconstruction chain at the cost of execution time.

The \acts track seed finding algorithm takes three-dimensional measurements from specified detector components as input, and applies selection criteria to prioritize measurements which are more likely to have originated from the same particle. These criteria must be optimized for a particular detector geometry and play an important role in determining the physics and technical performance of the track seed finding algorithm. As track parameters derived from triplets have limited resolution, approximations are used in their estimation, including a homogeneous approximation of the magnetic field. Information about the detector geometry is not required during execution because the track seed finder relies on global measurements.

The efficiency and computational performance of the track seed finding algorithm depends on the number of measurements and the event occupancy. The higher the measurement occupancy, the higher the computational cost of following all combinatorial paths and higher the number of fakes. This can be mitigated with tighter selection criteria at the cost of lower efficiency. Moreover, higher measurement occupancy results in a higher probability that a fake measurement, instead of the real one, is assigned to the track, which leads to additional efficiency loss. As the number of detector layers increases, more measurement points are available per particle, which increases the computational requirements and the duplicate rate, but also the efficiency. The accuracy of the detector alignment also impacts efficiency.

\subsection{Kalman Filter}
\label{subsec:kalmanfilter}
The Kalman filter technique processes a set of discrete measurements to determine the internal state of a linear dynamical system. In particular, random perturbations can be present in both the measurements and the system. It is commonly used for navigation, but has applications in many domains including charged particle reconstruction. The Kalman filter is an excellent choice of algorithm for charged particle reconstruction because it facilitates a straightforward treatment of the motion of charged particles in magnetic fields and the impact of the detector material on the particle trajectories including multiple scattering and energy loss. 

Kalman filter algorithms can be used both for track finding and track fitting.
In \acts, the Kalman filter algorithm estimates the parameters of a track by 
iteratively incorporating individual measurements assigned to the track by track finding algorithms. The implementation in \acts has the mathematical filtering and smoothing in configurable components which can be replaced at compile time.
The Kalman filter class includes a propagator instance which can be configured with different detector geometries and magnetic fields. The algorithm is primarily implemented in an actor that is fed into the propagator when the track fit is executed. This actor can access the transported track parameters and their associated covariance matrices, and operate on them.
It is also configurable in terms of the representation of the track parameters and measurements, and can include an \textit{outlier}~\footnote{A measurement which is not compatible with the predicted track parameters is term an outlier.} identification helper and a calibrator for the calibration of measurements using predicted track parameters during the fitting.

The Kalman filtering method creates a track state if the propagator reaches a surface with either material or a measurement. If a measurement is found, it is investigated by the outlier identification helper. Unless the measurement is tagged as an outlier, it is used to update the track parameters by applying the filtering procedure. A \textit{hole} track state is created on any traversed sensitive surface that does not have a measurement on it. Material effects can be included either before or after the Kalman filtering. When all the measurements have been processed or the navigation reaches the boundary of the tracking geometry, the Kalman smoothing procedure is triggered to obtain the smoothed track parameters either using the Rauch-Tung-Striebel smoothing formalism~\cite{RTS} starting from the last filtered track state or using the propagator but with the navigation direction reversed.

An extension of the Kalman filter (KF), the Combinatorial Kalman filter (CKF) technique~\cite{BILLOIR1989390,BILLOIR1990219,MANKEL1997169} is implemented within \acts to perform the measurement search at the same time as performing the fit. If multiple compatible measurements are found on a surface, the track propagation branches and is repeated for multiple sets of track parameters updated with each subsequent measurement. The search for compatible measurements is handled by a measurement selector, which supports custom implementation of the selection criteria. 

Both the KF and the CKF produce fitted track parameters at a user-defined target surface and a container object, which contains all the fitted track states. For a single seed, the KF and the CKF can provide one set and multiple sets of fitted track parameters and track states, respectively.

\subsection{Vertexing}
\label{subsec:vertexing}

\acts{} features a fast and flexible primary vertex reconstruction suite, comprising a range of components implementing a full
chain from vertex seeding to precision vertex parameter estimation. The vertexing module includes an iterative vertex finder (IVF) and an adaptive multi-vertex finder (AMVF)~\cite{ATL-PHYS-PUB-2019-015}. The IVF iteratively fits individual vertices starting from a vertex seed and a seed track collection. The AMVF fits multiple vertices simultaneously, while dynamically assigning tracks to candidate vertices during fitting. The AMVF exhibits good performance for high vertex-density environments such as the HL-LHC, and will be used as the default vertex reconstruction tool for the ATLAS experiment in Run-3. 

The input vertex seeds to both vertex finders are provided by four different vertex seed finding algorithms: a $z$-scan vertex seed finder based on a half-sample mode algorithm~\cite{BICKEL20063500}, a Gaussian track density vertex seed finder ~\cite{ATL-PHYS-PUB-2019-015} as well as a non-adaptive and adaptive version of a new fast and robust grid density vertex seed finder. Dedicated vertex fitters for the different vertex finding approaches, a Billoir fitter~\cite{BILLOIR1992139} and an adaptive multi-vertex Kalman fitter~\cite{Piacquadio:1243771}, as well as auxiliary vertexing tools such as impact point estimators and track linearizers complement the vertexing toolkit. 

The public interfaces of the vertexing components are designed to be highly configurable and flexible. The vertex finders accept a collection of representations of tracks or particles to be used for vertex finding. In addition, an option structure which allows the finder to supplied with a vertex constraint is provided as input. The output of the vertexing components is a list of all found vertices. Vertex seed finders are regarded as regular vertex finders in \acts, and therefore share the same interface. They have the special characteristic of returning a single-entry list of vertices, i.e.~the vertice obtained from the current vertex seed only, at a time. The vertexing can run on \acts bound track parameter objects as well as on any user-defined input track type in order to allow maximum flexibility. The only requirement for using an arbitrary input track type is to provide a \texttt{std::function} that unwraps and returns \acts bound track parameters.

%% file: acts_fig_impl_integration.tex
\begin{tikzpicture}[scale=1.0, node distance=2cm and 3cm,line width=1pt,rounded corners=5pt]
  

    \definecolor{tab1}{RGB}{31,119,180}
    \definecolor{tab2}{RGB}{255,127,14}
    \definecolor{tab3}{RGB}{44,160,44}
    \definecolor{tab4}{RGB}{214,39,40}
    \definecolor{tab5}{RGB}{148,103,189}
    \definecolor{tab6}{RGB}{140,86,75}
    \definecolor{tab7}{RGB}{227,119,194}
    \definecolor{tab8}{RGB}{127,127,127}
    \definecolor{tab9}{RGB}{188,189,34}
    \definecolor{tab10}{RGB}{23,190,207}

    \def\green{tab3!50}
    \def\greenm{tab3!50!black}
    \def\blue{tab1!50}
    \def\bluem{tab1!50!black}
    
    \coordinate (tr) at ({\linewidth-1pt},3);
    \node[\greenm,anchor=west] (explabel) at (0,0) {Experiment};
    \coordinate (bl) at (explabel.south west);




    \newcommand{\connbox}[2]{
      \node[] (conn) at (0,0) {#1};
      \draw[#2,tab4] ($(conn.south west) + (0.1,-0.05)$) -- ($(conn.south east) + (-0.1,-0.05)$);
      \begin{scope}[on background layer]
        \fill[tab2!15!white] ($(conn.south west) - (0,0.2)$) rectangle ($(conn.north east) + (0,0)$);
      \end{scope}
    }
    
    \def\buf{0.1};

    \node[\greenm,anchor=east] (explabel2) at (0,0-|tr) {Experiment};

    \node[\bluem] (actslable) at ($(explabel.east)!0.5!(explabel2.west)$) {ACTS};

    \node[rotate=90,below] at ($(bl)!0.5!(tr-|bl)$) {Detector};
    \node[rotate=90,above] at ($(bl-|tr)!0.5!(tr)$) {Analysis};

    \node[align=center] at ($(bl)!0.5!(tr)$) {
      Tracking,\\ 
      Vertexing, \\
      ...
    };

    \begin{scope}[on background layer]
      \fill[\green,draw=black,dashed,line width=1pt] (explabel.south west) rectangle (explabel.north east|-tr);

      \fill[\green,draw=black,dashed,line width=1pt] (explabel2.south west) rectangle (explabel2.north east|-tr);

      \fill[\blue,draw=black,dashed,line width=1pt] ($(explabel.south east)+(\buf,0)$) rectangle ($(explabel2.north west|-tr) - (\buf,0)$);
    \end{scope}

    \begin{scope}[shift={($(explabel.east)!0.85!(explabel.east|-tr) + ({\buf/2},0)$)}]
      \connbox{Context objects}{<->}
    \end{scope}
    \begin{scope}[shift={($(explabel.east)!0.55!(explabel.east|-tr) + ({\buf/2},0)$)}]
      \connbox{Source links}{<->}
    \end{scope}
    \begin{scope}[shift={($(explabel.east)!0.25!(explabel.east|-tr) + ({\buf/2},0)$)}]
      \connbox{Measurements}{->}
    \end{scope}

    \begin{scope}[shift={($(explabel2.west)!0.7!(explabel2.west|-tr) - ({\buf/2},0)$)}]
      \connbox{Tracks}{->}
    \end{scope}
    \begin{scope}[shift={($(explabel2.west)!0.3!(explabel2.west|-tr) - ({\buf/2},0)$)}]
      \connbox{Vertices}{->}
    \end{scope}


  \end{tikzpicture}

%% file: acts_fig_impl_components.tex
  \begin{tikzpicture}[scale=1.0, node distance=0.5cm and 1.5cm]
  
    \def\colorGeometry{green!15}
    \def\colorUtilities{blue!10}
    \def\colorMagneticField{red!15}
    \def\colorPropagation{cyan!15}
    \def\colorEDM{orange!10}
    \def\colorFitting{yellow!10}

    \definecolor{tab1}{RGB}{31,119,180}
    \definecolor{tab2}{RGB}{255,127,14}
    \definecolor{tab3}{RGB}{44,160,44}
    \definecolor{tab4}{RGB}{214,39,40}
    \definecolor{tab5}{RGB}{148,103,189}
    \definecolor{tab6}{RGB}{140,86,75}
    \definecolor{tab7}{RGB}{227,119,194}
    \definecolor{tab8}{RGB}{127,127,127}
    \definecolor{tab9}{RGB}{188,189,34}
    \definecolor{tab10}{RGB}{23,190,207}

    \tikzset{item/.append style={rectangle,rounded corners=5pt,draw,line width=0.7pt,anchor=center,fill=white}}
    \tikzset{group/.append style={item,dashed,line width=1pt}}


    \tikzset{arr/.append style={->,shorten <=0pt,shorten >=2pt}}

    \begin{scope}[]
      \node[item,minimum width=3cm,minimum height=1.5cm] (acts) at (0,0) {\sffamily \huge \textbf{ACTS}};
      \node[above left=2cm and 2cm of acts,item,fill=tab10!50] (util) {Utilities};
    \end{scope}

    \node[below left=1cm and 2.5cm of acts,item] (interp) {Interpolated};
    \node[below=of interp,item] (prov) {Experiment-specific};

    \begin{scope}[on background layer]
      \node[fit=(interp)(prov),group,fill=tab1!50,label={Magnetic field}] (mag) {};
    \end{scope}

    \node[above left=0cm and 2.5cm of acts,item] (tparam) {Track parameters};
    \node[below=of tparam,item] (vparam) {Vertex parameters};

    \begin{scope}[on background layer]
      \node[fit=(tparam)(vparam),group,fill=tab2!50,label={Event data model}] (edm) {};
    \end{scope}


    \node[below right=1.3cm and -0cm of acts,item] (eigenstep) {Eigen\-Stepper};
    \node[right=4mm of eigenstep,item,anchor=west] (atlstep) {Array-like Stepper};
    \node[below=of eigenstep,item] (actab) {Actors \& Aborters};
    \node[below=of actab,item] (strlstep) {Straight-line Stepper};
    \draw[->] (actab) -- (eigenstep);
    \node[below=of atlstep,item] (nav) {Navigator};

    \begin{scope}[on background layer]
      \node[fit=(eigenstep)(atlstep)(actab)(nav)(strlstep),group,fill=tab3!50,label={Propagation}] (prop) {};
    \end{scope}

    \node[above right=2cm and -0.8cm of acts,item] (mat) {Material};
    \node[right=2cm of mat,item] (sensitive) {Sensitive elements};
    \node[below=of sensitive,item] (simplegeo) {Simplified geometry};

    \begin{scope}[on background layer]
      \node[fit=(mat)(sensitive)(simplegeo),group,fill=tab4!50,label={Geometry}] (geo) {};
    \end{scope}

    \node[above right=-0.3cm and 2cm of acts,item] (seeding) {Seeding};
    \node[right=of seeding,item] (track) {Tracks};
    \coordinate (xx) at ($(seeding)!0.5!(track)$);
    \node[below=0.7cm of xx,item] (vertex) {Vertices};

    \begin{scope}[on background layer]
      \node[fit=(seeding)(track)(vertex),group,fill=tab5!50,label={Fitting \& Finding}] (fitfind) {};
    \end{scope}

    \foreach \n in {edm,util,fitfind,geo,prop,mag} {
      \draw[-] (acts) -- (\n);
    }

    \tikzset{conn/.append style={arr,line width=1.5pt}}


    \draw[conn] (nav.east) -| ($(fitfind.east)+(2mm,0)$) 
    node[pos=0.8,sloped,above] {uses}
    |- (geo.east) ;

    \draw[conn] (actab.west) -| ($(mag.east|-acts.west)!0.5!(acts.west)$)
    |- ($(mat.south)+(0,-5mm)$) -- (mat.south);


    \draw[conn] (prop.north west) -- (edm.south east) node[above,pos=0.3,sloped] {uses};

    \draw[conn] (prop.west) -- (mag.east) node[below,pos=0.5,sloped] {uses};


    \draw[conn] (fitfind.south) -- (atlstep|-prop.north) node[midway,right] {uses};

    \draw[conn] (vertex) -- ($(mag.south east)!0.7!(mag.north east)$) node[pos=0.3,above,sloped]{uses};

    \draw[conn] (geo.west) -- (util) node[midway,above,sloped] {uses};



  \end{tikzpicture}

%% file: acts_fig_impl_layer_geometry.tex
\begin{tikzpicture}[scale=1.0]

    \definecolor{tab1}{RGB}{31,119,180}
    \definecolor{tab2}{RGB}{255,127,14}
    \definecolor{tab3}{RGB}{44,160,44}
    \definecolor{tab4}{RGB}{214,39,40}
    \definecolor{tab5}{RGB}{148,103,189}
    \definecolor{tab6}{RGB}{140,86,75}
    \definecolor{tab7}{RGB}{227,119,194}
    \definecolor{tab8}{RGB}{127,127,127}
    \definecolor{tab9}{RGB}{188,189,34}
    \definecolor{tab10}{RGB}{23,190,207}

    \def\passivec{tab2!80}
    \def\sensitivec{tab3!80}

    \def\ao{90}
    \def\as{45}
    \pgfmathsetmacro{\al}{-\as+\ao}
    \pgfmathsetmacro{\ar}{\as+\ao}
    \def\ri{6}
    \def\ro{8}
    \def\rvi{5.5}
    \def\rvo{8.5}
    \pgfmathsetmacro{\rm}{(\ro+\ri)/2}
    
    \newcommand{\dbox}[4]{
      \def\w{#2}
      \def\h{#3}
      \node[#4,inner sep=0,minimum width=\w,minimum height=\h] (#1) at (0,0) {};
    }
    \def\n{5}

    \coordinate (gright) at ({\linewidth-2mm},0);
    \coordinate (gleft) at (0,0);

    \pgfmathsetmacro{\wcalc}{2*cos(\al)*\rvo * 1cm}
    \pgfmathsetmacro{\wscale}{0.4 / (\wcalc/\linewidth)}
    \def\gs{\wscale}

    \coordinate (gmid) at ({0.5*\linewidth},{\gs*\rm});

    

    \begin{scope}[shift={({0.25*\linewidth},0)}]
      \begin{scope}[scale=\gs]
        \coordinate (bl) at ({-cos(\al)*\rvo}, {sin(\al)*\rvi});
        \coordinate (tr) at ({cos(\al)*\rvo}, \rvo);
        \coordinate (pad) at (6pt,6pt);
        \clip ($(bl-|gleft) - (pad)$) rectangle ($(tr) + (pad)$);
      
        \node at (0,{sin(\al)*\ri}) {(a)};
      
        \foreach \r in {\ri,\ro} {
          \draw[line width=1pt] (0,0) ++ (\al:\r) arc (\al:\ar:\r);
        }
        
        \foreach \r in {\rvi,\rvo} {
          \draw[line width=1pt,black!70] (0,0) ++ (\al:\r) arc (\al:\ar:\r);
        }
      
        \foreach \i in {0,...,\n} {
          \pgfmathsetmacro{\a}{\al + (\ar-\al)/\n * \i}
          \draw[fill=\passivec] (\a:\ro) circle(6pt);
        }

        \foreach \i in {0,...,3} {
          \pgfmathsetmacro{\ma}{-\as + \i*25 + 5}
          \begin{scope}[rotate=\ma,transform shape]
            \begin{scope}[transform shape,shift={(0,\rm)},rotate=10]
              \begin{scope}[shift={(0,0.2)}]
                \dbox{b1}{8mm}{6mm}{draw,fill=\passivec,rounded corners=3pt}
              \end{scope}
              \begin{scope}[shift={($(b1.south) + (0,-2pt)$)}]
                \dbox{b2}{3cm}{2mm}{draw,anchor=north,fill=\sensitivec,rounded corners=1pt}
              \end{scope}
              \begin{scope}[shift={($(b2.south) + (0,-1pt)$)}]
                \dbox{b3}{2cm}{3mm}{draw,anchor=north,fill=\passivec}
              \end{scope}
      
              \begin{scope}[shift={($(b1.south east) + (2pt,0pt)$)}]
                \draw[fill=\passivec] (0,0) --(7mm,0) --(0,6mm) -- cycle;
                \draw[fill=white] (1.8mm,1.8mm) circle(3.5pt);
              \end{scope}
              \draw[line width=2pt,shorten >=2pt,\passivec] (b1.north west) -- (b2.north west);
            \end{scope}
          \end{scope}
        }
      
      \end{scope}
    \end{scope}
    
    \def\thec{black!5}

    \begin{scope}[shift={({0.75\linewidth},0)}]
      \begin{scope}[draw=black!50,scale=\gs]
        \coordinate (bl) at ({-cos(\al)*\rvo}, {sin(\al)*\rvi});
        \coordinate (tr) at ({cos(\al)*\rvo}, \rvo);
        \coordinate (pad) at (6pt,6pt);
        \clip ($(bl) - (pad)$) rectangle ($(tr) + (pad)$);
      
        \node at (0,{sin(\al)*\ri}) {(b)};
      
        \foreach \i in {0,...,\n} {
          \pgfmathsetmacro{\a}{\al + (\ar-\al)/\n * \i}
          \draw[fill=\thec] (\a:\ro) circle(6pt);
        }
        
        \foreach \r in {\ri,\ro} {
          \draw[line width=1.5pt,tab10,dashed] (0,0) ++ (\al:\r) arc (\al:\ar:\r);
        }
        
        \foreach \r[count=\ri] in {\rvi,\rvo} {
          \draw[line width=1.5pt,tab1,dashed] (0,0) ++ (\al:\r) arc (\al:\ar:\r) coordinate[at end] (vl\ri) coordinate[at start] (vr\ri);
        }
      
        \coordinate (appo) at (90:\ro);
      
        \foreach \i in {0,...,3} {
          \pgfmathsetmacro{\ma}{-\as + \i*25 + 5}
          \begin{scope}[rotate=\ma,transform shape]
            \begin{scope}[transform shape,shift={(0,\rm)},rotate=10]
              \begin{scope}[shift={(0,0.2)}]
                \dbox{b1}{8mm}{6mm}{draw,fill=\thec,rounded corners=3pt}
              \end{scope}
              \begin{scope}[shift={($(b1.south) + (0,-2pt)$)}]
                \dbox{b2}{3cm}{2mm}{draw,anchor=north,fill=\sensitivec,draw=black,rounded corners=1pt}
                \coordinate (sepo\i) at (15mm,0);
              \end{scope}
              \begin{scope}[shift={($(b2.south) + (0,-1pt)$)}]
                \dbox{b3}{2cm}{3mm}{draw,anchor=north,fill=\thec}
              \end{scope}
      
              \begin{scope}[shift={($(b1.south east) + (2pt,0pt)$)}]
                \draw[fill=black!15] (0,0) --(7mm,0) --(0,6mm) -- cycle;
                \draw[fill=white] (1.8mm,1.8mm) circle(3.5pt);
              \end{scope}
              \draw[line width=2pt,shorten >=2pt,\thec] (b1.north west) -- (b2.north west);
            \end{scope}
          \end{scope}
        }
        
        \def\aext{6}
        \draw[line width=1.5pt,tab4] (0,0) ++ ({\al-\aext}:\rm) arc ({\al-\aext}:{\ar+\aext}:\rm) coordinate[at end] (repo);
      
        \coordinate (label) at (0,{\rvo + 0.8});
        \coordinate (vlabloc) at (0,{\rvi - 0.8});
        \coordinate (bl2) at ({\ar}:\rvi);
      
      \end{scope}
    \end{scope}
    
    \node[anchor=base] (alabel) at (label) {Approach};

    \begin{scope}[line width=1pt,shorten >=2pt]

      \node[below] (rlab) at (bl2-|gmid) {Representative};
      \draw[->] (rlab) |- (repo);

      \node (vlab) at (rlab-|{\linewidth*0.75},0) {Volume bounds};

      \draw[->,line width=1pt,shorten >=1pt] (vlab) -| (vl1);
      \coordinate (lix) at ($(vr2)!0.6!(vr2-|gright)$);
      \draw[->,line width=1pt,shorten >=1pt] 
        (vlab) 
        -- (vlab-|vr2) 
        node[below,pos=0.6] (xx) {}
        -| (lix) 
        -- (vr2);

      \draw[->] (alabel) -- (appo);


    
      \coordinate (ml) at (0, {(\ro + sin(\al)*\ri)/2});

    \end{scope}

    \begin{scope}[rounded corners=2pt]
      \node[fill=\sensitivec,left=5pt] at (alabel-|{0.25*\linewidth},0) {Sensitive};
      \node[fill=\passivec,right=5pt] at (alabel-|{0.25*\linewidth},0) {Passive};
    \end{scope}
    
\end{tikzpicture}

%% file: acts_fig_impl_magfield.tex
  \begin{tikzpicture}[scale=1.3,line width=0.8pt]
    \definecolor{tab1}{RGB}{31,119,180}
    \definecolor{tab2}{RGB}{255,127,14}
    \definecolor{tab3}{RGB}{44,160,44}
    \definecolor{tab4}{RGB}{214,39,40}
    \definecolor{tab5}{RGB}{148,103,189}
    \definecolor{tab6}{RGB}{140,86,75}
    \definecolor{tab7}{RGB}{227,119,194}
    \definecolor{tab8}{RGB}{127,127,127}
    \definecolor{tab9}{RGB}{188,189,34}
    \definecolor{tab10}{RGB}{23,190,207}

    \tikzstyle{every node}=[font=\small]
    \begin{scope}
      \def\xmax{4}
      \def\ymax{4}
      \foreach \y in {0,...,\ymax}{
        \draw (0, \y) -- (\xmax, \y);
      }
      \foreach \x in {0,...,\xmax}{
        \draw [] (\x, 0) -- (\x, \ymax);
        \foreach \y in {0,...,\ymax}{
          \draw[fill=gray]  (\x, \y) circle (1pt);
        }
      }

      \draw [->] (-0.5,-0.5) -- (-0.5,\ymax) node [above] {$y$};
      \draw [->] (-0.5,-0.5) -- (\xmax,-0.5) node [right] {$x$};

      \draw [->] (-0.2,-0.2) to[out=65, in=185]%
        node foreach[evaluate={\pi*0.05 + 0.10} as \p] \pi in {1,...,20} [pos=\p] (n\pi) {}%
        (4.5, 2.9);

      \node (g1) at (0.5,0.2) {cell 1};
      \node (g2) at (0.4,1.8) {cell 2};
      \node (g3) at (1.6,1.2) {cell 3};
      \node (g4) at (1.5,2.7) {cell 4};
      \node (g5) at (2.5,1.5) {cell 5};
      \node (g6) at (3.5,1.5) {cell 6};

      \foreach \n in {1,...,2} {
        \draw [fill=tab1] (n\n) circle (1.5pt);
        \draw[->] (g1) -- (n\n);
      }
      \foreach \n in {3,...,4} {
        \draw [fill=tab2] (n\n) circle (1.5pt);
        \draw[->] (g2) -- (n\n);
      }
      \foreach \n in {5,...,7} {
        \draw [fill=tab3] (n\n) circle (1.5pt);
        \draw[->] (g3) -- (n\n);
      }
      \foreach \n in {8,...,9} {
        \draw [fill=tab4] (n\n) circle (1.5pt);
        \draw[->] (g4) -- (n\n);
      }
      \foreach \n in {10,...,13} {
        \draw [fill=tab5] (n\n) circle (1.5pt);
        \draw[->] (g5) -- (n\n);
      }
      \foreach \n in {14,...,16} {
        \draw [fill=tab6] (n\n) circle (1.5pt);
        \draw[->] (g6) -- (n\n);
      }

      \node[align=center] (gl) at (0.5, 3.5) {grid \\points};
      \foreach \x in {0,1}{
        \foreach \y in {3, 4} {
          \draw [->,shorten >=1mm,shorten <=-1mm] (gl) -- (\x, \y);
        }
      }

    \end{scope}

  \end{tikzpicture}

%% file: applications.tex
\section{Applications and Performance}
\label{sec:appperf}

\subsection{Selected Applications}
\label{subsec:application}
\begin{figure}[ht!]
  \centering
  \begin{subfigure}{\linewidth}
    \centering
    \includegraphics[width=\linewidth]{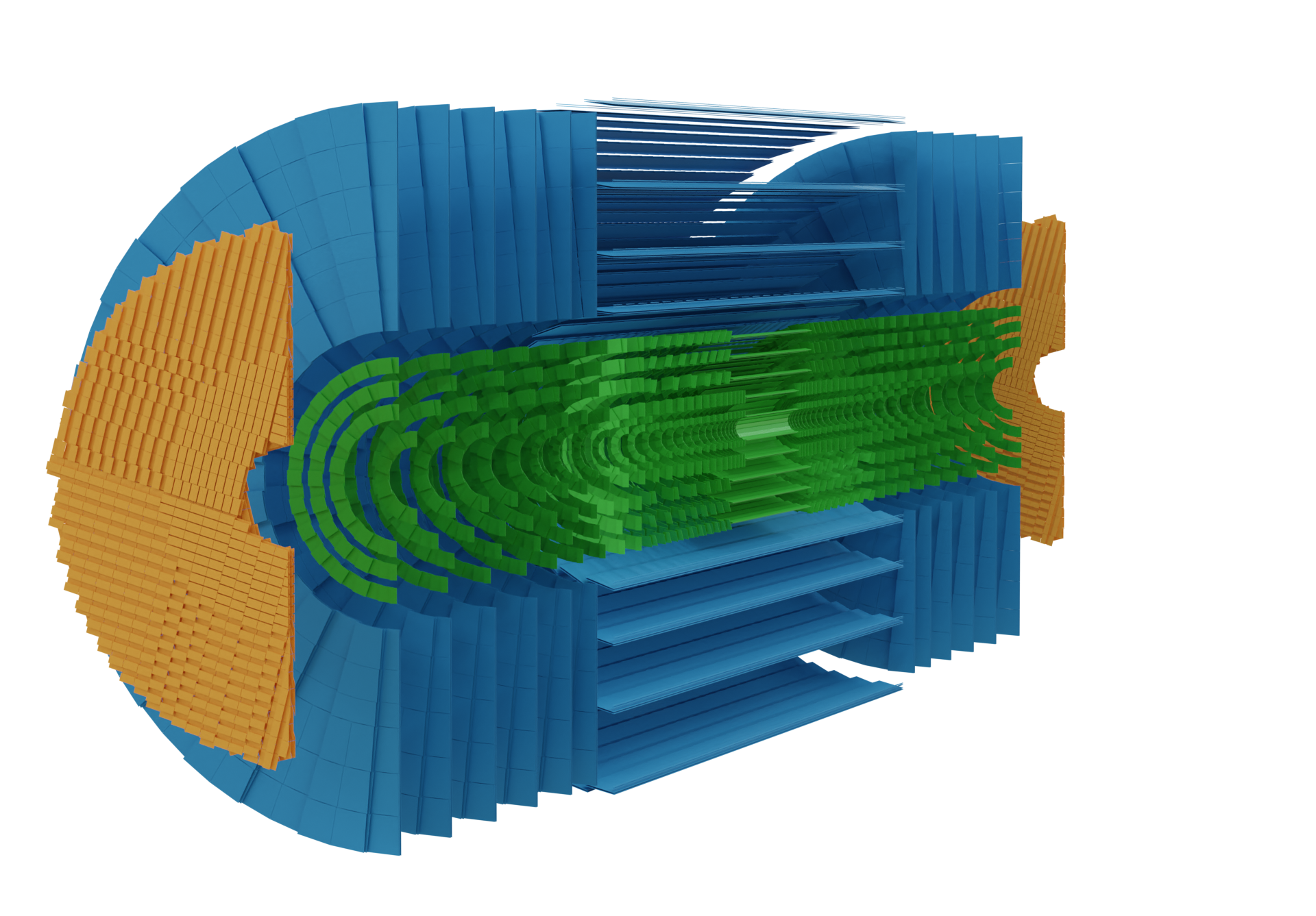}
    \caption{}
    \label{fig:det_geometry_itk}
  \end{subfigure}
  \begin{subfigure}{\linewidth}
    \centering
    \includegraphics[width=\linewidth]{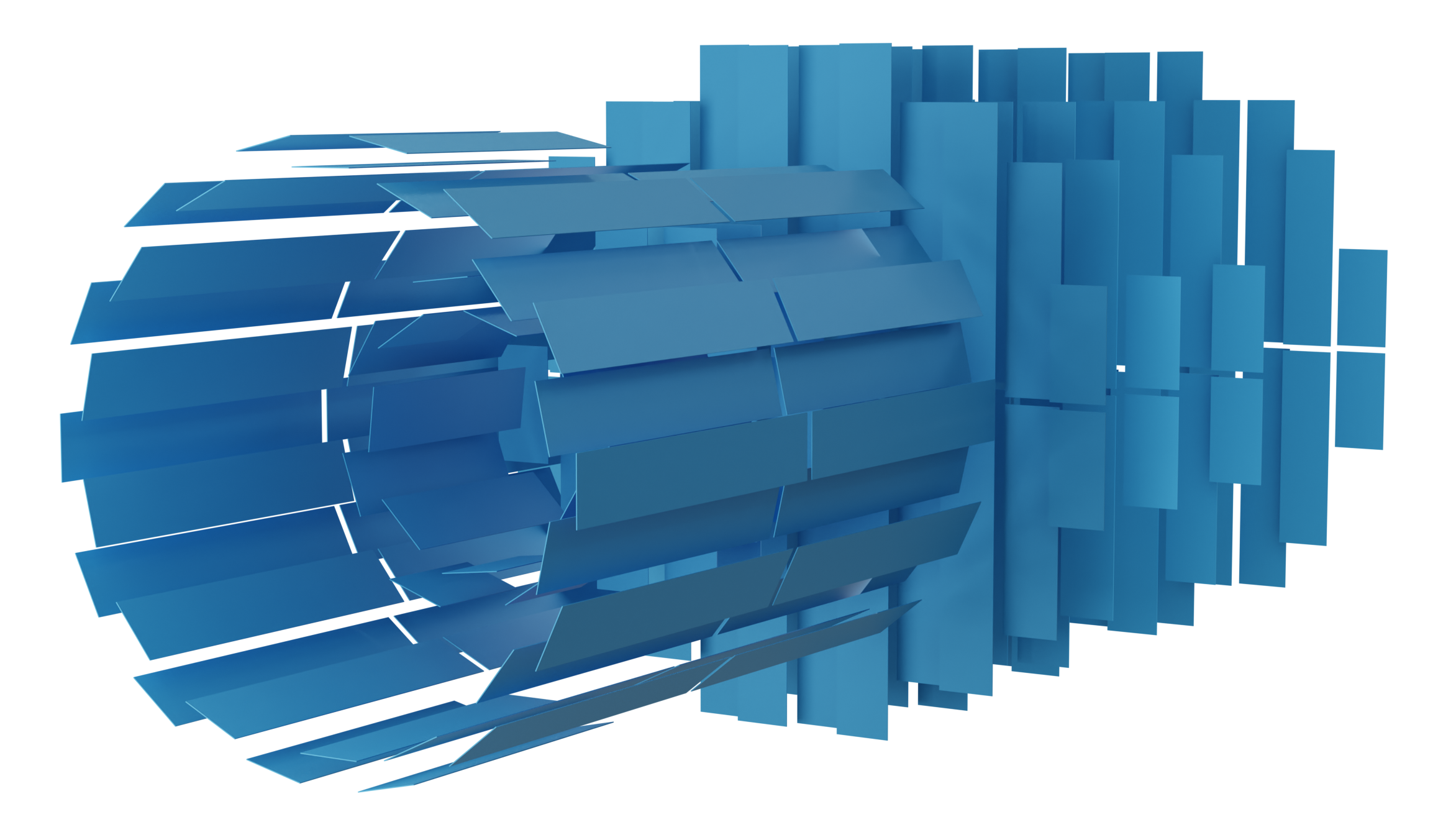}
    \caption{}
    \label{fig:det_geometry_panda}
  \end{subfigure}
  \begin{subfigure}{\linewidth}
    \centering
    \includegraphics[width=0.8\linewidth]{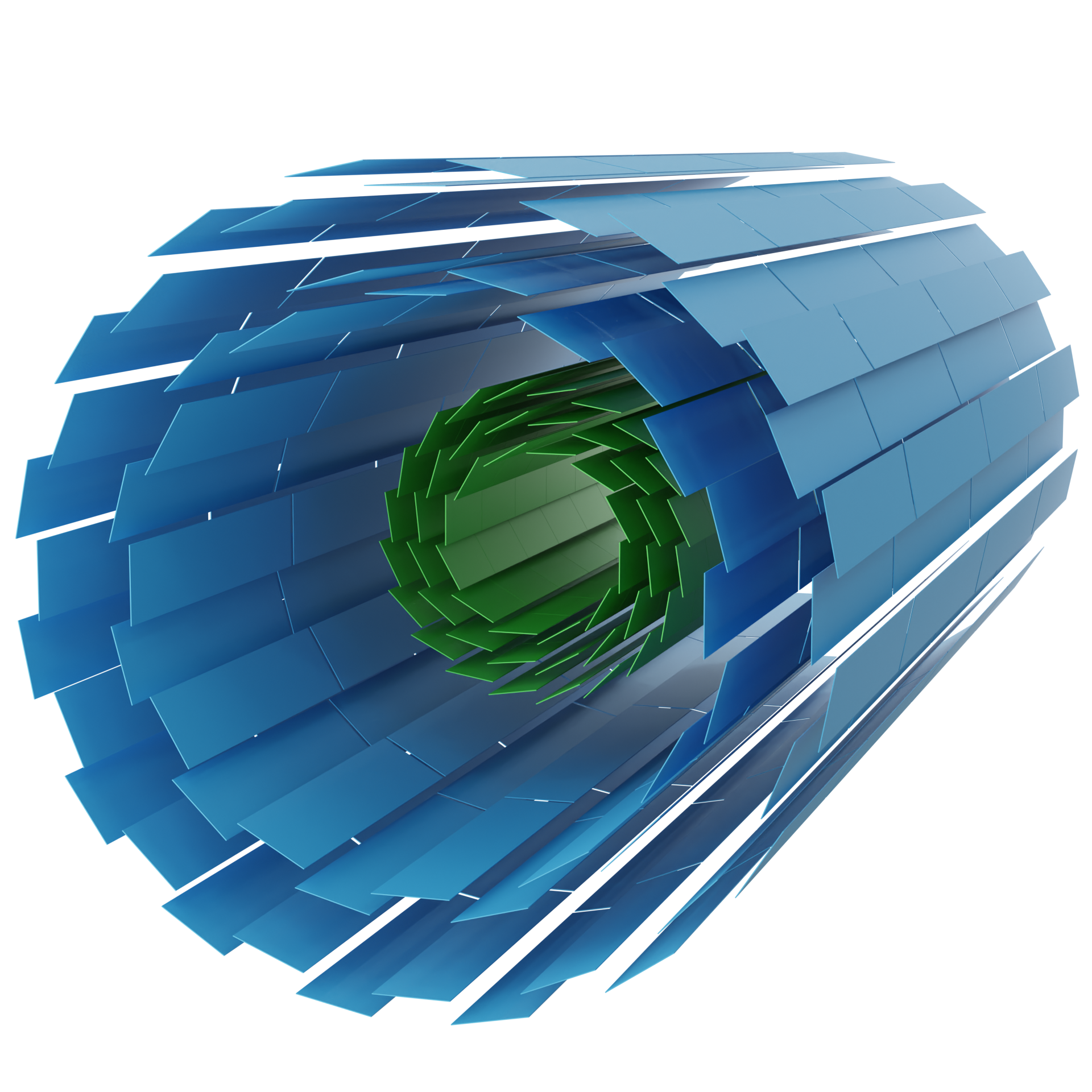}
    \caption{}
    \label{fig:det_geometry_sphenix}
  \end{subfigure}
  \caption{The geometry of the ATLAS ITk (\subref{fig:det_geometry_itk}), the PANDA silicon detector (\subref{fig:det_geometry_panda}) and the sPHENIX silicon tracking detectors (\subref{fig:det_geometry_sphenix}), implemented with \acts. Colors indicate different subsystems, in the top image, the High Granularity Timing Detector (HGTD)~\cite{CERN-LHCC-2020-007} is shown in orange.}

  \label{fig:det_geometry}
\end{figure}

\begin{figure}[!ht]
  \centering
   \begin{subfigure}{\linewidth}
    \centering
    \includegraphics[width=0.8\linewidth]{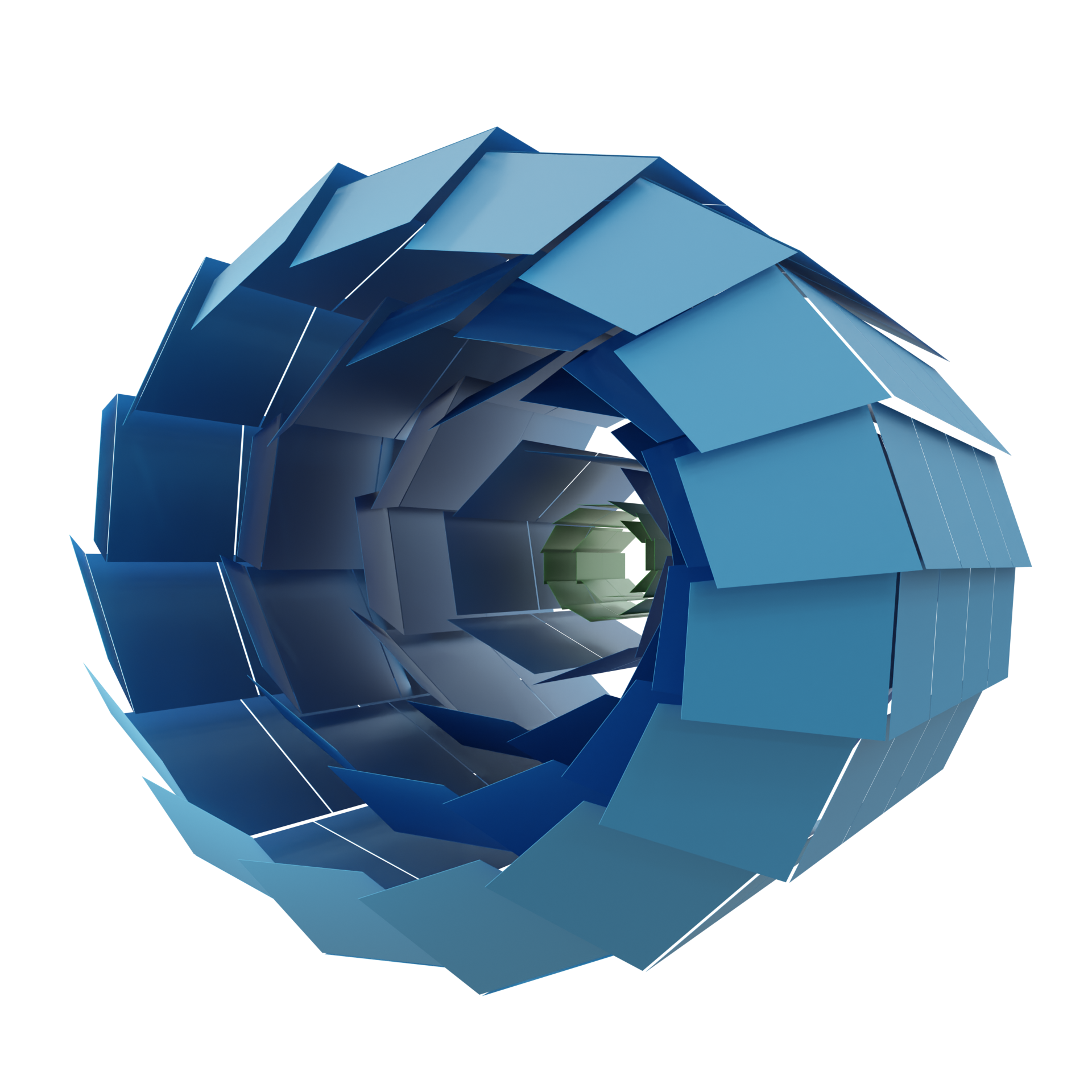}
    \caption{}
    \label{fig:det_geometry_belle2}
  \end{subfigure}
  \begin{subfigure}{\linewidth}
    \centering
    \includegraphics[width=\linewidth]{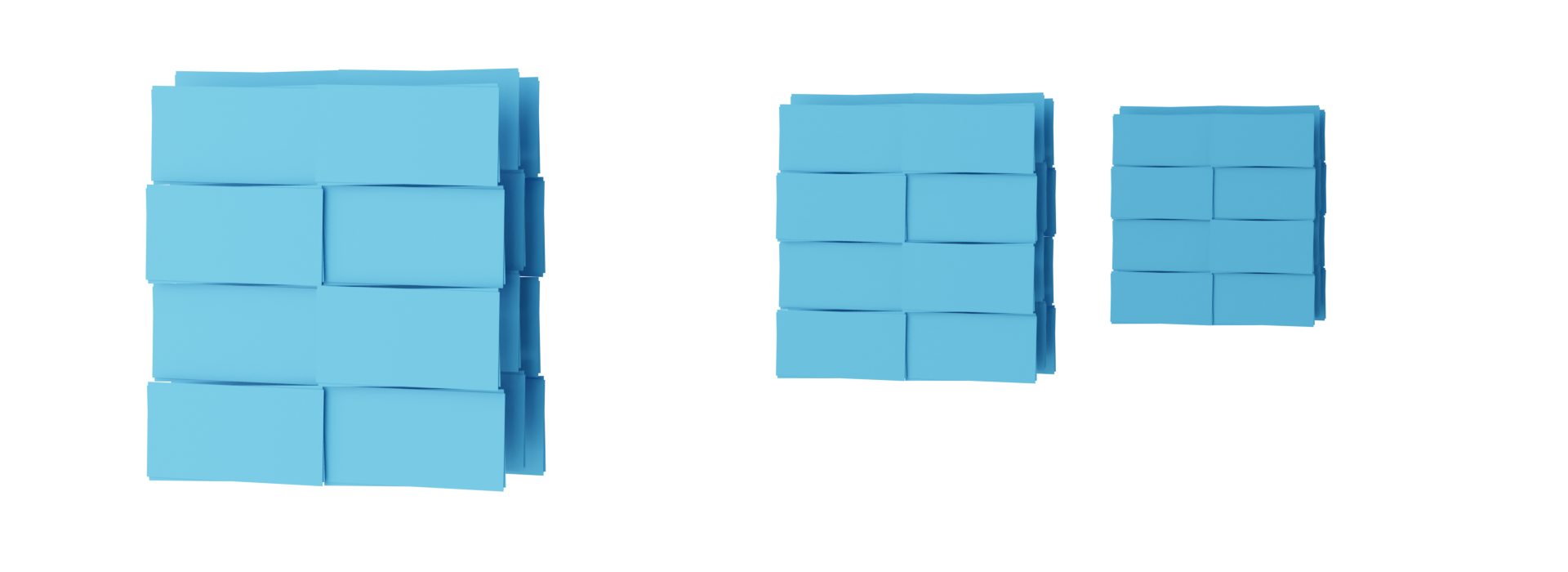}
    \caption{}
    \label{fig:det_geometry_faser}
  \end{subfigure}
  
  \caption{Geometries of Belle~II
  (\subref{fig:det_geometry_belle2}) and FASER (\subref{fig:det_geometry_faser}) implemented in ACTS. Colors indicate the different subsystems.}
  \label{fig:det_geometry_2}
\end{figure}

\acts is integrated or being integrated into a number of particle and nuclear physics experiments. Here we presented selected examples of experiments that either use or have explored the use of \acts.
\autoref{fig:det_geometry_itk} shows the geometry of the
ATLAS ITk. The \acts{} vertexing algorithms have already been integrated into the ATLAS Athena framework and the integration of the \acts{} tracking algorithms is ongoing. At the same time, preliminary optimization of the \acts{} tracking algorithms for ITk is in progress. \autoref{fig:det_geometry_panda} shows the geometry for the silicon tracker of the PANDA experiment, which is a planned particle physics experiment at the FAIR facility in Germany. 

The sPHENIX experiment is the next generation jet and heavy-flavor detector currently under construction at the Relativistic Heavy Ion Collider at Brookhaven National Laboratory. \autoref{fig:det_geometry_sphenix} shows the geometry for the
silicon tracker of sPHENIX. \acts{} components for seeding, track fitting and vertexing have been successfully deployed in the sPHENIX
production software chain.  

Belle~II is the next generation $B$-factory experiment located at the SuperKEKB
accelerator complex~\cite{Akai:2018mbz} in Japan.  A critical requirement is to reliably reconstruct low-momentum tracks with $p_T \approx 100 - 300 \,
\mathrm{MeV}$~\cite{Bertacchi:2020eez}. This is achieved with a combination of
silicon pixel and strip detectors, whose placement is shown in
\autoref{fig:det_geometry_belle2}. The Belle~II collaboration is currently
exploring in what form ACTS can supplement or replace existing tracking code.

Figure~\ref{fig:det_geometry_faser} shows the FASER detector, which is an experiment at the LHC, located $\simeq 480$\,m downstream the ATLAS
interaction point, featuring extremely forward acceptance ($\eta > 9.2$). The
FASER tracker is designed to detect two high-momentum charged tracks originating
from a decay vertex inside the decay volume, using three tracking stations with
silicon strip sensors, in a $0.55\,\mathrm{T}$ magnetic field. FASER will fully
rely on \acts{} for its track reconstruction and fitting. The implementation is
well progressed and first performance studies with the ACTS CKF are in
preparation.

\begin{figure}[htb!]
\centering
\begin{subfigure}{\linewidth}
  \centering
  \includegraphics[width=0.8\linewidth]{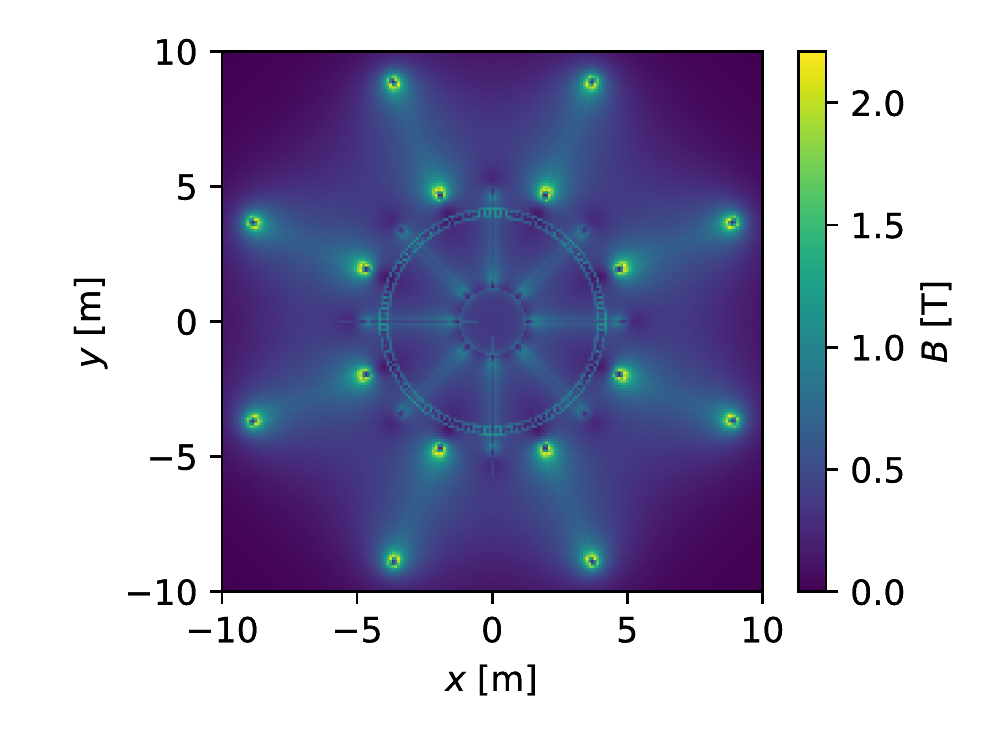}
  \caption{}
  \label{fig:atlas_bfield_xy}
\end{subfigure}
\begin{subfigure}{\linewidth}
  \centering
  \includegraphics[width=1.0\linewidth]{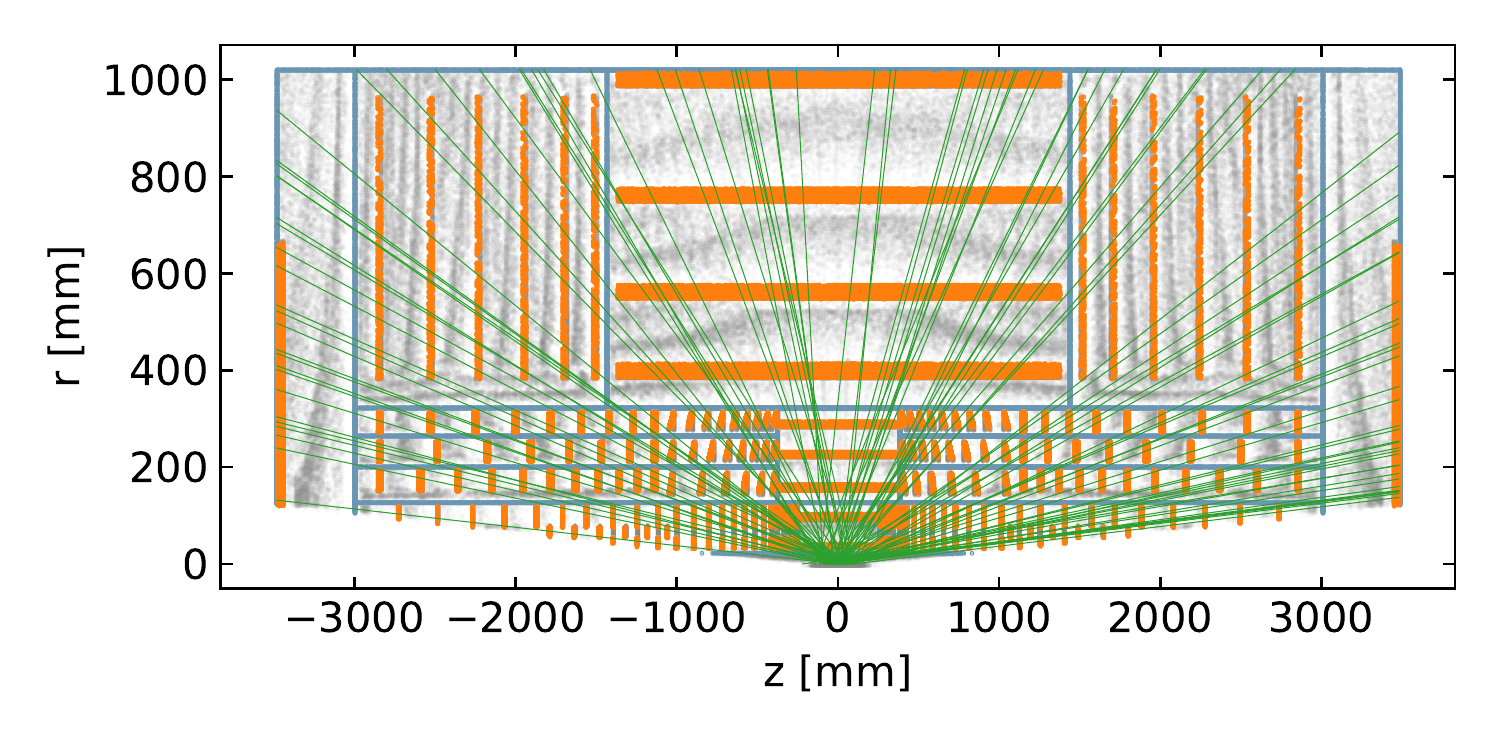}
  \caption{}
  \label{fig:atlas_bfield_itk_prop}
\end{subfigure}

\caption{A projection of the magnetic field implemented with \acts for the ATLAS tracking system into the $x-y$ plane (\subref{fig:atlas_bfield_xy}). The strength of the magnetic field at each point is indicated in color. The $r-z$ coordinates of the intersections of propagated pion tracks with the ATLAS ITk detector elements, using the ATLAS magnetic field (\subref{fig:atlas_bfield_itk_prop}). Boundary intersections are shown in blue, while intersections with sensors are shown in orange. Green lines indicate a subset of extrapolated particle tracks. Grey points are the intermediate integration steps, required within a predefined tolerance threshold.}
\label{fig:atlas_bfield_prop}
\end{figure}

\autoref{fig:atlas_bfield_xy} shows the magnetic field of the ATLAS experiment described using \acts.  Track parameter propagation based on
the detector geometry and magnetic field is used to determine the coordinates of
intersections of tracks with detectors. An example of track propagation with the
ATLAS ITk Detector is shown in \autoref{fig:atlas_bfield_itk_prop}.

\begin{figure}[htb]
\centering
\includegraphics[width=\linewidth]{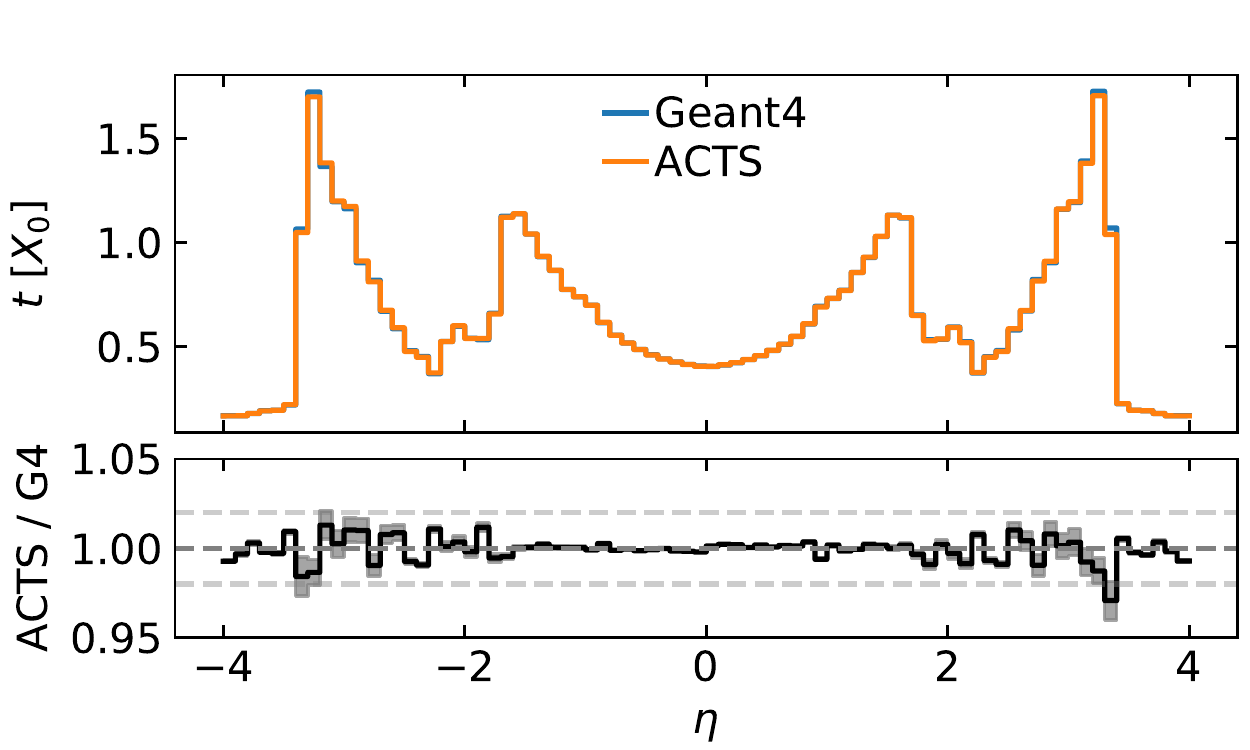}
\caption{Comparison of the mapped material obtained from \acts material mapping tool (orange line) and the \texttt{Geant4} material (blue line) as a function of $\eta$ for the Open Data Pixel Detector. The ratio of the material in ACTS to Geant4 is indicated in the panel below and the statistical uncertainty is indicated with the gray band. Agreement is within about 2\%, with excellent agreement seen in the central part of the detector. \label{fig:mat_mapping}}
\end{figure}

When using the simplified tracking
geometry described in \autoref{sec:geometry}, the detector material is modeled using a dedicated mapping algorithm that
remaps the detailed \texttt{Geant4} material. A comparison of the
mapped material with the material used in the full simulation geometry for the Open Data
Detector~\cite{ODD} is shown in \autoref{fig:mat_mapping}. The geometry of the Open Data Detector is described with a realistic passive material model based on DD4hep, which translates into a \texttt{Geant4} detector model. The agreement between the material budget described in \texttt{Geant4} and by the \acts{} geometry is within a few percent and can be further improved by using higher granular
binning of the material maps.

\subsection{Examples of a Track and Vertex Reconstruction Chain for the LHC}
\label{subsec:recochain}

At the LHC, track reconstruction typically proceeds through a multi-step process, which we briefly outline here. The procedure is largely similar for different experiments, but with some key differences in strategy. For example, the CMS experiment uses an iterative tracking approach~\cite{Chatrchyan:2014fea} in which the full track reconstruction pass is repeated a number of times, but with different configurations, and the measurements corresponding to tracks that have already been reconstructed are removed. ATLAS instead relies to a large extent on a single track reconstruction pass, but with loose track candidate search and an ambiguity resolution step to resolve between the multiple track candidates. Additional passes are used to target particular topologies, e.g. tracks produced at large radii.

As the first step, the energy deposited in the silicon detectors is grouped into \textit{clusters} with each cluster ideally corresponding to the energy deposited by a single particle. The clusters are three-dimensional space points formed from either a single pixel cluster or a pair of strip clusters with stereo angle between them from each side of a module, depending on the sensor technology.

Next, \textit{seed} finding algorithms are used to reconstruct the seeds. 
The seeds passing a set of selection cuts are used to initiate the track finding and following algorithms, such as the CKF. After track following, the ATLAS experiment runs an ambiguity resolution algorithm to resolve duplicate tracks and remove fakes~\cite{Cornelissen:2008zzc}. Track candidates are scored based on track properties such as the number of clusters, holes, shared clusters, and fitting quality. The scoring procedure is iterated until the selected set of track candidates are obtained. Next, the track candidates are extended into the Transition Radiation Tracker, consisting of gas-filled drift tubes, to search for additional measurements to improve the momentum resolution.

As the final track reconstruction step, if needed, a precise estimate of the track parameters is determined from \textit{track fitting} algorithms, including the KF and Global $\chi^2$ methods. The final track candidates are selected based on a set of track quality metrics, e.g. the number of clusters and holes, and the estimated track parameters. For example, the track candidates are usually required to satisfy a set of requirements for the momentum and impact parameters.

After track reconstruction, primary vertex candidates are reconstructed using the reconstructed tracks with estimated perigee track parameters at the beam line. Both ATLAS and CMS use an adaptive approach for primary vertex reconstruction~\cite{Borissov_2015,Chatrchyan:2014fea} similar to the \acts Adaptive Multi-Vertex Finder (AMVF).

\subsection{Track and Vertex Reconstruction Performance}
\label{subsec:physicsperf}

\begin{figure}[!htb]
  \centering
  \includegraphics[width=\linewidth]{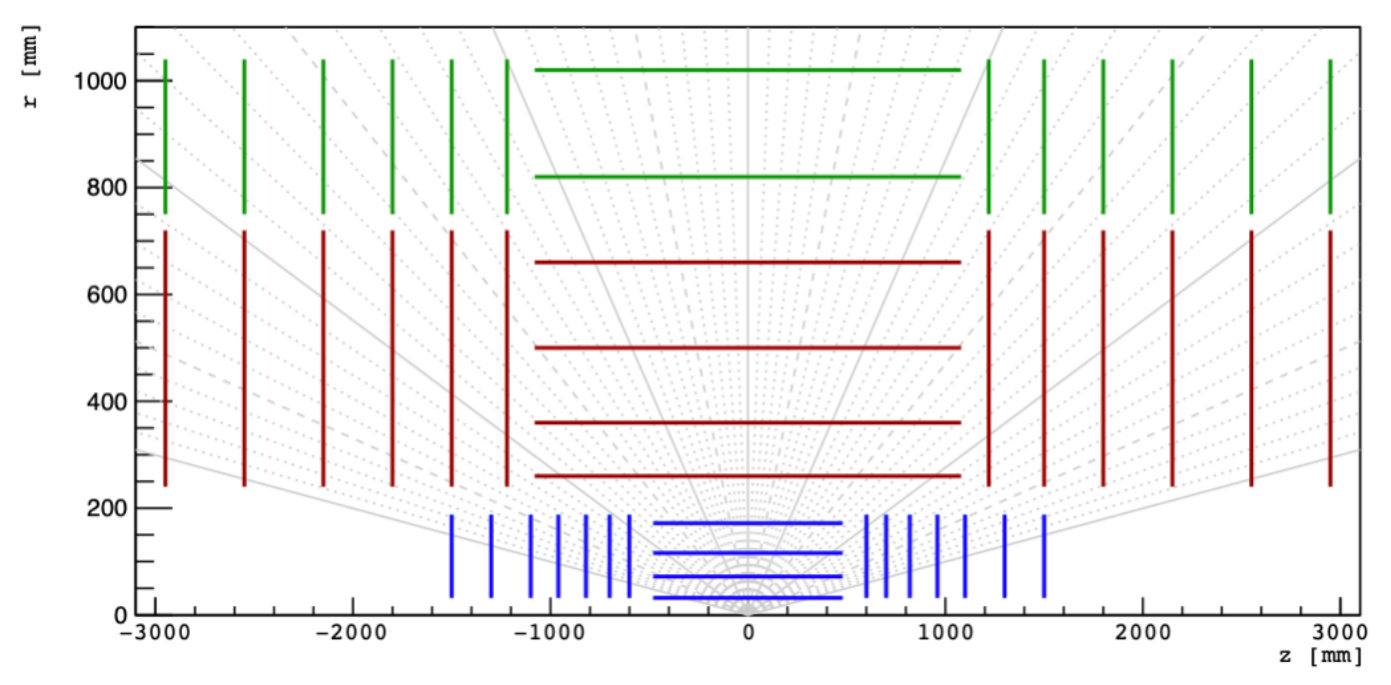}
  \caption{Schematic layout of the TrackML detector showing the coverage of the pixel detector in blue, short strip detector in red and long strip detector in green. \label{fig:trackML_layout}}
\end{figure}

As discussed in Section~\ref{subsec:recochain}, experimental applications of track reconstruction usually include many steps depending on the algorithms used, the collision environment, and the required precision. Here, an example of simplified track reconstruction chain based on a combined effort of track seeding and the CKF is discussed. The detector used for the TrackML challenge has the layout as shown in \autoref{fig:trackML_layout}, and a solenoidal magnetic field with a strength of 2\,T centered on the beam line, is used to demonstrate the \acts track and vertex reconstruction performance.

Particles generated using a particle gun, both muons and charged pions, and particles from the \ttbar physics process generated in $pp$ collisions at a center-of-mass energy of 14\,TeV, the energy target for the HL-LHC, with the PYTHIA 8 generator~\cite{SJOSTRAND2008852, Sj_strand_2006} are used for the performance studies. The single particle samples either contain a single particle per event for physics performance studies or a thousand particles within a single event for timing performance studies. As muons have little sensitivity to detector material, they are used to study the technical performance of the track reconstruction algorithms, while the pions are used to study the sensitivity of the track reconstruction algorithms to material. No pile-up is included in the single particle events. 
Two \ttbar samples are produced: one with $\langle \mu \rangle = 200$ to match the highest pile-up foreseen for the HL-LHC and the other with $\langle \mu \rangle$ varying from 0 to 300 to allow the dependence of the performance on pile-up to be studied. The interactions of the generated particles with transverse momentum, $\pT > 400$\,MeV and pseudorapidity, $\eta$\footnote{Pseudorapidity is an angular quantity calculated as follows $\eta = -\ln\left[\tan\left(\frac \theta 2\right)\right]$ from the polar angle $\theta$. $\eta = \pm\infty$ corresponds to directions along the beam axis.}, within $|\eta|<2.5$ with the detector are simulated with Fatras, the \acts{} fast simulation library. 

Detector readout and measurement creation are detector-specific\footnote{For silicon detectors, this often requires finding connected readout cells and either processing or emulating the detector signal.}, hence a smearing algorithm applies module-specific resolutions to emulate the input measurements based on the simulated hits. The space-points constructed from the emulated measurements in the innermost four pixel layers are grouped into seeds using the \acts seeding algorithm as described in \autoref{subsec:seedfinder}. Both truth and reconstructed seeds are used. Truth seeds eliminate the pattern recognition step in the seed finding, i.e. the truth information is used to identify the hits for a seed corresponding to a true particle. \textit{Truth-generated seeds} are produced by smearing the particle properties at its point of generation. \textit{Truth-propagated seeds} are produced by smearing the true particle information at the first detector layer. \textit{Reconstructed seeds} are the output of the seed finding algorithm based on the simulated hits.

While each truth seed is a set of initial track parameters with associated covariance matrix, estimation of the track parameters with associated covariance matrix at the surface of the innermost space point is performed for each reconstructed seed. These initial track parameters based on either the truth seeds or the reconstructed seeds and are used to seed the CKF algorithm. After the CKF algorithm has been run, the reconstructed track candidates must satisfy a set of track quality cuts. The reconstructed tracks are required to have at least six measurements based on expectations from the TrackML detector layout and to allow initial track parameters to be located in any of the first three layers of the pixel detector. Four different types of tracks are studied, which allows the effects of the different steps in the track reconstruction sequence to be disentangled. The \textit{truth tracks} ignore the pattern recognition entirely and are based on the properties of the simulated hits of the true particles. The \textit{truth-generated-seeded tracks}, \textit{truth-propagated-seeded tracks} and \textit{reco-seeded tracks} are reconstructed by running the CKF on the truth-generated seeds, the truth-propagated seeds and the reconstructed seeds, respectively.

The performance of the \acts primary vertex reconstruction module is evaluated using truth tracks with fitted perigee track parameters defined at the beam line using the same detector, magnetic field and simulation configuration as used for the studies of the track reconstruction.

\subsubsection{Track Reconstruction Efficiency and Fake and Duplicate Rates}

\begin{figure}[htbp!]
\centering
\includegraphics[width=0.8\linewidth]{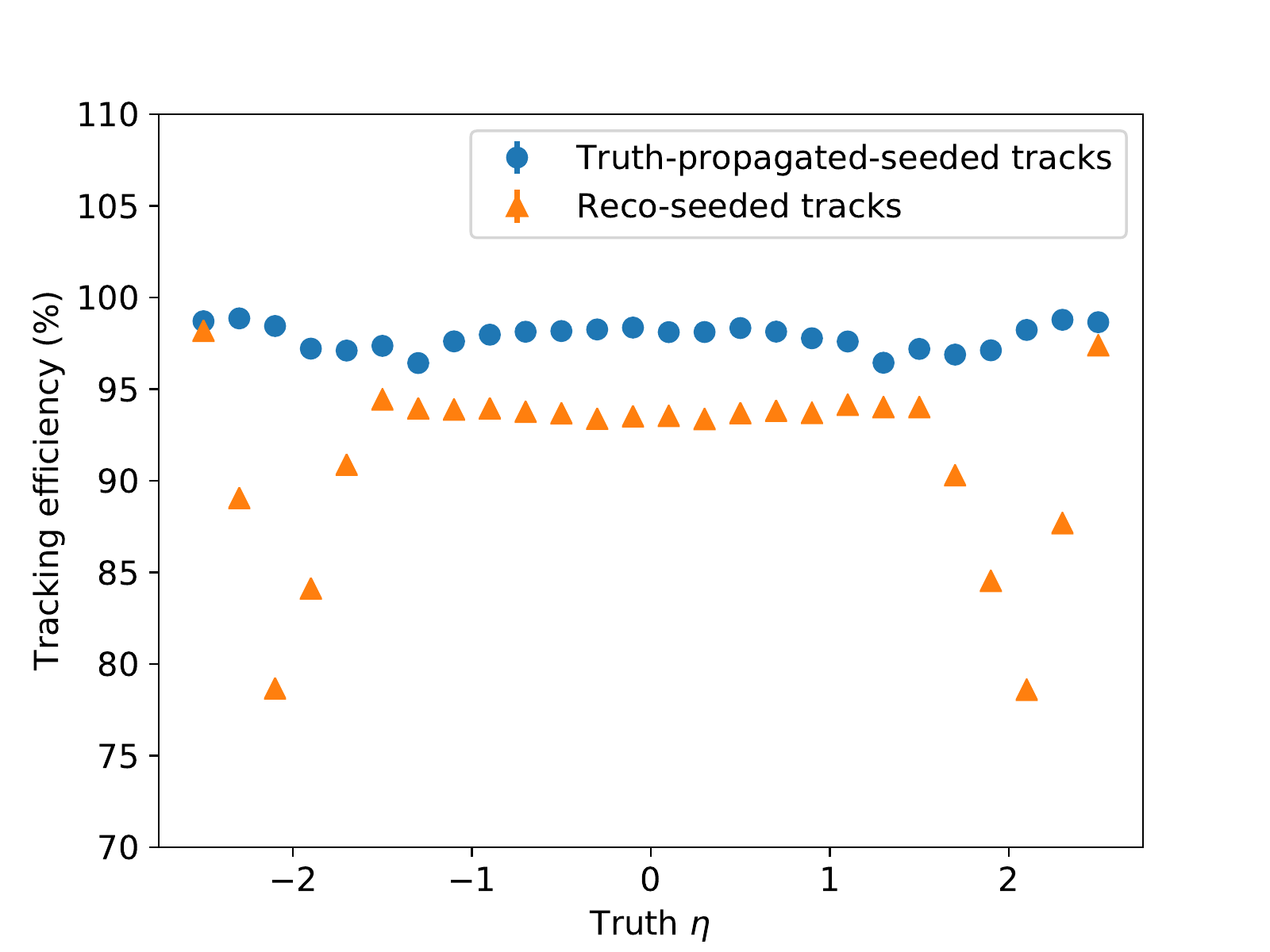}
\includegraphics[width=0.8\linewidth]{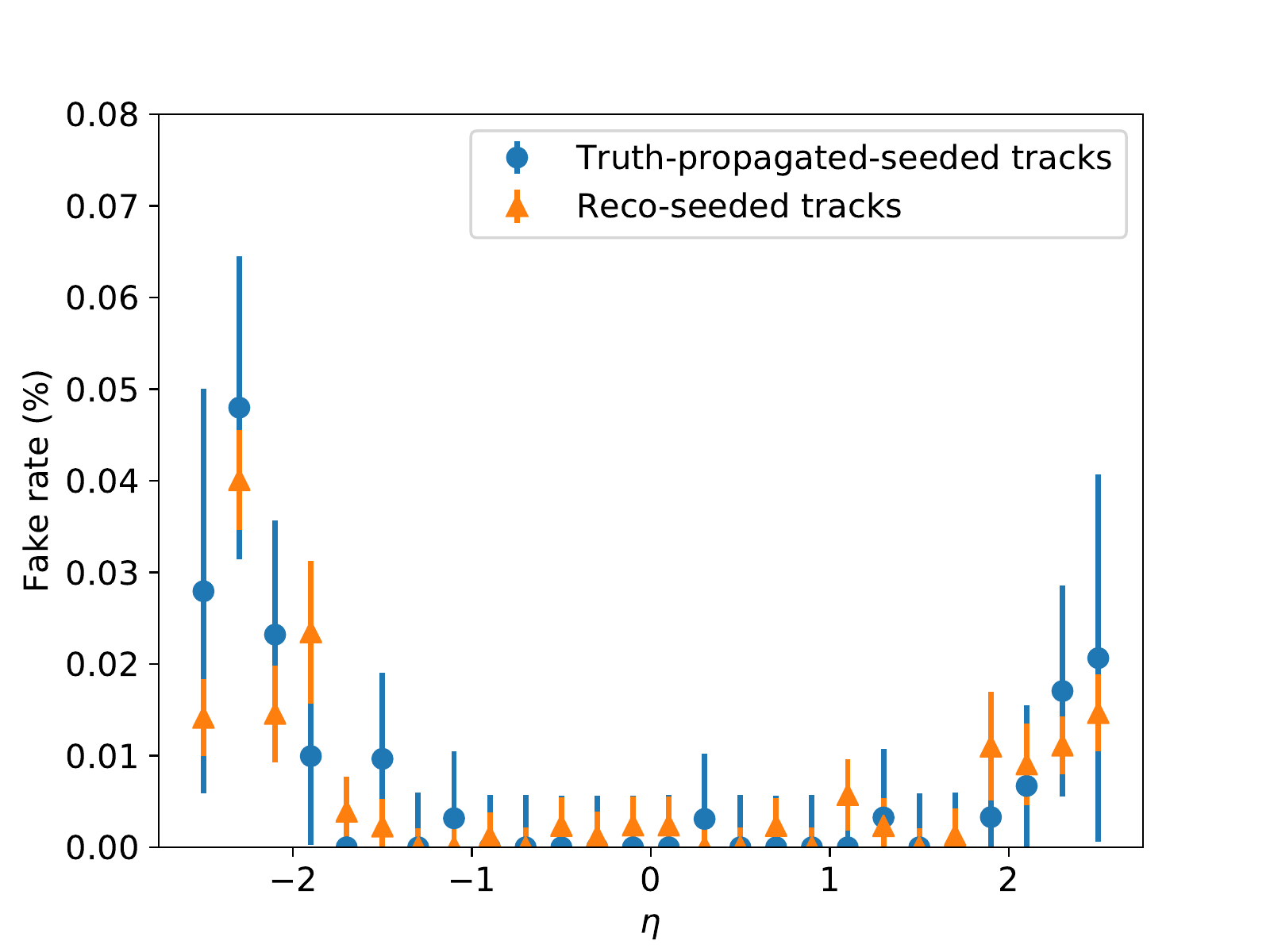}
\includegraphics[width=0.8\linewidth]{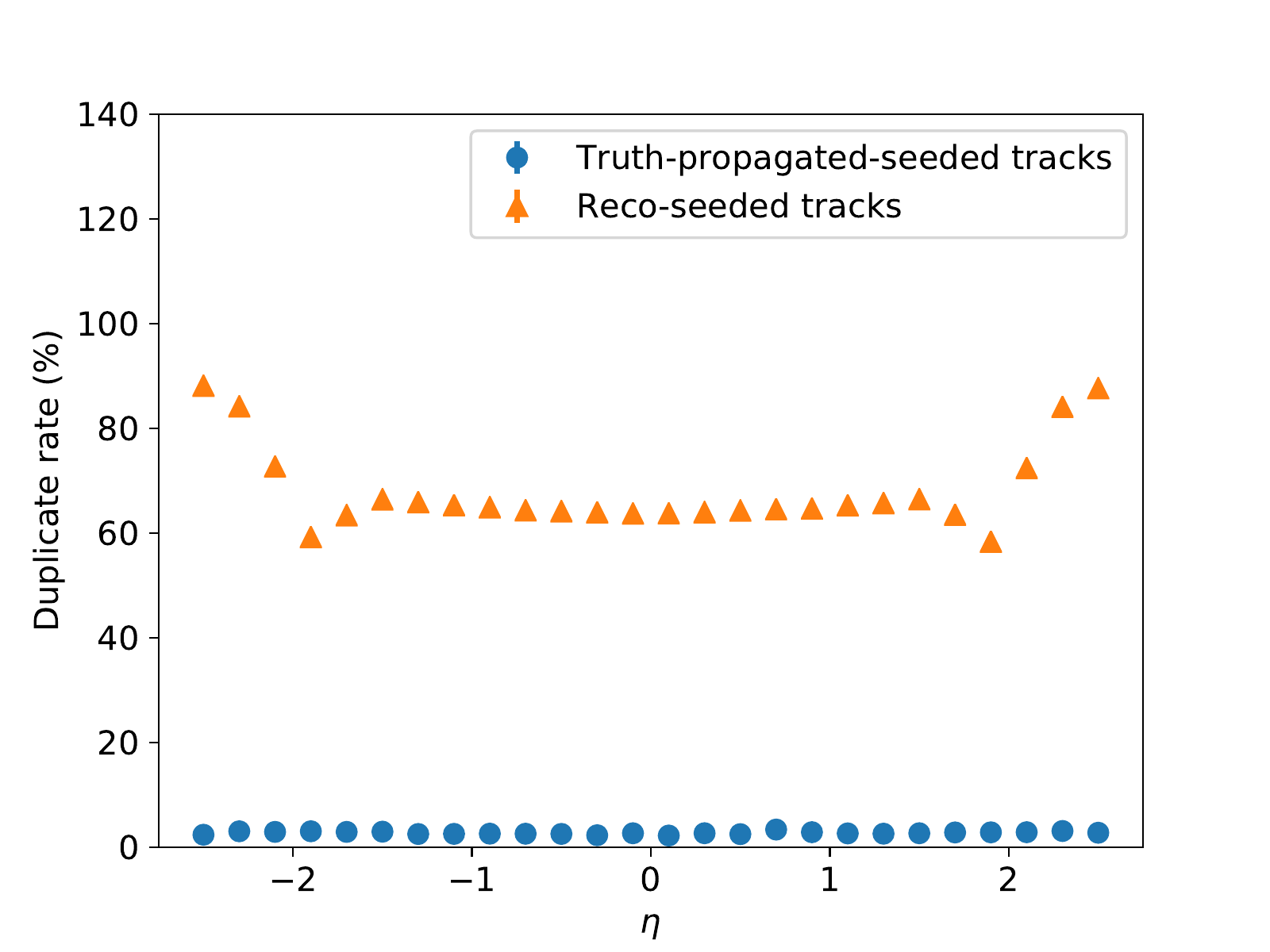}
\caption{The track reconstruction efficiency (top), fake rate (middle) and duplicate rate (bottom) for 1,000 $t\bar{t}$ events with \pu = 200 obtained using \acts CKF on the TrackML detector.
The blue dots and orange triangles represent results using starting parameters based on truth track parameters and those estimated from seeds found the \acts seed finding algorithm, respectively.
The truth particles used to calculate the track reconstruction efficiency are required to have $p_T>$ 1 \GeV and have nine measurements on the traversed detectors. The reconstructed tracks are required to have $p_T>$ 1 \GeV and have six measurements in the detectors. \label{fig:effdupfake_trackML}}
\end{figure}

Key indicators of the performance of a track reconstruction algorithm are the track reconstruction efficiency, the track duplicate rate and the rate at which fake tracks are reconstructed. Their definitions require reconstructed tracks to be associated to generated particles. A reconstructed track is associated to a generated particle if the largest fraction of measurements on the track is from this simulated particle and the fraction of associated measurements is at least 50\%. A track that is not associated to any simulated particle is considered to be a fake track. Duplicate tracks occur when multiple tracks are associated to the same generated particle. 
 
The track reconstruction efficiency is defined as the fraction of generated particles which have made at least nine measurements on the traversed detectors and are associated with tracks. The fake rate and duplicate rate of the tracks are defined as the fraction of \textit{fake} and \textit{duplicate} tracks among all the reconstructed tracks, respectively. 
\autoref{fig:effdupfake_trackML} shows the preliminary track reconstruction efficiency as a function of the $\eta$ of the simulated true particle as well as the fraction of fakes and duplicates as a function of the $\eta$ of the reconstructed track with the CKF for 1,000 $t\bar{t}$ events with \pu = 200. Results are shown for both the truth-propagated-seeded tracks and the reco-seeded tracks. The results for the truth-propagated-seeded tracks are excellent however inefficiencies and high duplicate rates are observed for the reconstructed tracks. This is because no detector specific tuning has been performed for the TrackML detector and the performance would be improved by tuning the seed finding criteria as a function of $\eta$. The tuning of track reconstruction algorithms for a particular geometry is typically performed with several iterations and is beyond the scope of this paper.

\subsubsection{Track Parameter Resolution}

\begin{figure*}[!htbp]
\centering
\includegraphics[width=0.9\textwidth]{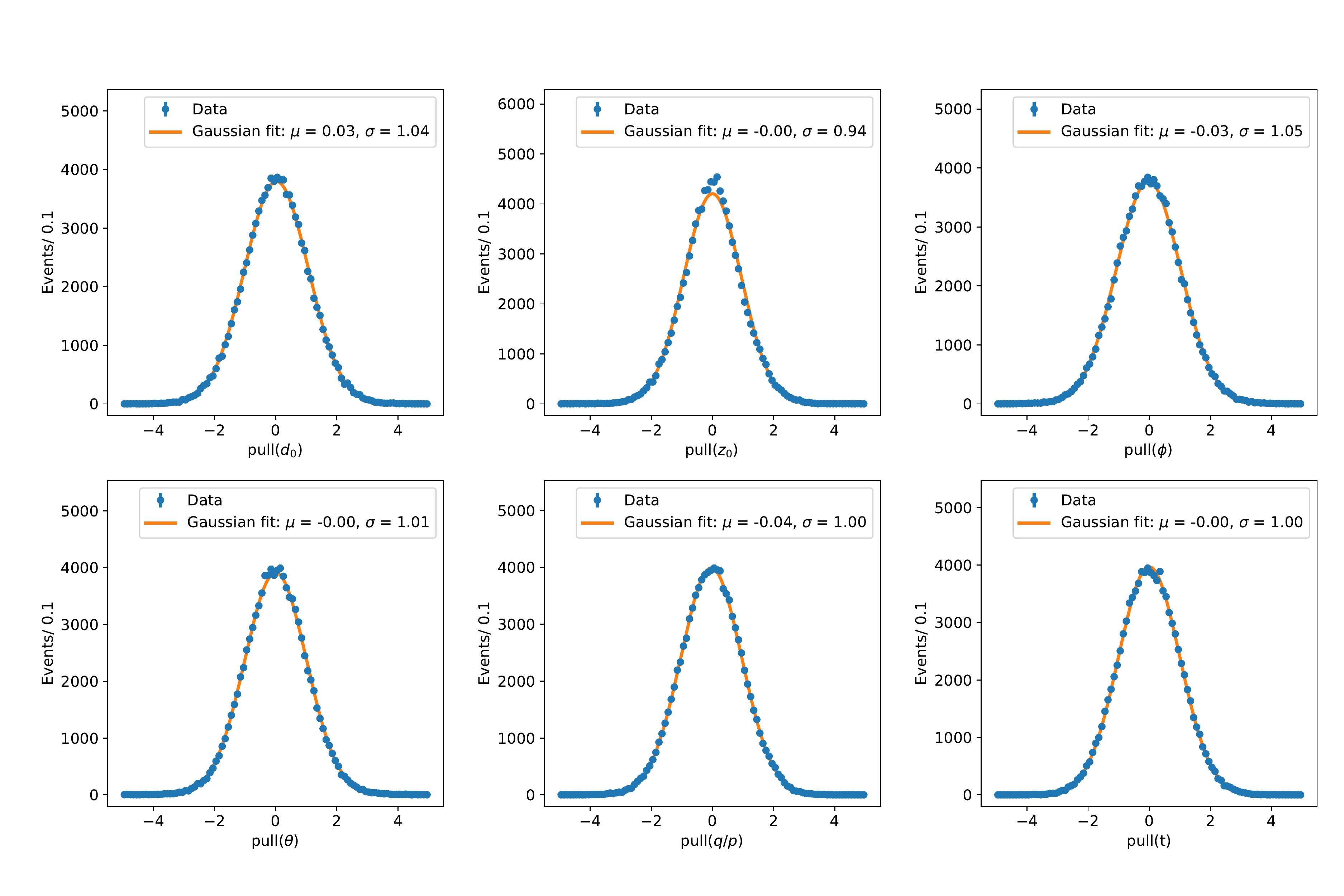}
\caption{The pull distributions of the six bound track parameters, $d_0$, $z_0$, $\phi$, $\theta$, $\frac{q}{p}$, and $t$, as obtained with the KF on the TrackML detector. The blue dots are the obtained pull values and the orange lines are the fitted Gaussian curves. For each Gaussian fit, the fitted values (with negligible uncertainties) for the parameters mean ($\mu$) and standard deviation ($\sigma$) are shown in the legend. Truth-generated seeds are used for the KF. A sample of 100,000 single muons with 500 \MeV $<p_{T}<10$ \GeV and at least nine measurements on the detector is used.
\label{fig:pulls}}
\end{figure*}

Track fitting in \acts can be performed using either the KF or the CKF. Here we study the track parameter resolution using the KF based on the truth tracks to remove the impact of any fake or duplicate tracks. Single muons are used to minimize the impact of detector material.

Gaussian fits are performed to the distributions of the pull values defined as the $(v_\mathrm{fit} - v_\mathrm{truth})/\sigma_{v}$. Here, $v_\mathrm{fit}$ and $v_\mathrm{truth}$ are the estimated value of the track parameter and its true simulated value, and $\sigma_{v}$ is the estimated uncertainty of the reconstructed track parameter.
The distributions of the pulls of the track parameters at the perigee surface defined at the beam line are shown in \autoref{fig:pulls}. The parameters of the Gaussian distributions are approximately consistent with standard normal distributions, which demonstrates that the track parameters and their uncertainties are estimated properly by the \acts KF. The fitted standard deviations of the Gaussian curves deviate slightly from one for the impact parameters and the momentum direction angle $\phi$ due to the impact of non-linear effect of the measurement model.

\subsubsection{Primary Vertex Reconstruction Efficiency}
\label{subsubsec:vertexperf}

\begin{figure}[htbp!]
\centering
\includegraphics[width=0.9\linewidth]{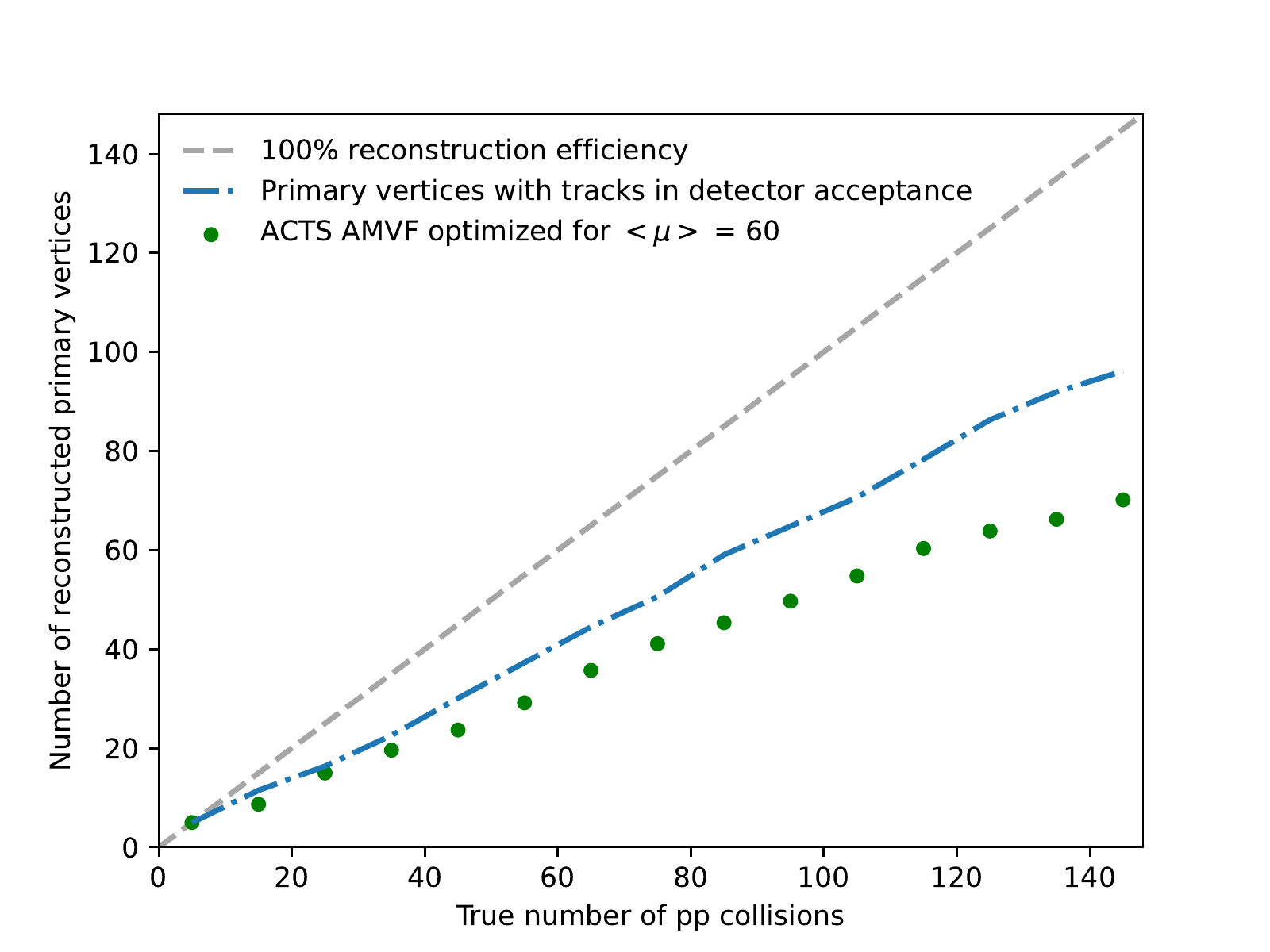}
\caption{Number of reconstructed primary vertices with the \acts AMVF for different numbers of true $pp$ collisions in simulated $t\bar{t}$ events. For reference, the gray dashed line indicates a $100\%$ vertex reconstruction efficiency and the blue dots indicate the vertex reconstruction efficiency given a detector acceptance of $|\eta|<2.5$ and $p_{T}>400$ \MeV
\label{fig:privtxefficiency}}
\end{figure}

\autoref{fig:privtxefficiency} shows the number of reconstructed primary vertices as a function of \pu of the $t\bar{t}$ sample using the \acts AMVF based on the truth tracks. The AMVF efficiency is optimized for a mid-range working point of expected pile-up conditions for the upcoming data-taking run of the LHC, Run-3. These have $\big<\mu\big> \approx 60$ but the performance extrapolates well to higher numbers of simultaneous $pp$ interactions. When used by an experiment, the AMVF configuration would be optimized for the small pile-up range targeting the experiment's needs and accelerator conditions.

\subsection{CPU Performance}
\label{subsec:cpuperf}
The CPU performance, including the CPU utilization and time performance, was tested on a Haswell node at the National Energy Research Scientific Computing Center (NERSC)~\cite{cori-specs} (Cori-Haswell). The node has 32 physical cores and 64 threads at a clock rate of 2.3 GHz. 

The TrackML detector is used to benchmark the CPU performance. The pion samples are used to evaluate the CPU utilization and the timing performance of the propagator with different numerical integration methods, and the $t\bar{t}$ samples with $\langle \mu \rangle$ varying from 0 to 300 are used to evaluate the time performance of the seed finder and CKF.

\subsubsection{CPU Utilization}
\label{subsubsec:cpuutil}

\begin{figure}[htbp!]
\centering
\includegraphics[width=\linewidth]{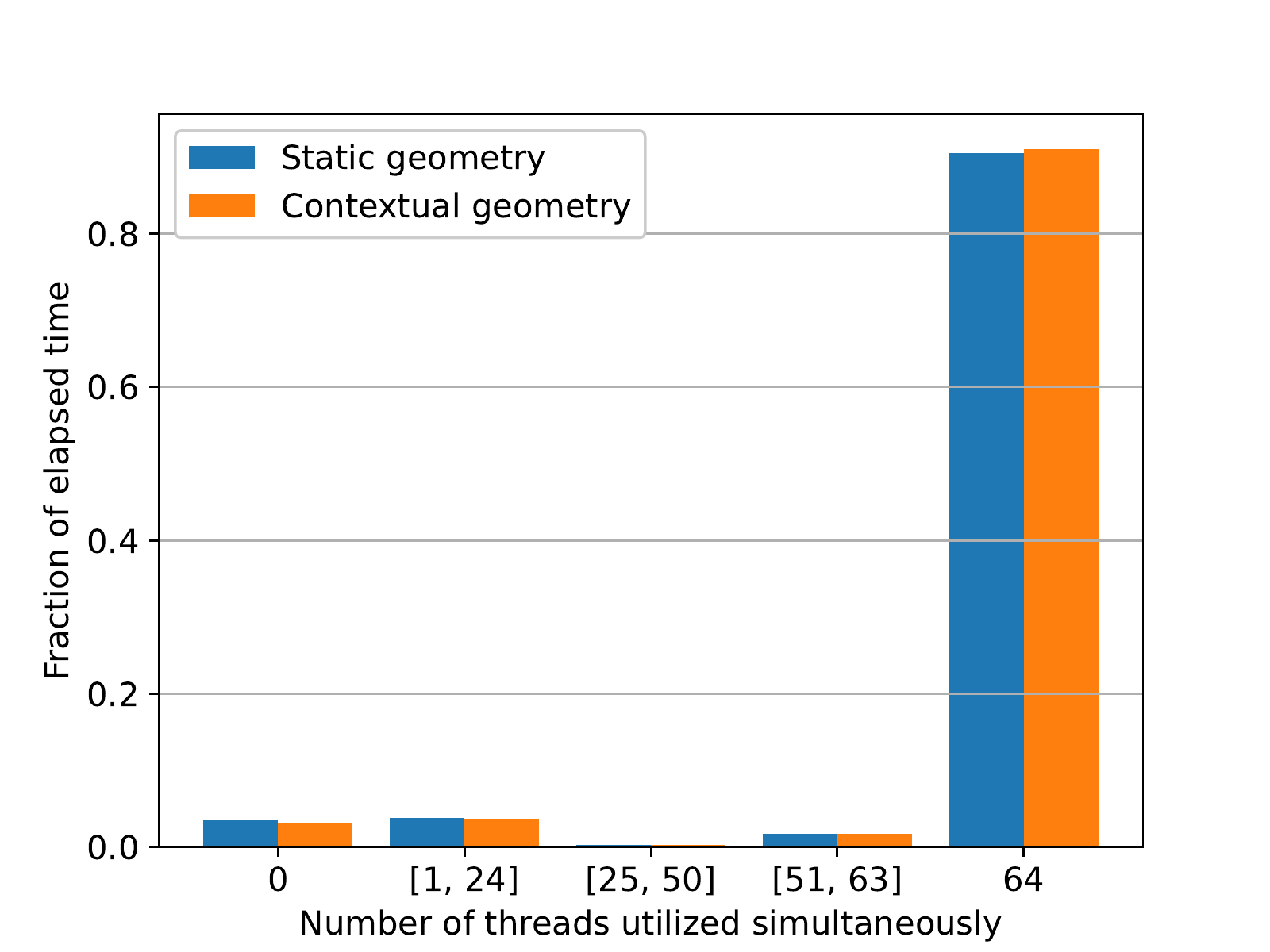}
\caption{The fraction of wall time during which different numbers of threads were running simultaneously while running track propagation through the TrackML detector. Either a static (blue) or contextual (orange) geometry is used for 100,000 events with 1,000 pions per event using multi-threads on a Cori-Haswell node. \label{fig:cpuutil_prop}}
\end{figure}

The CPU utilization of \acts was analyzed using the \texttt{Intel VTune profiler}~\cite{vtune}. \autoref{fig:cpuutil_prop} shows the CPU utilization for running track propagation through the TrackML detector with a static geometry, and a contextual geometry in which the detector alignment changes for 100,000 events with 1,000 pions per event using multiple threads on the Cori-Haswell node. All 64 threads are utilized during 91\% of the total execution time for both the static and the contextual geometry. This demonstrates highly efficient multi-threaded execution.

\subsubsection{Timing Performance}
\label{subsubsec:timingperf}

\begin{figure}[htbp!]
\centering
\includegraphics[width=\linewidth]{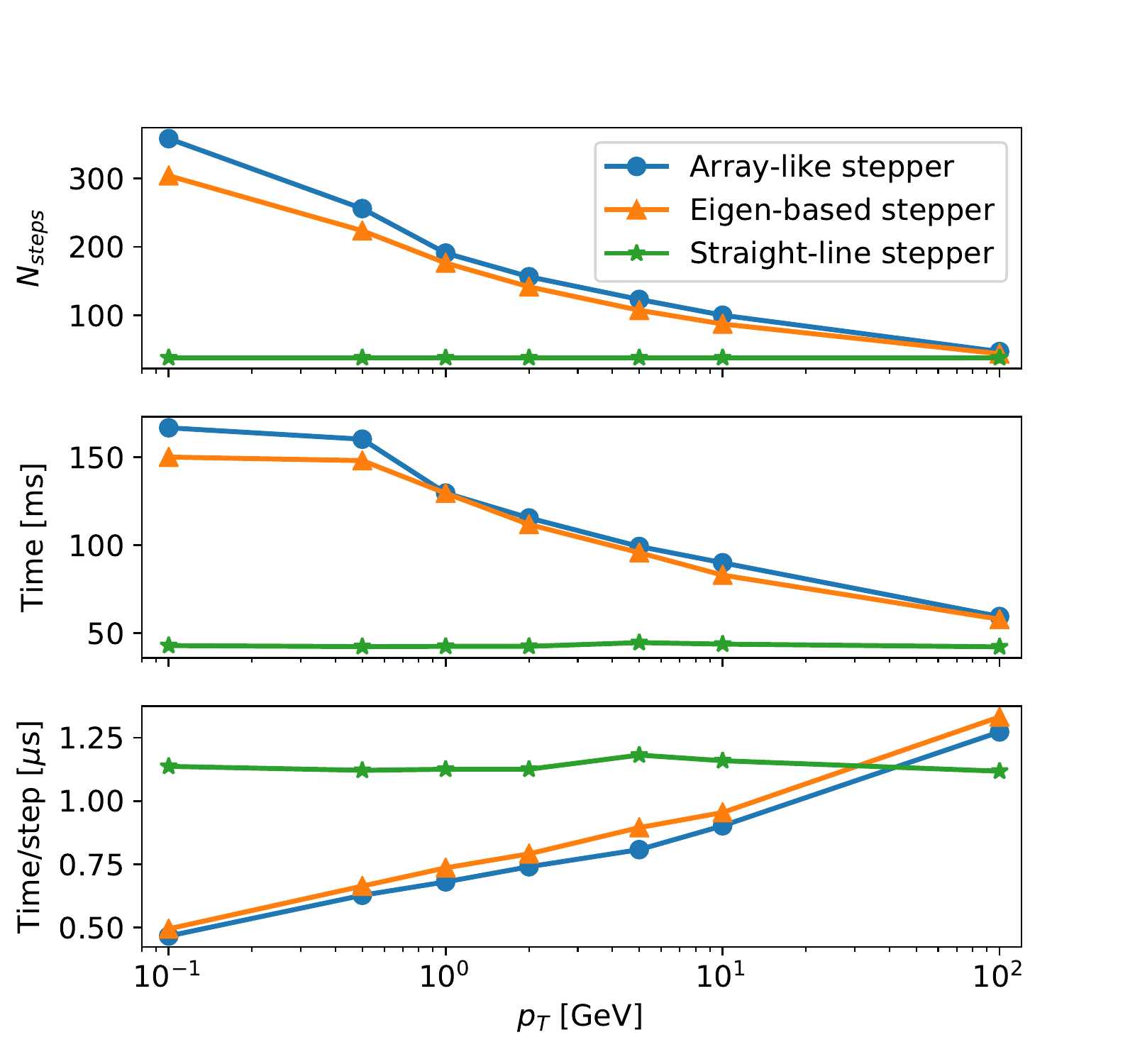}
\caption{The mean number of propagation steps (top), propagation time for 1,000 pions (middle) and mean propagation time per step (bottom) of the track parameter propagation as a function of the $p_T$ of the pions ($|\eta|<2.5$) with the array-like math implementation (blue dots), the main Eigen-based stepper (orange triangles) and the straight-line stepper (green stars). \label{fig:cpuperf_propagator}}
\end{figure}

\autoref{fig:cpuperf_propagator} shows the CPU time of the \acts track parameter propagation through the TrackML detector as a function of pion $p_T$. Three different steppers to perform the numerical integration are shown: the main \texttt{EigenStepper} in \acts, the stepper using manual array mathematical operations, and the straight-line stepper.

The straight-line stepper is used as the baseline because it executes the minimal number of integration steps even though it yields a geometrically incorrect solution, due to presence of the 2\,T magnetic field. For increasing transverse momentum, the CPU time of the other two steppers approach this theoretical best-case scenario. The other two steppers have very similar results, demonstrating that the \texttt{Eigen}-based implementation has nearly optimal computational performance.

\begin{figure}[htbp!]
\centering
\includegraphics[width=\linewidth]{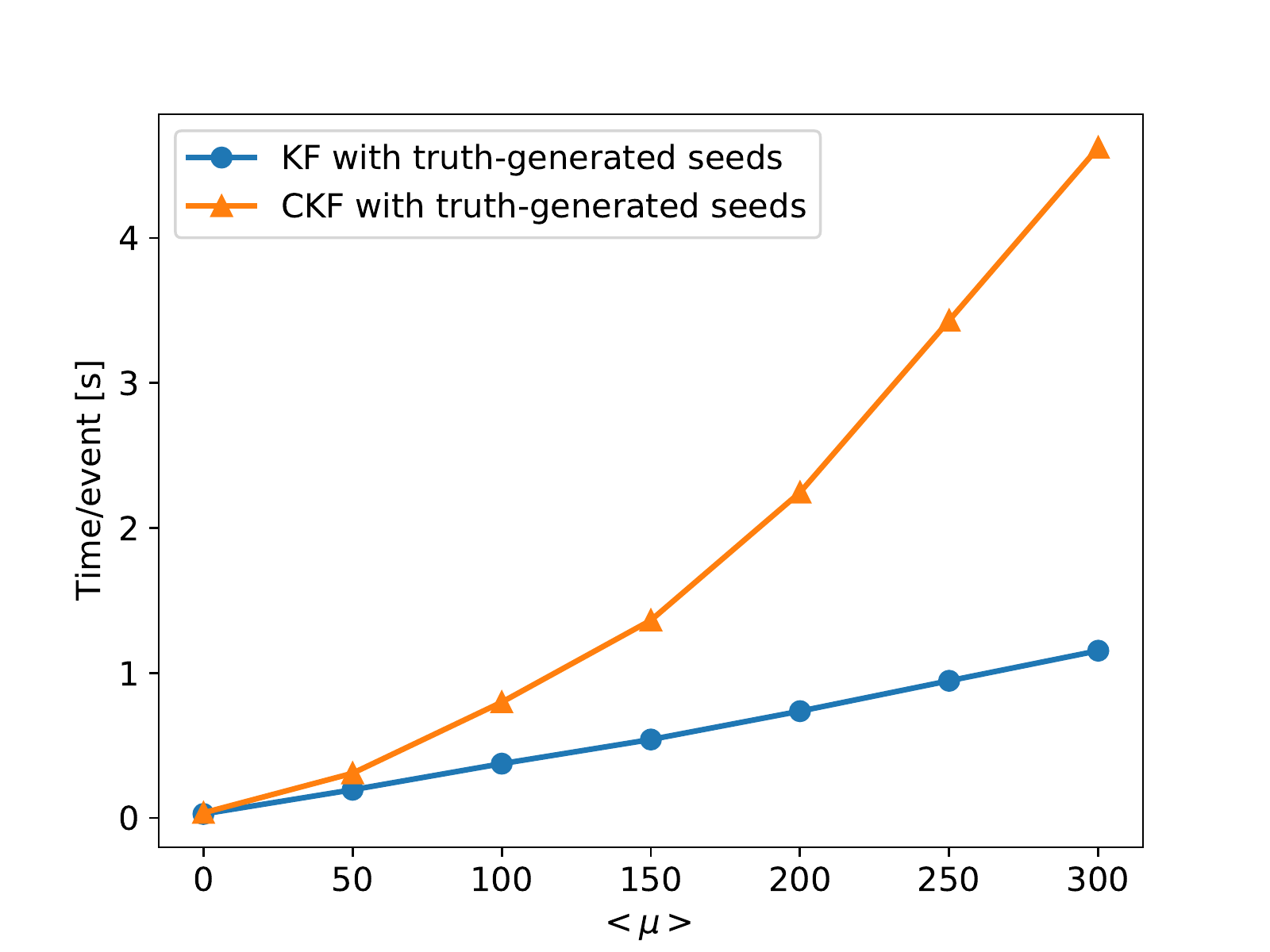}
\caption{The CPU time of track fitting per event using \acts KF (blue dots) and combined track finding and track fitting (orange triangles) per event using the \acts CKF as a function of the \pu of the $t\bar{t}$ sample. Truth-generated seeds are used for both the KF and the CKF. Only simulated particles with $p_T>$ 500 \MeV and having at least nine measurements on the detector are considered.\label{fig:cpuperf_fitting}}
\end{figure}

\begin{figure}[htbp!]
\centering
\includegraphics[width=\linewidth]{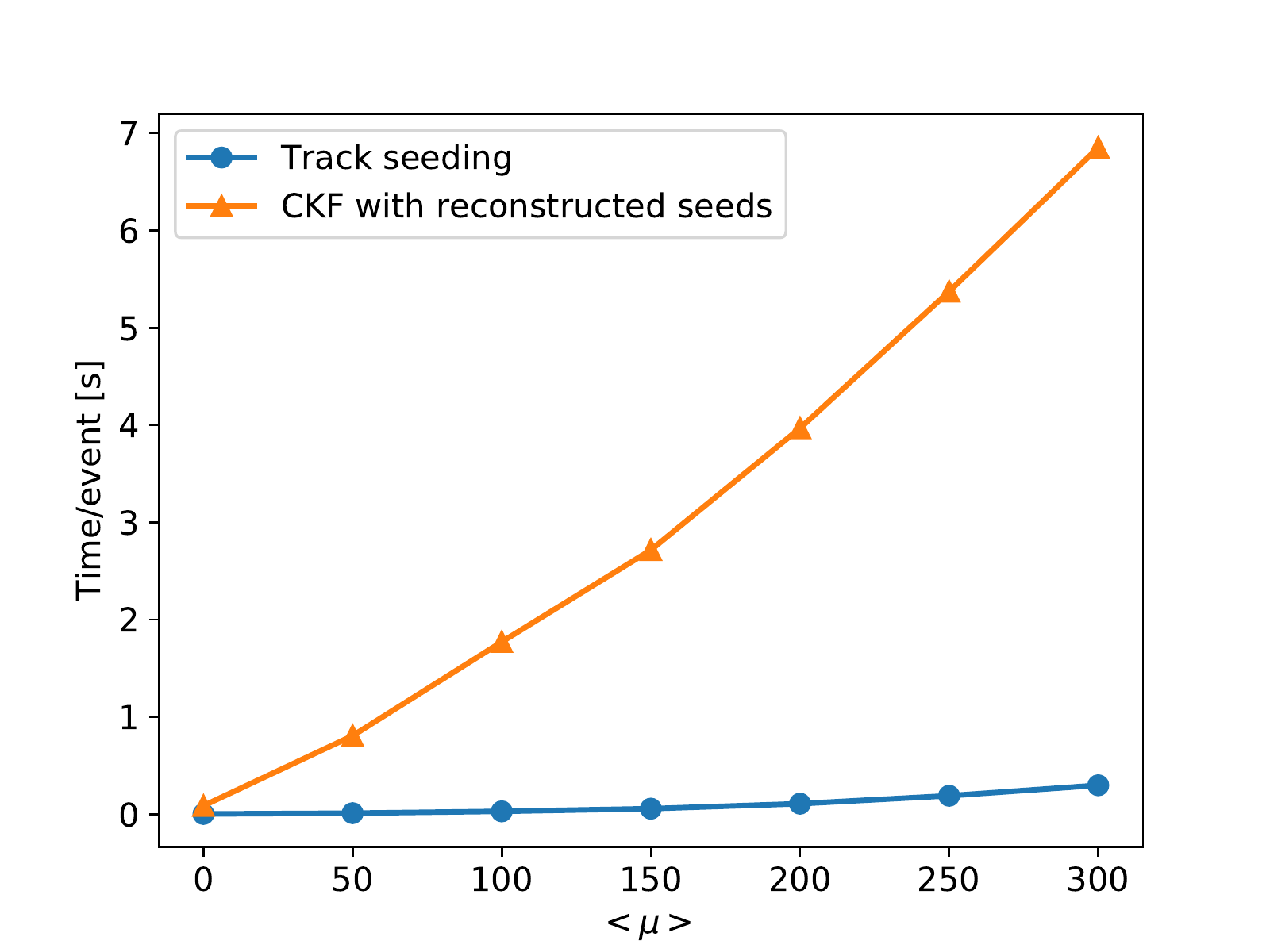}
\caption{The average CPU time for seed finding (blue dots) per event, and combined track finding and track fitting (orange triangles) per event with \acts CKF using reconstructed seeds as a function of \pu of the $t\bar{t}$ sample. Only seeds with $p_T>$ 500 MeV are considered.\label{fig:cpuperf_reco}}
\end{figure}

\autoref{fig:cpuperf_fitting} shows the CPU time as a function of the \pu of the $t\bar{t}$ samples for track fitting with the KF, and the combined track finding and track fitting with the CKF. Both the KF and CKF are using truth-generated seeds. Only simulated particles with transverse momentum greater than 500\,MeV are included. The KF and CKF require 0.74 seconds and 2.25 seconds per event, respectively, for a $t\bar{t}$ sample at \pu = 200. The additional amount of time for the CKF is spent on the search of the compatible measurements on each measurement surface and the possible branching of the track propagation into multiple branches when more than one compatible measurements is found.

\autoref{fig:cpuperf_reco} shows the CPU time as a function of the \pu of the $t\bar{t}$ samples for seed finding, and combined track finding and fitting with CKF using the reconstructed seeds. The seed finding and CKF require 0.11 seconds and 3.97 seconds per event, respectively, for a $t\bar{t}$ sample at \pu = 200. The increase in time for the CKF in \autoref{fig:cpuperf_reco} compared to \autoref{fig:cpuperf_fitting} is due to the presence of duplicate seeds among the reconstructed seeds, which would be improved by dedicated tuning of the track reconstruction algorithms.

%% file: experience.tex
\section{Experience}
\label{sec:experience}

\acts has been initiated to preserve and evolve the well-tested track reconstruction software of the LHC era that has been used for many outstanding physics results, while also creating a research and development toolkit for algorithm optimization and design. Through the use of \acts as the fast track simulation engine for the TrackML challenge, a wider community has become familiar with the \acts project. The resulting R\&D projects sparked by the TrackML challenge are still ongoing and have introduced new concepts and algorithms into the core \acts software. This section discusses selected topics from experience of the development of the \acts project.

Successful and less successful design choices typically become evident when integrating the software within experiments' software stacks. The less restrictive the initial software design, the easier such an integration is. However, this needs to be balanced against the performance of the track reconstruction software.

An example of successful design choice for \acts is the contextual data handling: detector conditional data, such as alignment parameters, calibration constants or other changing parameters are usually very specific to the experiment code and a common solution for these data objects is hard to find. Due to the evolution and aging of running experiments, details of calibration data may not necessarily be known when an experiment begins. Within \acts the implementation of the contextual data and the data flow through the software are split. This allows experiments to implement specific data objects for conditions and the \acts software handles them throughout the entire call chain. This also allows the conditions to be unpacked at the appropriate time by the detector specific code. As the object type is known and specified on both ends of the call chain, this guarantees minimum conversion overhead. The contextual data handling was first demonstrated within the \acts examples, and has also been demonstrated while integrating \acts in Athena. 

A similar example of a successful design choice is the implementation of screen logging. Messages output on the screen are commonly for debugging and quality control with particle and nuclear physics software, hence a seamless integration of \acts with the experiments logging infrastructure has been a priority during development. The integration has been achieved by allowing the logging instance in \acts to be replaced with a custom logger connected to the experiments framework logging facility. The logging has been proven to work within the Gaudi-based software frameworks of ATLAS and FCC-hh. In addition, a generic demonstrator showing how to change the logging instance is included in the \acts test suite.

The choice of using C++ was easy, given the current landscape of particle and nuclear physics software. C++ is an extremely powerful language, but comes, like any language choice, with its shortcomings. Initially, extensive use of template expressions in the \acts core software led to huge resource requirements during compilation. Therefore, this has been revised to reduce the resource requirements. Care is needed to maintain simplicity within the code, which will also be important for an eventual re-use of parts of the \acts software on heterogeneous hardware. While writing code for heterogeneous hardware has not been an immediate target of the \acts project, compatibility should be foreseen, allowing \acts to adapt to future particle and nuclear physics computing landscapes.

Given that the origin of \acts lies in the ATLAS Common Tracking Software, several initial design choices focused towards general-purpose collider experiments. Weaknesses relating to the use of \acts for different geometry types, particularly for forward detectors, time projection chambers, drift tube and telescope setups have been identified. While some of them have already been resolved, these remain active areas of development. 

%% file: conclusion.tex
\section{Conclusion}
\label{sec:conc}

The development of efficient and maintainable track reconstruction is a challenge for current and future particle and nuclear physics experiments. We have introduced the \acts toolkit, which provides a set of open-source, experiment-independent and framework-independent reconstruction algorithms for tracking detectors. The high-level track reconstruction tools do not depend on the details of the detection technologies and magnetic field configuration and have been tested for thread-safety to support concurrent event processing.

We have demonstrated that \acts is maturing as a toolkit and currently provides a range of algorithms for track and vertex reconstruction, which have been or are actively being integrated into a range of experimental frameworks. Geometries for a number of particle and nuclear physics detectors have been included in \acts. Initial studies of the physics and computational performance of the track and vertex reconstruction algorithms using the TrackML detector were presented.  A detailed tuning of the algorithms is required to achieve full performance for any specific detector, which is beyond the scope of this paper and is expected to be performed by the individual experimental collaborations.

A discussion of selected experiences obtained during the \acts project has been presented. Future development directions for the \acts project are expected to include further deployment into experimental frameworks and increasing investment into the R\&D lines. The authors would welcome contact from other experiments interested in exploring the use of and contributions to \acts. 

%% file: references.bib
@misc{Ai:2021kzk,
	archiveprefix = {arXiv},
	author = {Ai, Xiaocong and Mania, Georgiana and Gray, Heather M. and Kuhn, Michael and Styles, Nicholas},
	eprint = {2105.01796},
	month = {5},
	primaryclass = {physics.ins-det},
	reportnumber = {DESY-21-058},
	title = {{A GPU-based Kalman Filter for Track Fitting}},
	year = {2021}}

@article{Aad:2008zzm,
	author = {Aad, G. and others},
	collaboration = {ATLAS},
	doi = {10.1088/1748-0221/3/08/S08003},
	journal = {JINST},
	pages = {S08003},
	title = {{The ATLAS Experiment at the CERN Large Hadron Collider}},
	volume = {3},
	year = {2008}}

@article{cms_2008,
	author = {Chatrchyan, S. and others},
    collaboration = {CMS},
    doi = {10.1088/1748-0221/3/08/s08004},
    journal = {JINST},
	pages = {S08004},
	title = {The {CMS} experiment at the {CERN} {LHC}},
	volume = {3},
	year = {2008}}

@article{lhcb_2008,
	author = {Alves, A. and others},
    collaboration = {LHCb},
    doi = {10.1088/1748-0221/3/08/s08005},
    journal = {JINST},
	pages = {S08005},
	title = {The {LHCb} Detector at the {LHC}},
	volume = {3},
	year = {2008}}

@techreport{ATL-PHYS-PUB-2019-014,
	address = {Geneva},
	collaboration = {ATLAS},
	institution = {CERN},
	month = {Mar},
	reportnumber = {ATL-PHYS-PUB-2019-014},
	title = {{Expected Tracking Performance of the ATLAS Inner Tracker at the HL-LHC}},
	url = {https://cds.cern.ch/record/2669540},
	year = {2019}}

@article{Bertacchi:2020eez,
	archiveprefix = {arXiv},
	author = {Bertacchi, Valerio and others},
	collaboration = {Belle II Tracking Group},
	doi = {10.1016/j.cpc.2020.107610},
	eprint = {2003.12466},
	journal = {Comput. Phys. Commun.},
	pages = {107610},
	primaryclass = {physics.ins-det},
	title = {{Track finding at Belle II}},
	volume = {259},
	year = {2021}}

@misc{Ariga:2019ufm,
	archiveprefix = {arXiv},
	author = {Ariga, Akitaka and others},
	collaboration = {FASER},
	eprint = {1901.04468},
	month = {1},
	primaryclass = {hep-ex},
	reportnumber = {UCI-TR-2019-01, KYUSHU-RCAPP-2018-08},
	title = {{FASER: ForwArd Search ExpeRiment at the LHC}},
	year = {2019}}

@article{Akai:2018mbz,
	archiveprefix = {arXiv},
	author = {Akai, Kazunori and Furukawa, Kazuro and Koiso, Haruyo},
	collaboration = {SuperKEKB},
	doi = {10.1016/j.nima.2018.08.017},
	eprint = {1809.01958},
	journal = {Nucl. Instrum. Meth. A},
	pages = {188--199},
	primaryclass = {physics.acc-ph},
	title = {{SuperKEKB Collider}},
	volume = {907},
	year = {2018}}

@techreport{Calafiura:2729668,
	address = {Geneva},
	author = {Calafiura, P and Catmore, J and Costanzo, D and Di Girolamo, A},
	institution = {CERN},
	month = {Sep},
	number = {CERN-LHCC-2020-015, LHCC-G-178},
	reportnumber = {CERN-LHCC-2020-015, LHCC-G-178},
	title = {{ATLAS HL-LHC Computing Conceptual Design Report}},
	url = {http://cds.cern.ch/record/2729668},
	year = {2020}}

@techreport{Software:2751565,
	address = {Geneva},
	collaboration = {CMS},
	institution = {CERN},
	month = {Jan},
	reportnumber = {CMS-NOTE-2021-001, CERN-CMS-NOTE-2021-001},
	title = {{Evolution of the CMS Computing Model towards Phase-2}},
	url = {https://cds.cern.ch/record/2751565},
	year = {2021}}

@misc{Amrouche:2021tio,
	archiveprefix = {arXiv},
	author = {Amrouche, Sabrina and others},
	eprint = {2105.01160},
	month = {5},
	primaryclass = {cs.LG},
	title = {{The Tracking Machine Learning challenge : Throughput phase}},
	year = {2021}}

@techreport{CERN-LHCC-2020-007,
	address = {Geneva},
	collaboration = {ATLAS},
	institution = {CERN},
	month = {6},
	number = {CERN-LHCC-2020-007; ATLAS-TDR-031},
	reportnumber = {CERN-LHCC-2020-007},
	title = {{Technical Design Report: A High-Granularity Timing Detector for the ATLAS Phase-II Upgrade}},
	url = {https://cds.cern.ch/record/2719855},
	year = {2020}}

@misc{hepspec,
	howpublished = {\url{https://w3.hepix.org/benchmarking.html}},
	title = {HEP-SPEC06 Benchmark},
	urldate = {2021-05-17}}

@unpublished{geomodel,
	author = {Tsulaia, Vakho and Boudreau, Joe},
	eventtitle = {{{CHEP}} 2004},
	howpublished = {\url{https://indico.cern.ch/event/0/contributions/1294152/}},
	month = {9},
	title = {The GeoModel Toolkit for Detector Description},
	venue = {{Interlaken}},
	year = {2004}}

@misc{acts_on_github,
	howpublished = {\url{https://github.com/acts-project/acts}},
	title = {ACTS on Github},
	urldate = {2021-05-04}}

@inproceedings{Amrouche:2021tlm,
	archiveprefix = {arXiv},
	author = {Amrouche, Sabrina and Kiehn, Moritz and Golling, Tobias and Salzburger, Andreas},
	booktitle = {{33rd Annual Conference on Neural Information Processing Systems}},
	eprint = {2101.06428},
	month = {1},
	primaryclass = {hep-ex},
	title = {{Hashing and metric learning for charged particle tracking}},
	year = {2021}}

@inproceedings{Ai:2019kze,
	archiveprefix = {arXiv},
	author = {Ai, Xiaocong},
	booktitle = {{Meeting of the Division of Particles and Fields of the American Physical Society}},
	eprint = {1910.03128},
	month = {10},
	primaryclass = {physics.ins-det},
	title = {{Acts: A common tracking software}},
	year = {2019}}

@misc{Ai:2020jbw,
	archiveprefix = {arXiv},
	author = {Ai, Xiaocong},
	eprint = {2007.01239},
	month = {7},
	primaryclass = {physics.ins-det},
	title = {{Tracking with A Common Tracking Software}},
	year = {2020}}

@inproceedings{Osborn:2021zlr,
	archiveprefix = {arXiv},
	author = {Osborn, Joseph D.},
	booktitle = {{25th International Conference on Computing in High-Energy and Nuclear Physics}},
	collaboration = {sPHENIX},
	eprint = {2103.06703},
	month = {3},
	primaryclass = {physics.ins-det},
	title = {{Implementation of ACTS into sPHENIX track reconstruction}},
	year = {2021}}

@article{Benedikt:2018csr,
	author = {Abada, A. and others},
	collaboration = {FCC},
	doi = {10.1140/epjst/e2019-900087-0},
	journal = {Eur. Phys. J. ST},
	number = {4},
	pages = {755-1107},
	reportnumber = {CERN-ACC-2018-0058},
	slaccitation = {%%CITATION = 00619,228,755;%%},
	title = {{FCC-hh: The Hadron Collider}},
	volume = {228},
	year = {2019}}

@misc{mplv2,
	howpublished = {\url{https://www.mozilla.org/en-US/MPL/2.0/}},
	title = {Mozilla Public License Version 2},
	urldate = {2021-02-03}}

@software{ODD,
	author = {Corentin Allaire and Paul Gessinger and Julia Hdrinka and Moritz Kiehn and Fabian Kimpel and Joana Niermann and Andreas Salzburger and Stanislava Sevova},
	doi = {10.5281/zenodo.4674401},
	month = apr,
	publisher = {Zenodo},
	title = {OpenDataDetector},
	version = {v1},
	year = 2021}

@techreport{panda_vtx_tdr:2012,
	archiveprefix = {arXiv},
	collaboration = {PANDA},
	eprint = {1207.6581},
	primaryclass = {physics.ins-det},
	title = {Technical Design Report for the: PANDA Micro Vertex Detector},
	year = {2012}}

@article{Bettoni:2007ms,
	archiveprefix = {arXiv},
	author = {Bettoni, Diego},
	editor = {Mahlke, Hanna and Napolitano, Jim},
	eprint = {0710.5664},
	journal = {eConf},
	pages = {39},
	primaryclass = {hep-ex},
	reportnumber = {CHARM-2007-39},
	title = {{The PANDA experiment at FAIR}},
	volume = {C070805},
	year = {2007}}

@article{Brun:1997pa,
	author = {Brun, R. and Rademakers, F.},
	doi = {10.1016/S0168-9002(97)00048-X},
	editor = {Werlen, M. and Perret-Gallix, D.},
	journal = {Nucl. Instrum. Meth. A},
	pages = {81--86},
	title = {{ROOT: An object oriented data analysis framework}},
	volume = {389},
	year = {1997}}

@techreport{CMS:2005aa,
	address = {Geneva},
	collaboration = {CMS},
	institution = {CERN},
	month = {6},
	reportnumber = {CERN-LHCC-2005-023},
	title = {{CMS: The computing project. Technical design report}},
	url = {http://cds.cern.ch/record/838359},
	year = {2005}}

@techreport{Duckeck:2005rb,
	address = {Geneva},
	collaboration = {ATLAS},
	editor = {Duckeck, G. and Barberis, D. and Hawkings, R. and Jones, R. and McCubbin, N. and Poulard, G. and Quarrie, D. and Wenaus, T. and Obreshkov, E.},
	institution = {CERN},
	month = {6},
	reportnumber = {CERN-LHCC-2005-022, ATLAS-TRD-017},
	title = {{ATLAS computing: Technical design report}},
	url = {http://cds.cern.ch/record/837738},
	year = {2005}}

@article{Myrheim:1979ng,
	author = {Myrheim, Jan and Bugge, Lars},
	doi = {10.1016/0029-554X(79)90163-0},
	journal = {Nucl.\ Instrum.\ Meth.},
	number = {1},
	pages = {43--48},
	title = {{A fast Runge-Kutta method for fitting tracks in a magnetic field}},
	volume = {160},
	year = {1979}}

@article{RTS,
	author = {Rauch, H. E. and Tung, F. and Striebel, C. T.},
	doi = {10.2514/3.3166},
	journal = {AIAA Journal},
	number = {8},
	pages = {1445-1450},
	title = {Maximum likelihood estimates of linear dynamic systems},
	volume = {3},
	year = {1965}}

@article{Lund:2009zzc,
	author = {Lund, Esben and Bugge, Lars and Gavrilenko, Igor and Strandlie, Are},
	doi = {10.1088/1748-0221/4/04/P04001},
	journal = {JINST},
	pages = {P04001},
	title = {{Track parameter propagation through the application of a new adaptive Runge-Kutta-Nystr{\"o}m method in the ATLAS experiment}},
	volume = {4},
	year = {2009}}

@article{Agostinelli:2002hh,
	author = {Agostinelli, Stefano and others},
	collaboration = {GEANT4},
	doi = {10.1016/S0168-9002(03)01368-8},
	journal = {Nucl. Instrum. Meth. A},
	pages = {250--303},
	reportnumber = {SLAC-PUB-9350, FERMILAB-PUB-03-339},
	title = {{GEANT4: A Simulation toolkit}},
	volume = {506},
	year = {2003}}

@misc{autodiff,
	howpublished = {\url{https://autodiff.github.io/}},
	title = {autodiff},
	urldate = {2021-02-03}}

@misc{onnx,
	howpublished = {\url{https://github.com/onnx}},
	title = {Open Neural Network Exchange},
	urldate = {2021-05-17}}

@misc{sycl,
	howpublished = {\url{https://www.khronos.org/sycl/}},
	title = {{SYCL: C++ Single-source Heterogeneous Programming for OpenCL}},
	urldate = {2021-02-03}}

@article{cuda,
	address = {New York, NY, USA},
	author = {Nickolls, John and Buck, Ian and Garland, Michael and Skadron, Kevin},
	doi = {10.1145/1365490.1365500},
	issn = {1542-7730},
	issue_date = {March/April 2008},
	journal = {Queue},
	month = mar,
	number = {2},
	numpages = {14},
	pages = {40--53},
	publisher = {Association for Computing Machinery},
	title = {Scalable Parallel Programming with CUDA: Is CUDA the Parallel Programming Model That Application Developers Have Been Waiting For?},
	volume = {6},
	year = {2008}}

@misc{cmake,
	howpublished = {\url{https://cmake.org}},
	title = {CMake},
	urldate = {2021-02-03}}

@misc{boost,
	howpublished = {\url{https://www.boost.org}},
	title = {Boost},
	urldate = {2021-02-03}}

@misc{Guennebaud:2010aa,
	author = {Ga\"{e}l Guennebaud and Beno\^{i}t Jacob and others},
	howpublished = {http://eigen.tuxfamily.org},
	title = {Eigen v3},
	year = {2010}}

@article{Kiehn:2019tbl,
	author = {Kiehn, Moritz and others},
	doi = {10.1051/epjconf/201921406037},
	editor = {Forti, A. and Betev, L. and Litmaath, M. and Smirnova, O. and Hristov, P.},
	journal = {EPJ Web Conf.},
	pages = {06037},
	title = {{The TrackML high-energy physics tracking challenge on Kaggle}},
	volume = {214},
	year = {2019}}

@misc{tbb,
	howpublished = {\url{https://github.com/oneapi-src/oneTBB}},
	title = {Threading Building Blocks},
	urldate = {2021-02-03}}

@misc{CMSSW,
	collaboration = {CMS},
	howpublished = {\url{http://cms-sw.github.io/}},
	title = {{CMS Offline Software}},
	urldate = {2021-02-03}}

@misc{athena,
	collaboration = {ATLAS},
	doi = {10.5281/zenodo.2641996},
	month = apr,
	publisher = {Zenodo},
	title = {Athena},
	version = {22.0.1},
	year = 2019}

@article{Barrand:2001ny,
	author = {Barrand, Guy and others},
	doi = {10.1016/S0010-4655(01)00254-5},
	journal = {Comput. Phys. Commun.},
	pages = {45--55},
	title = {{GAUDI - A software architecture and framework for building HEP data processing applications}},
	volume = {140},
	year = {2001}}

@phdthesis{Pantaleo:2293435,
	author = {Pantaleo, Felice},
	reportnumber = {CMS-TS-2017-028; CERN-THESIS-2017-242},
	school = {Hamburg U.},
	title = {{New Track Seeding Techniques for the CMS Experiment}},
	url = {https://cds.cern.ch/record/2293435},
	year = {2017}}

@inproceedings{Osborn:2020soo,
	archiveprefix = {arXiv},
	author = {Osborn, Joseph D.},
	booktitle = {{Proceedings for the Connecting The Dots 2020 Workshop}},
	collaboration = {sPHENIX},
	eprint = {2007.00771},
	month = {6},
	primaryclass = {physics.ins-det},
	title = {{Requirements, Status, and Plans for Track Reconstruction at the sPHENIX Experiment}},
	year = {2020}}

@article{kalman1960,
	author = {Kalman, R. E.},
	doi = {10.1115/1.3662552},
	journal = {Journal of Basic Engineering},
	month = {03},
	number = {1},
	pages = {35-45},
	title = {{A New Approach to Linear Filtering and Prediction Problems}},
	volume = {82},
	year = {1960}}

@inproceedings{Amrouche:2019wmx,
	author = {Amrouche, Sabrina and Basara, Laurent and Calafiura, Paolo and Estrade, Victor and Farrell, Steven and Ferreira, Diogo R. and Finnie, Liam and Finnie, Nicole and Germain, C{\'e}cile and Gligorov, Vladimir Vava and Golling, Tobias and Gorbunov, Sergey and Gray, Heather and Guyon, Isabelle and Hushchyn, Mikhail and Innocente, Vincenzo and Kiehn, Moritz and Moyse, Edward and Puget, Jean-Fran{\c{c}}ois and Reina, Yuval and Rousseau, David and Salzburger, Andreas and Ustyuzhanin, Andrey and Vlimant, Jean-Roch and Wind, Johan Sokrates and Xylouris, Trian and Yilmaz, Yetkin},
	booktitle = {The NeurIPS '18 Competition},
	doi = {10.1007/978-3-030-29135-8_9},
	pages = {231--264},
	title = {The Tracking Machine Learning Challenge: Accuracy Phase},
	year = {2020}}

@article{Bapst:2019llh,
	archiveprefix = {arXiv},
	author = {Bapst, Frederic and Bhimji, Wahid and Calafiura, Paolo and Gray, Heather and Lavrijsen, Wim and Linder, Lucy},
	doi = {10.1007/s41781-019-0032-5},
	eprint = {1902.08324},
	journal = {Comput. Softw. Big Sci.},
	number = {1},
	pages = {1},
	primaryclass = {quant-ph},
	title = {{A Pattern Recognition Algorithm for Quantum Annealers}},
	volume = {4},
	year = {2020}}

@inproceedings{Amrouche:2019yxv,
	author = {Amrouche, Sabrina and Golling, Tobias and Kiehn, Moritz and Plant, Claudia and Salzburger, Andreas},
	booktitle = {{IEEE International Conference on Big Data 2019}},
	doi = {10.1109/BigData47090.2019.9006316},
	pages = {1595--1600},
	title = {{Similarity hashing for charged particle tracking}},
	year = {2019}}

@inproceedings{Heintz:2020soy,
	archiveprefix = {arXiv},
	author = {Heintz, Aneesh and others},
	booktitle = {{NeurIPS 2020}},
	eprint = {2012.01563},
	month = {11},
	primaryclass = {physics.ins-det},
	reportnumber = {FERMILAB-CONF-20-622-CMS-SCD},
	title = {{Accelerated Charged Particle Tracking with Graph Neural Networks on FPGAs}},
	year = {2020}}

@misc{Fox:2020hfm,
	archiveprefix = {arXiv},
	author = {Fox, Patrick J. and Huang, Shangqing and Isaacson, Joshua and Ju, Xiangyang and Nachman, Benjamin},
	eprint = {2012.04533},
	month = {12},
	primaryclass = {physics.ins-det},
	reportnumber = {FERMILAB-PUB-20-650-T},
	title = {{Beyond 4D Tracking: Using Cluster Shapes for Track Seeding}},
	year = {2020}}

@inproceedings{Tuysuz:2020eaa,
	archiveprefix = {arXiv},
	author = {T{\"u}ys{\"u}z, Cenk and Novotny, Kristiane and Rieger, Carla and Carminati, Federico and Demirk\"oz, Bilge and Dobos, Daniel and Fracas, Fabio and Potamianos, Karolos and Vallecorsa, Sofia and Vlimant, Jean-Roch},
	booktitle = {{BASARIM 2020 conference proceedings}},
	eprint = {2012.01379},
	month = {12},
	primaryclass = {quant-ph},
	title = {{Performance of Particle Tracking Using a Quantum Graph Neural Network}},
	year = {2020}}

@article{Gessinger:2020nne,
	author = {Gessinger, Paul and Grasland, Hadrien and Gray, Heather and Kiehn, Moritz and Klimpel, Fabian and Langenberg, Robert and Salzburger, Andreas and Schlag, Bastian and Zhang, Jin and Ai, Xiaocong},
	doi = {10.1051/epjconf/202024510003},
	editor = {Doglioni, C. and Kim, D. and Stewart, G. A. and Silvestris, L. and Jackson, P. and Kamle, W.},
	journal = {EPJ Web Conf.},
	pages = {10003},
	title = {{The Acts project: track reconstruction software for HL-LHC and beyond}},
	volume = {245},
	year = {2020}}

@inproceedings{Ju:2020xty,
	archiveprefix = {arXiv},
	author = {Ju, Xiangyang and others},
	booktitle = {{NeurIPS 2019}},
	eprint = {2003.11603},
	month = {3},
	primaryclass = {physics.ins-det},
	reportnumber = {FERMILAB-CONF-20-163-PPD-QIS-SCD},
	title = {{Graph Neural Networks for Particle Reconstruction in High Energy Physics detectors}},
	year = {2020}}

@article{moore,
	author = {{G.E. Moore}},
	doi = {10.1109/JPROC.1998.658762},
	journal = {Electronics 38},
	title = {{Cramming More Components onto Integrated Circuits}},
	volume = {8},
	year = {1965}}

@misc{ApollinariG.:2017ojx,
	doi = {10.23731/CYRM-2017-004},
	editor = {Apollinari, G. and B\'ejar Alonso, I. and Br\"uning, O. and Fessia, P. and Lamont, M. and Rossi, L. and Tavian, L.},
	month = {11},
	reportnumber = {CERN-2017-007-M},
	title = {{High-Luminosity Large Hadron Collider (HL-LHC)}: {Technical Design Report V. 0.1}},
	volume = {4/2017},
	year = {2017}}

@software{xiaocong_ai_2020_3937454,
	author = {Xiaocong Ai and Corentin Allaire and Noemi Calace and Paul Gessinger and Hadrien Grasland and Heather Gray and Christian Gumpert and Julia Hrdinka and Moritz Kiehn and Fabian Klimpel and Robert Langenberg and Andreas Salzburger and Bastian Schlag},
	doi = {10.5281/zenodo.3937454},
	month = nov,
	publisher = {Zenodo},
	title = {Acts Project: v3.0.0},
	version = {v3.0.0},
	year = 2020}

@article{Hopfield:1982pe,
	author = {Hopfield, J.J.},
	doi = {10.1073/pnas.79.8.2554},
	journal = {Proc. Nat. Acad. Sci.},
	pages = {2554--2558},
	title = {{Neural networks and physical systems with emergent collective computational abilities}},
	volume = {79},
	year = {1982}}

@article{Hough:1959qva,
	author = {Hough, P.V.C.},
	editor = {Kowarski, L.},
	journal = {Conf. Proc. C},
	pages = {554--558},
	title = {{Machine Analysis of Bubble Chamber Pictures}},
	volume = {590914},
	year = {1959}}

@article{Chatrchyan:2014fea,
	collaboration = {CMS},
	doi = {10.1088/1748-0221/9/10/P10009},
	issn = {1748-0221},
	journal = {JINST},
	month = oct,
	number = {10},
	pages = {P10009},
	title = {Description and Performance of Track and Primary-Vertex Reconstruction with the {{CMS}} Tracker},
	volume = {9},
	year = {2014}}

@article{Cornelissen:2008zzc,
	author = {Cornelissen, T and Elsing, M and Gavrilenko, I and Liebig, W and Moyse, E and Salzburger, A},
	doi = {10.1088/1742-6596/119/3/032014},
	issn = {1742-6596},
	journal = {Journal of Physics: Conference Series},
	month = jul,
	number = {3},
	pages = {032014},
	title = {The New {{ATLAS}} Track Reconstruction ({{NEWT}})},
	volume = {119},
	year = {2008}}

@techreport{ATLAS_ITPixel_Phase2_TDR,
	collaboration = {ATLAS},
	number = {CERN-LHCC-2017-021, ATLAS-TDR-030},
	reportnumber = {CERN-LHCC-2017-021},
	title = {{Technical Design Report for the ATLAS Inner Tracker Pixel Detector}},
	url = {https://cds.cern.ch/record/2285585},
	year = {2017}}

@techreport{Edmonds:1091969,
	address = {Geneva},
	author = {Edmonds, K and Fleischmann, S and Lenz, T and Magass, C and Mechnich, J and Salzburger, A},
	institution = {CERN},
	month = {Mar},
	number = {ATL-SOFT-PUB-2008-001. ATL-COM-SOFT-2008-002},
	reportnumber = {ATL-SOFT-PUB-2008-001},
	title = {{The Fast ATLAS Track Simulation (FATRAS)}},
	url = {https://cds.cern.ch/record/1091969},
	year = {2008}}

@techreport{ATLAS_ITStrip_Phase2TDR,
	collaboration = {ATLAS},
	number = {CERN-LHCC-2017-005, ATLAS-TDR-025},
	reportnumber = {CERN-LHCC-2017-005},
	title = {{Technical Design Report for the ATLAS Inner Tracker Strip Detector}},
	url = {https://cds.cern.ch/record/2257755},
	year = {2017}}

@misc{Adare:2015kwa,
	archiveprefix = {arXiv},
	author = {Adare, A. and others},
	eprint = {1501.06197},
	month = {1},
	primaryclass = {nucl-ex},
	title = {{An Upgrade Proposal from the PHENIX Collaboration}},
	year = {2015}}

@techreport{ISO14882,
	address = {Geneva},
	institution = {International Organization for Standardization},
	key = {ISO/IEC 14882:2017},
	month = dec,
	title = {{ISO/IEC 14882:2017, Programming languages --- C++}},
	type = {Standard},
	url = {https://www.iso.org/standard/68564.html},
	urldate = {2021-05-17},
	volume = {2017},
	year = {2017}}

@article{10.1145/3282307,
	address = {New York, NY, USA},
	author = {Hennessy, John L. and Patterson, David A.},
	doi = {10.1145/3282307},
	issn = {0001-0782},
	issue_date = {February 2019},
	journal = {Commun. ACM},
	month = jan,
	number = {2},
	numpages = {13},
	pages = {48--60},
	publisher = {Association for Computing Machinery},
	title = {A New Golden Age for Computer Architecture},
	volume = {62},
	year = {2019}}

@techreport{Gumpert:2243297,
	address = {Geneva},
	author = {Gumpert, Christian and Salzburger, Andreas and Kiehn, Moritz and Hrdinka, Julia and Calace, Noemi},
	doi = {10.1088/1742-6596/898/4/042011},
	institution = {CERN},
	month = {Jan},
	number = {ATL-SOFT-PROC-2017-030. 4},
	reportnumber = {ATL-SOFT-PROC-2017-030},
	title = {{ACTS: from ATLAS software towards a common track reconstruction software}},
	year = {2017}}

@article{RevModPhys.82.1419,
	author = {Strandlie, Are and Fr\"uhwirth, Rudolf},
	doi = {10.1103/RevModPhys.82.1419},
	issue = {2},
	journal = {Rev. Mod. Phys.},
	month = {May},
	numpages = {0},
	pages = {1419--1458},
	publisher = {American Physical Society},
	title = {Track and vertex reconstruction: From classical to adaptive methods},
	volume = {82},
	year = {2010}}

@article{Billoir:1983mz,
	author = {Billoir, Pierre},
	doi = {10.1016/0167-5087(84)90274-6},
	journal = {Nucl. Instrum. Meth. a},
	pages = {352-366},
	reportnumber = {LPC-83-12},
	slaccitation = {%%CITATION = NUIMA,A225,352;%%},
	title = {{Track Fitting With Multiple Scattering: A New Method}},
	volume = {225},
	year = {1984}}

@article{Fruhwirth:1987fm,
	author = {Fr{\"u}hwirth, R.},
	doi = {10.1016/0168-9002(87)90887-4},
	journal = {Nucl. Instrum. Meth.},
	pages = {444-450},
	reportnumber = {HEPHY-PUB-87-503},
	slaccitation = {%%CITATION = NUIMA,A262,444;%%},
	title = {{Application of Kalman filtering to track and vertex fitting}},
	volume = {A262},
	year = {1987}}

@article{BILLOIR1992139,
	author = {P. Billoir and S. Qian},
	doi = {10.1016/0168-9002(92)90859-3},
	issn = {0168-9002},
	journal = {Nucl. Instrum. Methods. Phys. Res. A},
	number = {1},
	pages = {139-150},
	title = {Fast vertex fitting with a local parametrization of tracks},
	volume = {311},
	year = {1992}}

@techreport{ATL-PHYS-PUB-2019-015,
	address = {Geneva},
	collaboration = {ATLAS},
	institution = {CERN},
	month = {Apr},
	number = {ATL-PHYS-PUB-2019-015},
	reportnumber = {ATL-PHYS-PUB-2019-015},
	title = {{Development of ATLAS Primary Vertex Reconstruction for LHC Run 3}},
	url = {https://cds.cern.ch/record/2670380},
	year = {2019}}

@phdthesis{Piacquadio:1243771,
	author = {Piacquadio, Giacinto},
	reportnumber = {CERN-THESIS-2010-027},
	school = {Freiburg U.},
	title = {{Identification of b-jets and investigation of the discovery potential of a Higgs boson in the $WH \to l \nu b \bar{b}$ channel with the ATLAS experiment}},
	url = {https://cds.cern.ch/record/1243771},
	year = {2010}}

@article{BICKEL20063500,
	author = {David R. Bickel and Rudolf Fr{\"u}hwirth},
	doi = {10.1016/j.csda.2005.07.011},
	issn = {0167-9473},
	journal = {Computational Statistics \& Data Analysis},
	number = {12},
	pages = {3500-3530},
	title = {On a fast, robust estimator of the mode: Comparisons to other robust estimators with applications},
	volume = {50},
	year = {2006}}

@misc{cori-specs,
	howpublished = {\url{https://docs.nersc.gov/systems/cori/\#system-specification}},
	title = {{NERSC Cori System Specification}},
	urldate = {2021-02-04}}

@misc{clang-format,
	howpublished = {\url{https://clang.llvm.org/docs/ClangFormat.html}},
	title = {{clang-format}},
	urldate = {2021-06-14}}

@misc{llvm,
	howpublished = {\url{https://llvm.org}},
	title = {{The LLVM compiler infrastructure}},
	urldate = {2021-06-14}}

@article{SJOSTRAND2008852,
	author = {Torbj{\"o}rn Sj{\"o}strand and Stephen Mrenna and Peter Skands},
	doi = {10.1016/j.cpc.2008.01.036},
	issn = {0010-4655},
	journal = {Computer Physics Communications},
	number = {11},
	pages = {852-867},
	title = {A brief introduction to PYTHIA 8.1},
	volume = {178},
	year = {2008}}

@article{Sj_strand_2006,
	author = {Torbj{\"o}rn Sj{\"o}strand and Stephen Mrenna and Peter Skands},
	doi = {10.1088/1126-6708/2006/05/026},
	journal = {Journal of High Energy Physics},
	month = {may},
	number = {05},
	pages = {026--026},
	publisher = {Springer Science and Business Media {LLC}},
	title = {{PYTHIA} 6.4 physics and manual},
	volume = {2006},
	year = 2006}

@article{Duda1972UseOT,
	author = {R. Duda and P. Hart},
	journal = {Commun. ACM},
	pages = {11-15},
	title = {Use of the Hough transformation to detect lines and curves in pictures},
	volume = {15},
	year = {1972}}

@misc{vtune,
	howpublished = {\url{https://software.intel.com/content/www/us/en/develop/tools/vtune-profiler.html}},
	title = {Intel VTune Profiler},
	urldate = {2021-05-17}}

@article{BRUN2003676,
	author = {R Brun and A Gheata and M Gheata},
	doi = {10.1016/S0168-9002(03)00541-2},
	issn = {0168-9002},
	journal = {Nucl. Instrum. Methods. Phys. Res. A},
	number = {2},
	pages = {676-680},
	title = {The ROOT geometry package},
	volume = {502},
	year = {2003}}

@article{Petri__2017,
	author = {M Petri{\v{c}} and M Frank and F Gaede and S Lu and N Nikiforou and A Sailer},
	doi = {10.1088/1742-6596/898/4/042015},
	journal = {Journal of Physics: Conference Series},
	month = {oct},
	pages = {042015},
	publisher = {{IOP} Publishing},
	title = {Detector Simulations with {DD}4hep},
	volume = {898},
	year = 2017}

@misc{sv,
	howpublished = {\url{https://semver.org/}},
	title = {Semantic Versioning},
	urldate = {2021-05-17}}

@article{BILLOIR1989390,
	author = {Pierre Billoir},
	doi = {10.1016/0010-4655(89)90249-X},
	issn = {0010-4655},
	journal = {Comput. Phys. Commun.},
	number = {1},
	pages = {390-394},
	title = {Progressive track recognition with a Kalman-like fitting procedure},
	volume = {57},
	year = {1989}}

@article{BILLOIR1990219,
	author = {P. Billoir and S. Qian},
	doi = {10.1016/0168-9002(90)91835-Y},
	issn = {0168-9002},
	journal = {Nucl. Instrum. Methods. Phys. Res. A},
	number = {1},
	pages = {219-228},
	title = {Simultaneous pattern recognition and track fitting by the Kalman filtering method},
	volume = {294},
	year = {1990}}

@article{MANKEL1997169,
	author = {Rainer Mankel},
	doi = {10.1016/S0168-9002(97)00705-5},
	issn = {0168-9002},
	journal = {Nucl. Instrum. Methods. Phys. Res. A},
	number = {2},
	pages = {169-184},
	title = {A concurrent track evolution algorithm for pattern recognition in the HERA-B main tracking system},
	volume = {395},
	year = {1997}}

@misc{thecepcstudygroup2018cepc_v1,
	archiveprefix = {arXiv},
	author = {{The CEPC Study Group}},
	eprint = {1809.00285},
	primaryclass = {physics.acc-ph},
	title = {CEPC Conceptual Design Report: Volume 1 - Accelerator},
	year = {2018}}

@misc{thecepcstudygroup2018cepc_v2,
	archiveprefix = {arXiv},
	author = {{The CEPC Study Group}},
	eprint = {1811.10545},
	primaryclass = {hep-ex},
	title = {CEPC Conceptual Design Report: Volume 2 - Physics \& Detector},
	year = {2018}}

@misc{abe2010belle,
	archiveprefix = {arXiv},
	author = {T. Abe and I. Adachi and K. Adamczyk and S. Ahn and H. Aihara and K. Akai and M. Aloi and L. Andricek and K. Aoki and Y. Arai and A. Arefiev and K. Arinstein and Y. Arita and D. M. Asner and V. Aulchenko and T. Aushev and T. Aziz and A. M. Bakich and V. Balagura and Y. Ban and E. Barberio and T. Barvich and K. Belous and T. Bergauer and V. Bhardwaj and B. Bhuyan and S. Blyth and A. Bondar and G. Bonvicini and A. Bozek and M. Bracko and J. Brodzicka and O. Brovchenko and T. E. Browder and G. Cao and M. -C. Chang and P. Chang and Y. Chao and V. Chekelian and A. Chen and K. -F. Chen and P. Chen and B. G. Cheon and C. -C. Chiang and R. Chistov and K. Cho and S. -K. Choi and K. Chung and A. Comerma and M. Cooney and D. E. Cowley and T. Critchlow and J. Dalseno and M. Danilov and A. Dieguez and A. Dierlamm and M. Dillon and J. Dingfelder and R. Dolenec and Z. Dolezal and Z. Drasal and A. Drutskoy and W. Dungel and D. Dutta and S. Eidelman and A. Enomoto and D. Epifanov and S. Esen and J. E. Fast and M. Feindt and M. Fernandez Garcia and T. Fifield and P. Fischer and J. Flanagan and S. Fourletov and J. Fourletova and L. Freixas and A. Frey and M. Friedl and R. Fruehwirth and H. Fujii and M. Fujikawa and Y. Fukuma and Y. Funakoshi and K. Furukawa and J. Fuster and N. Gabyshev and A. Gaspar de Valenzuela Cueto and A. Garmash and L. Garrido and Ch. Geisler and I. Gfall and Y. M. Goh and B. Golob and I. Gorton and R. Grzymkowski and H. Guo and H. Ha and J. Haba and K. Hara and T. Hara and T. Haruyama and K. Hayasaka and K. Hayashi and H. Hayashii and M. Heck and S. Heindl and C. Heller and T. Hemperek and T. Higuchi and Y. Horii and W. -S. Hou and Y. B. Hsiung and C. -H. Huang and S. Hwang and H. J. Hyun and Y. Igarashi and C. Iglesias and Y. Iida and T. Iijima and M. Imamura and K. Inami and C. Irmler and M. Ishizuka and K. Itagaki and R. Itoh and M. Iwabuchi and G. Iwai and M. Iwai and M. Iwasaki and M. Iwasaki and Y. Iwasaki and T. Iwashita and S. Iwata and H. Jang and X. Ji and T. Jinno and M. Jones and T. Julius and T. Kageyama and D. H. Kah and H. Kakuno and T. Kamitani and K. Kanazawa and P. Kapusta and S. U. Kataoka and N. Katayama and M. Kawai and Y. Kawai and T. Kawasaki and J. Kennedy and H. Kichimi and M. Kikuchi and C. Kiesling and B. K. Kim and G. N. Kim and H. J. Kim and H. O. Kim and J. -B. Kim and J. H. Kim and M. J. Kim and S. K. Kim and K. T. Kim and T. Y. Kim and K. Kinoshita and K. Kishi and B. Kisielewski and K. Kleese van Dam and J. Knopf and B. R. Ko and M. Koch and P. Kodys and C. Koffmane and Y. Koga and T. Kohriki and S. Koike and H. Koiso and Y. Kondo and S. Korpar and R. T. Kouzes and Ch. Kreidl and M. Kreps and P. Krizan and P. Krokovny and H. Krueger and A. Kruth and W. Kuhn and T. Kuhr and R. Kumar and T. Kumita and S. Kupper and A. Kuzmin and P. Kvasnicka and Y. -J. Kwon and C. Lacasta and J. S. Lange and I. -S. Lee and M. J. Lee and M. W. Lee and S. -H. Lee and M. Lemarenko and J. Li and W. D. Li and Y. Li and J. Libby and A. Limosani and C. Liu and H. Liu and Y. Liu and Z. Liu and D. Liventsev and A. Lopez Virto and Y. Makida and Z. P. Mao and C. Marinas and M. Masuzawa and D. Matvienko and W. Mitaroff and K. Miyabayashi and H. Miyata and Y. Miyazaki and T. Miyoshi and R. Mizuk and G. B. Mohanty and D. Mohapatra and A. Moll and T. Mori and A. Morita and Y. Morita and H. -G. Moser and D. Moya Martin and T. Mueller and D. Muenchow and J. Murakami and S. S. Myung and T. Nagamine and I. Nakamura and T. T. Nakamura and E. Nakano and H. Nakano and M. Nakao and H. Nakazawa and S. -H. Nam and Z. Natkaniec and E. Nedelkovska and K. Negishi and S. Neubauer and C. Ng and J. Ninkovic and S. Nishida and K. Nishimura and E. Novikov and T. Nozaki and S. Ogawa and K. Ohmi and Y. Ohnishi and T. Ohshima and N. Ohuchi and K. Oide and S. L. Olsen and M. Ono and Y. Ono and Y. Onuki and W. Ostrowicz and H. Ozaki and P. Pakhlov and G. Pakhlova and H. Palka and H. Park and H. K. Park and L. S. Peak and T. Peng and I. Peric and M. Pernicka and R. Pestotnik and M. Petric and L. E. Piilonen and A. Poluektov and M. Prim and K. Prothmann and K. Regimbal and B. Reisert and R. H. Richter and J. Riera-Babures and A. Ritter and A. Ritter and M. Ritter and M. Roehrken and J. Rorie and M. Rosen and M. Rozanska and L. Ruckman and S. Rummel and V. Rusinov and R. M. Russell and S. Ryu and H. Sahoo and K. Sakai and Y. Sakai and L. Santelj and T. Sasaki and N. Sato and Y. Sato and J. Scheirich and J. Schieck and C. Schwanda and A. J. Schwartz and B. Schwenker and A. Seljak and K. Senyo and O. -S. Seon and M. E. Sevior and M. Shapkin and V. Shebalin and C. P. Shen and H. Shibuya and S. Shiizuka and J. -G. Shiu and B. Shwartz and F. Simon and H. J. Simonis and J. B. Singh and R. Sinha and M. Sitarz and P. Smerkol and A. Sokolov and E. Solovieva and S. Stanic and M. Staric and J. Stypula and Y. Suetsugu and S. Sugihara and T. Sugimura and K. Sumisawa and T. Sumiyoshi and K. Suzuki and S. Y. Suzuki and H. Takagaki and F. Takasaki and H. Takeichi and Y. Takubo and M. Tanaka and S. Tanaka and N. Taniguchi and E. Tarkovsky and G. Tatishvili and M. Tawada and G. N. Taylor and Y. Teramoto and I. Tikhomirov and K. Trabelsi and T. Tsuboyama and K. Tsunada and Y. -C. Tu and T. Uchida and S. Uehara and K. Ueno and T. Uglov and Y. Unno and S. Uno and P. Urquijo and Y. Ushiroda and Y. Usov and S. Vahsen and M. Valentan and P. Vanhoefer and G. Varner and K. E. Varvell and P. Vazquez and I. Vila and E. Vilella and A. Vinokurova and J. Visniakov and M. Vos and C. H. Wang and J. Wang and M. -Z. Wang and P. Wang and A. Wassatch and M. Watanabe and Y. Watase and T. Weiler and N. Wermes and R. E. Wescott and E. White and J. Wicht and L. Widhalm and K. M. Williams and E. Won and H. Xu and B. D. Yabsley and H. Yamamoto and H. Yamaoka and Y. Yamaoka and M. Yamauchi and Y. Yin and H. Yoon and J. Yu and C. Z. Yuan and Y. Yusa and D. Zander and M. Zdybal and Z. P. Zhang and J. Zhao and L. Zhao and Z. Zhao and V. Zhilich and P. Zhou and V. Zhulanov and T. Zivko and A. Zupanc and O. Zyukova},
	eprint = {1011.0352},
	primaryclass = {physics.ins-det},
	title = {Belle II Technical Design Report},
	year = {2010}}

@article{Borissov_2015,
	doi = {10.1088/1742-6596/664/7/072041},
	url = {https://doi.org/10.1088/1742-6596/664/7/072041},
	year = 2015,
	month = {dec},
	publisher = {{IOP} Publishing},
	volume = {664},
	number = {7},
	pages = {072041},
	author = {G Borissov and D Casper and K Grimm and S Pagan Griso and L Egholm Pedersen and K Prokofiev and M Rudolph and A Wharton},
	title = {{ATLAS} strategy for primary vertex reconstruction during Run-2 of the {LHC}},
	journal = {Journal of Physics: Conference Series}
}
